\documentclass{nature}

\usepackage{mathtools}
\usepackage{siunitx}
\usepackage{textcomp}
\usepackage{float}
\usepackage{ulem}

\usepackage{braket}
\usepackage{amsmath}
\usepackage{amsfonts}
\usepackage{amssymb}
\usepackage{extarrows}
\usepackage{mathrsfs} 

\usepackage{cancel}
\usepackage{graphicx}
\usepackage{color}
\usepackage{pagecolor}

\usepackage[margin=1.0in]{geometry}
\usepackage{authblk}
\usepackage{longtable}
\usepackage{mdwlist}
\usepackage{parskip}
\usepackage{tcolorbox} 

\usepackage{caption}

\DeclareMathAlphabet{\mathtensor}{OT1}{cmss}{bx}{n}

\usepackage{tabularx}
\usepackage{multirow}
\usepackage{colortbl}
\usepackage{soul}

\title{Enantiospecificity in NMR Enabled by Chirality-Induced Spin Selectivity}

\author{T. Georgiou\textsuperscript{1}, J.L. Palma\textsuperscript{2}, V. Mujica\textsuperscript{3}, S. Varela\textsuperscript{4},
M. Galante\textsuperscript{3}, 
V. Santamar\'{i}a Garc\'{i}a\textsuperscript{5,6},
L. Mboning\textsuperscript{7}, R.N. Schwartz\textsuperscript{8}, G. Cuniberti\textsuperscript{3,9}, L.-S. Bouchard\textsuperscript{1,7,10,11}}

\makeatletter
\let\saved@includegraphics\includegraphics
\AtBeginDocument{\let\includegraphics\saved@includegraphics}
\renewenvironment*{figure}{\@float{figure}}{\end@float}
\makeatother

\renewcommand{\hl}[1]{#1}

\begin{document}

\pagecolor{white}

\maketitle

\begin{affiliations}
\item Molecular Biology Interdepartmental Program (MBIDP), The Molecular Biology Institute, University of California Los Angeles, 611 Charles E. Young Drive East, Los Angeles, CA 90095-1570, USA.
\item Department of Chemistry, Penn State University, 2201 University Drive, Lemont Furnace, PA 15456
\item School of Molecular Sciences, Arizona State University, 551 E University Dr, Tempe, AZ 85281, USA.
\item Institute for Materials Science and Max Bergmann Center of Biomaterials, TU Dresden, 01062 Dresden, Germany.
\item Department of Mechanical Engineering, Massachusetts Institute of Technology, 77 Massachusetts Avenue, Cambridge, Massachusetts 02139, US
\item Tecnologico de Monterrey, Escuela de Ingeniería y Ciencias, Ave. Eugenio Garza Sada 2501, Monterrey 64849, Mexico
\item Department of Chemistry and Biochemistry, University of California Los Angeles, 607 Charles E. Young Dr. East, Los Angeles, CA 90095-1569, USA.
\item Department of Electrical and Computer Engineering, University of California Los Angeles, 420 Westwood Plaza, Los Angeles, CA 90095-1594, USA.
\item Dresden Center for Computational Materials Science (DCMS), TU Dresden, 01062 Dresden, Germany.
\item California NanoSystems Institute, University of California Los Angeles, 607 Charles E. Young Dr. East, Los Angeles, CA 90095-1569, USA.
\item Department of Bioengineering, University of California Los Angeles, 607 Charles E. Young Dr. East, Los Angeles, CA 90095-1569, USA. \\

\end{affiliations}
\par 

{\bf Abstract} 


Spin polarization in chiral molecules is a magnetic molecular response associated with electron transport and enantioselective bond polarization that occurs even in the absence of an external magnetic field. An unexpected finding by Santos and co-workers reported enantiospecific NMR responses in solid-state cross-polarization (CP) experiments, suggesting a possible additional contribution to the indirect nuclear spin-spin coupling in chiral molecules induced by bond polarization in the presence of spin-orbit coupling.  Herein we provide a theoretical treatment for this phenomenon, presenting an effective spin-Hamiltonian for helical molecules like DNA and density functional theory (DFT) results on amino acids that confirm the dependence of J-couplings on the choice of enantiomer.  The connection between nuclear spin dynamics and chirality could offer insights for molecular sensing and quantum information sciences. These results establish NMR as a potential tool for chiral discrimination without external agents.

{\bf Introduction} 

Chirality is a structural property integral to various chemical and biological processes. It plays a significant role in diverse research fields, including asymmetric synthesis and drug design. The investigation of electron transport, electron transfer, and photo-ionization in chiral molecules has led to the discovery of the Chiral-Induced Spin Selectivity (CISS) effect. This discovery was made by Naaman and colleagues in 1999~\cite{ray1999asymmetric}. Electrons traversing chiral molecules experience a momentum-dependent effective magnetic field due to spin-orbit coupling (SOC). This leads to spin selectivity and polarization under conditions where time-reversal symmetry is not conserved. The CISS effect not only provides new perspectives on electron transport but also actively modifies it by limiting backscattering and altering electron flow rules. Different spin components have distinct transmission probabilities, resulting in unique distance and temperature dependencies. These dependencies are governed by the interactions between electrons and phonons, as well as electron-electron interactions. When chiral molecules interact with other structures, charge polarization occurs, resulting in distinct spin orientations with respect to the electric and magnetic fields. The CISS effect provides insights into electron transfer in chiral molecules and has implications for chemical reactions and biorecognition. It also emphasizes the spin-filtering capabilities of chiral structures, including DNA and peptides; for additional information, see Ref.~\cite{bib:reviewCISSacsnano}.

Cross polarization (CP), although seemingly unrelated, is a key technique in solid-state NMR used primarily to enhance the signals of less abundant nuclei with low gyromagnetic ratios, such as $^{31}\text{P}$, $^{13}\text{C}$ and $^{15}\text{N}$. Magnetization is transferred from more abundant nuclei like $^1\text{H}$ using an RF field. The Hartmann-Hahn condition enables this transfer to occur in the presence of pairwise spin couplings. The primary mechanism involves nuclear magnetic dipole interactions, which are not influenced by chirality. CP is sensitive to the distance between nuclei and the dynamics of participating molecules or functional groups. It is thus valuable for identifying linked nuclei and observing molecular dynamics in solid structures. When augmented with techniques like magic-angle spinning (MAS), CP's sensitivity to molecular geometry and dynamics is enhanced.

Another avenue for nuclear spin-spin coupling leading to CP is indirect coupling via electrons~\cite{bib:ernstJCP}. Two notable papers by Santos and colleagues~\cite{santos2018chirality,san2020enantiospecific} reported enantiospecific NMR responses in CP MAS solid-state NMR experiments. While through-space dipolar coupling is likely the dominant contribution to the transfer of polarization in their experiments, it does not explain enantioselectivity in the measurement. While $J$ couplings have been known to generate CP for quite some time~\cite{bib:ernstJCP}, these findings were unexpected, as there are no established links between nuclear magnetism and molecular chirality. The authors proposed a mechanism to explain these observations in terms of the CISS effect giving rise to or influencing the indirect $J$-coupling. In this scenario, bond polarization via a chiral center or a helical structure could lead to distinct contributions from different enantiomers. Such a mechanism would create a unique magnetic environment for the nuclei participating in these CP experiments. Despite these considerations, the exact mechanism for chirality-dependent indirect coupling remains unclear. These results generated controversy in the literature~\cite{bib:rossini}.  For example, particle size, sample preparation and impurity content have been argued to contribute to the observed effect~\cite{bib:rossini}.

In this study, we investigate the coupling between nuclear spins and electronic states in chiral molecules. We find that remote nuclear spins can couple effectively via conduction electrons, thereby creating a mechanism for chirality-dependent indirect spin-spin coupling between nuclei. A theoretical framework is introduced to  assess the plausibility of potential spin-dependent mechanisms responsible for this effect and their role in probing enantioselectivity in CP.  Our theoretical analysis addresses the DNA helix. We also present a more quantitative analysis via DFT, which suggests an underlying mechanism for the experimental observations of Santos and colleagues on amino acids~\cite{santos2018chirality,san2020enantiospecific}. These results help establish the plausibility of enantioselective bond polarization-mediated indirect nuclear spin-spin couplings involving either a chiral center or a helical structure. This uniquely described mechanism bridges nuclear spins and the CISS effect, augmenting our understanding of chiral molecular systems. The study thus contributes to our understanding of chirality-induced phenomena, and to the possible development of applications in NMR-based sensing and quantum information processing at the molecular level.

{\bf Results}

{\bf NMR and Chirality.} NMR, as described by D. Buckingham~\cite{buckingham2004chirality}, is ``blind'' to chirality since none of its standard parameters appear to be sensitive to it. Enantiomers display identical NMR spectra in an achiral environment. Thus, differentiating enantiomers using standard NMR techniques in the absence of a chiral resolvent or probe is challenging. We are aware of three methods to indirectly detect chirality by NMR: (1) Chiral Derivatizing Agents (CDAs)~\cite{bib:chiralderivatizingagents,silva2017recent}: These compounds react with a chiral substrate to produce diastereomers, which have distinct NMR spectra. For example, when a chiral alcohol reacts with a chiral derivatizing agent like Mosher's acid, the resultant diastereomeric esters can be distinguished by their NMR chemical shifts, revealing the absolute configuration of the alcohol. (2) Chiral Solvents~\cite{bib:chiralsolventsNMR,silva2017recent}: In these solvents, enantiomers present slight differences in their NMR spectra due to unique interactions with the chiral environment. These differences can help deduce enantiomeric excess and sometimes the absolute configuration. (3) Chiral Lanthanide Shift Reagents~\cite{parker2004excitement,parker1991nmr,parker2002being,morrill1986lanthanide}: These metal complexes can cause shifts in the NMR spectra of chiral compounds. Lanthanide ions, especially Eu, Yb, and Dy, have been used to distinguish the NMR signals of enantiomers by forming diastereomeric complexes detectable due to their differing chemical shifts.  However, each of these methods has limitations.  Mainly, they are indirect molecular effects that rely on external agents. To determine chirality conclusively, complementary analytical methods are often necessary. Alternatively, Buckingham, Harris, Jameson, and colleagues have proposed using electric fields for chiral discrimination~\cite{garbacz2016chirality,buckingham2014communication,buckingham2006direct,buckingham2004chirality,walls2008measuring,harris2006note}, though this remains to be demonstrated in experiments.

Indirect NMR methods to distinguish enantiomers are less accessible and often more cumbersome than non-NMR methods such as chiral chromatography, high-performance liquid chromatography, gas chromatography, capillary electrophoresis, circular dichroism spectroscopy, optical rotatory dispersion, X-ray crystallography or vibrational circular dichroism. The development of a method to directly probe the chirality of a molecule using NMR, without the reliance on external agents or indirect techniques, would be an important development in the fields of stereochemistry and analytical chemistry. 
Direct enantiomeric detection via NMR would uniquely combine non-destructive, quantitative capabilities with  reproducibility, while concurrently bypassing the need for reactive chiral derivatizing agents, chiral solvents, and chromatography columns.

{\bf Cross-Polarization and Enantiospecificity.} The experiments performed in Refs.~\cite{santos2018chirality,san2020enantiospecific} demonstrated the existence of an enantiospecific response in CP. This effect was also observed recently by Bryce and co-workers~\cite{bib:bryce}, but the authors argued that experimental artifacts such as  particle size could contribute. Rossini and colleagues~\cite{bib:rossini} suggested that impurities, crystallization, and particle size effects likely contribute to the observation. Although such factors may influence the measurements, the data presented in~\cite{bib:bryce,bib:rossini} does not rule out the contribution from CISS. As to CP, it is the bread-and-butter of solid-state NMR thanks to its ability to  dramatically increase the sensitivity of experiments involving nuclei in low concentrations.  CP is a technique where magnetization is transferred from an abundant, high gamma nucleus ($\mathbf{I}_1$) to a low gamma, dilute nucleus ($\mathbf{I}_2$) that is coupled to the $\mathbf{I}_1$ spin bath during a certain ``contact'' period~\cite{pines1973proton,pines1972proton1,pines1972proton2}. During the contact time, radiofrequency (r.f.) fields for both $\mathbf{I}_1$ and $\mathbf{I}_2$ are turned on.   Usually, the dominant magnetic coupling between pairs of nuclei is due to the magnetic dipole interaction.  In the simplest solid-state NMR experiment, the enhanced magnetization of the dilute isotope is then detected while the abundant protons, or any other reference nuclei, are decoupled. The maximum gain in sensitivity is equal to the ratio of gyromagnetic ratios between the two nuclei (e.g., for $^1$H and $^{13}$C this ratio is approximately 4:1; a factor of 4 implies 16-fold SNR gains).  

The method of using heteronuclear double resonance to transfer coherence between nuclei in a two-spin system was introduced by Hartmann and Hahn~\cite{hartmann1962nuclear,pines1973proton,pines1972proton1,pines1972proton2} and has since become widely employed in solid-state NMR.  It is possible to do highly selective recoupling among nuclei~\cite{baldus1998cross,lewandowski2007proton}. Spectroscopists can also modulate the amplitude of spin-locking pulses to enhance CP dynamics, perform Lee-Goldburg decoupling to reduce homonuclear proton couplings during spin-locking, or apply multiple-quantum CP to half-integer quadrupole systems~\cite{ashbrook1998multiple,ashbrook2000multiple,rovnyak2000multiple}. CP is a highly useful experiment that facilitates high-resolution NMR in the solid state encompassing key principles of dipolar coupling (decoupling/recoupling) and MAS~\cite{rovnyak2008tutorial}.

\begin{figure}[H]
\begin{center}
\includegraphics[width=0.78\textwidth]{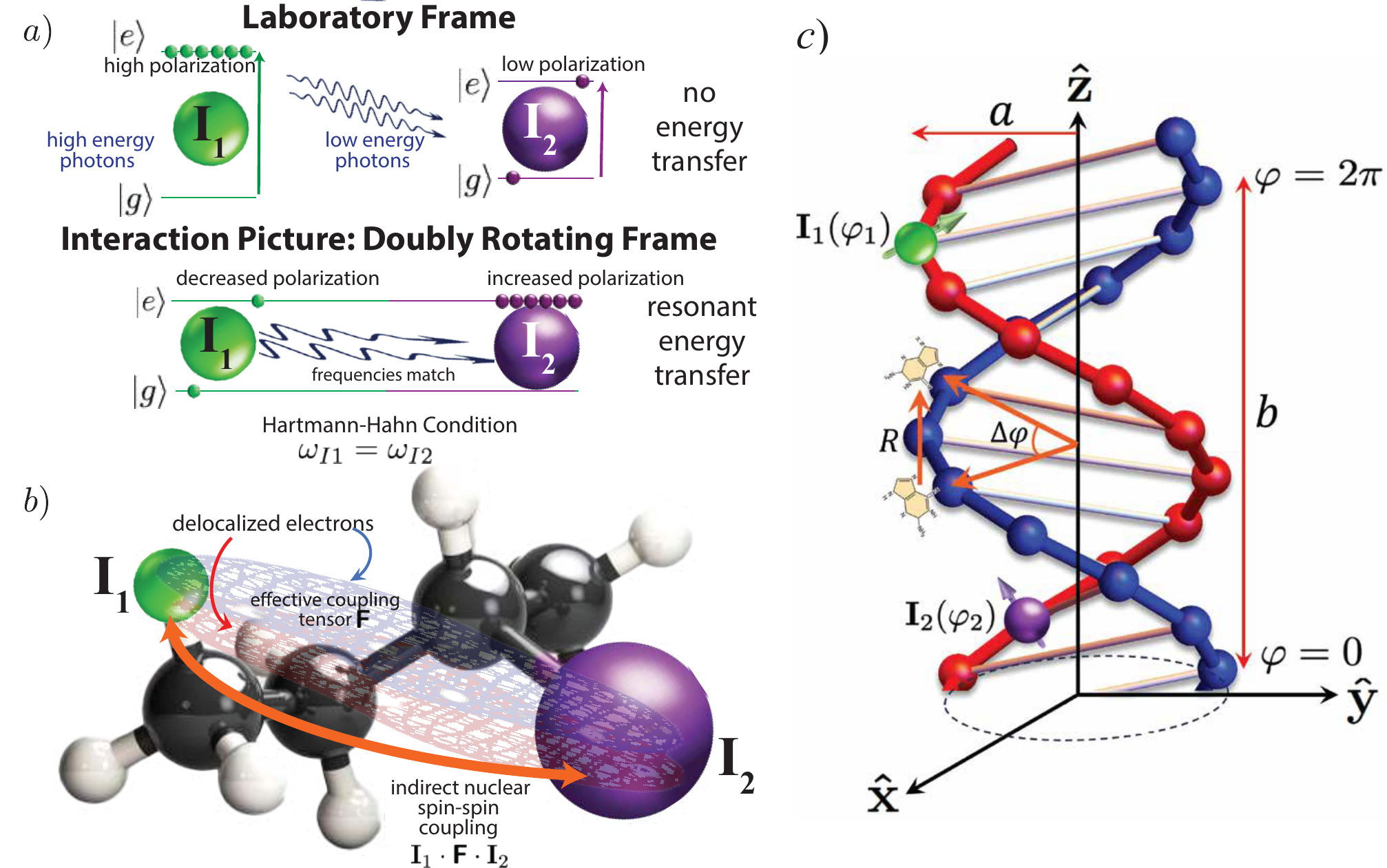}
\end{center}
\caption{Indirect nuclear spin-spin ($J$) coupling enables cross-polarization in NMR.  The CISS effect gives rise to delocalized conduction bands.  Delocalized electrons can in turn mediate indirect nuclear spin-spin couplings.  (a) In the cross-polarization experiment of solid-state NMR energy transfers between heteronuclei are forbidden in the lab frame.  Application of a bichromatic RF field oscillating at the resonance frequencies of both nuclei, enables energy transfer.   At the Hartmann-Hahn condition $\gamma_{I1} B_{1,I1} = \gamma_{I2} B_{1,I2}$, resonant energy transfer will lead to transfer of polarization from the cold to hot spin systems.  (b) Indirect spin-spin coupling between two nuclei is mediated by conduction electrons. (c) Model for DNA helix, helicoidal coordinates $(a,b,\varphi)$ and two nuclear spins $\mathbf{I}_1$, $\mathbf{I}_2$ and their corresponding positions $\varphi_1$, $\varphi_2$ along the helix. $R$ is the  distance whereas $\Delta \varphi$ is the angle between consecutive nucleotides. $a$ is the helix radius and $b$ is its pitch. }
\label{fig:cp}
\end{figure}

The working principle of CP is illustrated in Fig.~\ref{fig:cp}a.  If two nuclear spins $\mathbf{I}_1$ and $\mathbf{I}_2$ with gyromagnetic ratios $\gamma_{I1}$ and $\gamma_{I2}$, respectively, are placed in an external magnetic field $B_0$, they will be able to absorb r.f. photons at frequencies $\gamma_{I1} B_0$ and $\gamma_{I2} B_0$, respectively, according to the Zeeman effect. They will not be able to exchange energy spontaneously, since the two frequencies $\gamma_{I1} B_0$ and $\gamma_{I2} B_0$ are different. If instead a bimodal oscillating r.f. field is applied at these two frequencies, with amplitudes such that $\omega_{I1} = \omega_{I2}$ (Hartmann-Hahn condition~\cite{hartmann1962nuclear}), where $\omega_{I1} = \gamma_{I1} B_{1,I1}$ and  $\omega_{I2} = \gamma_{I2} B_{1,I2}$.  In the ``doubly rotating frame'' generated by these frequencies both nuclear spins $\mathbf{I}_1$ and $\mathbf{I}_2$ appear stationary.   Photons can now be absorbed by either of these spins at the same frequency $\omega_{I1} = \omega_{I2}$, the condition for resonant energy transfer.

The CP experiment is often described using the concept of spin temperature~\cite{stejskal1994high,abragam1961principles,goldman1970spin}. The abundant spin system is prepared with an artificially low temperature. This is typically done by applying a $\pi/2$-pulse on the abundant nuclei, followed by a spin-locking field~\cite{stejskal1994high}.  One then allows the dilute system to come into thermal contact with the cold system of abundant spins.   Contact is typically established through the magnetic dipole-dipole interaction between nuclei. Heat flows from the sparse spin system to the cold abundant spins, which produces a drop in temperature of the sparse spins.  Physically, we observe resonance energy transfer if the natural frequencies of the two systems are close.   This was Hahn's ingenious concept~\cite{hartmann1962nuclear}.   This experiment requires the heat capacity of the abundant system to be larger than that of dilute spins.  In the context of such experiments, to say that the spin temperature has dropped is nearly equivalent to the statement that population difference between the ground $\ket{g}$ and excited $\ket{e}$ states is increased, which leads to an increased sensitivity of the NMR experiment.

In the references~\cite{santos2018chirality,san2020enantiospecific} different efficiencies of CP were obtained depending on the choice of enantiomer.   The existence of an enantiospecific bilinear coupling (see Fig.~\ref{fig:cp}b) of the form $\mathbf{I}_1 \cdot \mathtensor{F} \cdot \mathbf{I}_2$ was postulated in those papers, where the coupling tensor $\mathtensor{F}$ depends on Rashba SOC, an interaction which is itself enantiospecific.  Herein we argue that bond polarization and SOC provides a possible mechanism to mediate the interaction between two nuclear spins through the creation of enantiospecific delocalized electron conduction bands which, in turn, enable these electrons to couple to both nuclear spins simultaneously via magnetic dipole interaction.

{
\scriptsize
\begin{longtable}{|p{1.1in}|p{1.0in}|p{0.8in}|p{0.45in}|p{1.4in}|p{0.3in}|}
\hline
  \cellcolor{black}\textcolor{white}{Molecule / System} & \cellcolor{black}\textcolor{white}{Experiment}   & \cellcolor{black}\textcolor{white}{$\delta$ (ppm)} & \cellcolor{black}\textcolor{white}{CT (ms)}  & \cellcolor{black}\textcolor{white}{ I(\textsf{D})/I(\textsf{L}) } & \cellcolor{black}\textcolor{white}{Ref.}  \\ \hline
 \textsf{D} vs  \textsf{L}-aspartic acid & $^{15}$N $\{ {}^1 \mbox{H} \}$ CP-MAS  & 0.14  & 2 & 1.95 & \cite{santos2018chirality} \\ \hline
  \textsf{D} vs  \textsf{L}-cysteine & $^{15}$N  $\{ {}^1 \mbox{H} \}$ CP-MAS  & 8.92  & 2 & 1.88 & \cite{santos2018chirality} \\ \hline
  \textsf{D} vs  \textsf{L}-phenylalanine & $^{15}$N  $\{ {}^1 \mbox{H} \}$ CP-MAS  & 3.1  & 0.5 & 1.3 & \cite{santos2018chirality} \\ \hline
  \textsf{D} vs  \textsf{L}-phenylglycine & $^{15}$N  $\{ {}^1 \mbox{H} \}$ CP-MAS  & 10.1; 1.7  & 1.5 & 1.05; 1.07 & \cite{santos2018chirality}  \\ \hline
  \textsf{D} vs  \textsf{L}-threonine & $^{15}$N  $\{ {}^1 \mbox{H} \}$ CP-MAS  & -0.64  & 2 & 1.09 & \cite{santos2018chirality} \\ \hline
  \textsf{D} vs  \textsf{L}-tyrosine & $^{15}$N  $\{ {}^1 \mbox{H} \}$ CP-MAS  & 0.80  & 1.5 & 1.14 & \cite{santos2018chirality}  \\ \hline
  \textsf{D} vs  \textsf{L}-serine & $^{15}$N  $\{ {}^1 \mbox{H} \}$ CP-MAS  & -3.01  & 2.0 & 1.05 & \cite{santos2018chirality} \\ \hline
  \textsf{D} vs  \textsf{L}-valine & $^{15}$N  $\{ {}^1 \mbox{H} \}$ CP-MAS  & -1.81  & 0.5 & 1.22 & \cite{santos2018chirality} \\ \hline
  \textsf{D} vs  \textsf{L}-TAR & $^{13}$C  $\{ {}^1 \mbox{H} \}$ DP-MAS & 176; 171; 74; 72   & 0.5 & 0.943; 0.982; 0.971; 0.985 & \cite{san2020enantiospecific} \\ \hline
  \textsf{D} vs  \textsf{L}-TAR & $^{13}$C  $\{ {}^1 \mbox{H} \}$ CP-MAS & 176; 171; 74; 72   & 0.5 & 1.15; 1.17; 1.19; 1.23 & \cite{san2020enantiospecific} \\ \hline
 \textsf{D} vs  \textsf{L}-1 & $^{13}$C  $\{ {}^1 \mbox{H} \}$ DP-MAS & 189-173; 80-69   & 0.5 & 1.06; 1.03 & \cite{san2020enantiospecific} \\ \hline
  \textsf{D} vs  \textsf{L}-1 & $^{13}$C  $\{ {}^1 \mbox{H} \}$ CP-MAS & 189-173; 80-69   & 0.5 & 1.28; 1.38 & \cite{san2020enantiospecific} \\ \hline
\caption{Summary of solid-state NMR experimental results from Refs.~\cite{santos2018chirality} and~\cite{san2020enantiospecific} on CP of enantiomers of several different molecular structures. DP-MAS: direct polarization MAS.  Chemical shift of the resonance ($\delta$), contact time (CT), I(\textsf{D})/I(\textsf{L}) intensity ratio. 
\textsf{D}- and \textsf{L}-TAR refers to organic ligands  \textsf{D}- and  \textsf{L}-tartaric acids.
 \textsf{D}- and  \textsf{L}-1 refer to 3D metal-organic frameworks \{[Y$_2$($\mu_4$-\textsf{L}-TAR)$_2$($\mu$-\textsf{L}-TAR)(H$_2$O)$_2$]$\cdot$4H$_2$O\}$_n$ (\textsf{L}-1) and \{[Y$_2$($\mu_4$-\textsf{D}-TAR)$_2$($\mu$-\textsf{D}-TAR)(H$_2$O)$_2$]$\cdot$4H$_2$O\}$_n$ (\textsf{D}-1).  (See Ref.~\cite{san2020enantiospecific} for details.)  For CP-MAS the rule I(\textsf{D})/I(\textsf{L}) $>1$ is observed.  \label{tab:resultssummary} }
\end{longtable}
}

A summary of all known CP results on enantiomers published to date (see Refs.~\cite{santos2018chirality,san2020enantiospecific}) is shown in Table~\ref{tab:resultssummary}. According to the traditional view, NMR parameters are not supposed to depend on the handedness of enantiomers; and therefore, the ratio I(\textsf{D})/I(\textsf{L}) should be 1.    Instead, 
for all CP-MAS results a clear trend I(\textsf{D})/I(\textsf{L}) $>1$ is observed indicating that the 
\textsf{D} enantiomer consistently inherits more polarization compared to the \textsf{L} enantiomer.  This goes against all known mechanisms describing nuclear spin interactions in a diamagnetic molecule.

Santos and co-workers ~\cite{santos2018chirality,san2020enantiospecific} postulated the existence of an effective nuclear spin-spin interaction mediated by SOC because the effective strength of the SOC interaction in molecules exhibiting CISS is enantiospecific (Fig.~\ref{fig:cp}b), leading to different transmission probabilities for the two values of electronic spin.  However, the precise mechanism remains elusive, as SOC itself does not couple directly to nuclear spins, as far as fundamental interactions are concerned.   SOC only directly affects the electronic wavefunction.  We must then turn our attention to the nature of effective interactions affecting these nuclear spins. The hyperfine interaction defines the manner in which nuclear spins couple to electron spins.  From this emerges a possible mechanism.  The electron-nuclear hyperfine interaction is made up of three contributions: Fermi contact, electron-nuclear dipole interaction, and nuclear spin-electron orbital angular momentum. The first two interactions provide a possible mechanism for spin-spin coupling, albeit indirectly. Indirect couplings in NMR, also known as J couplings, were discovered independently by Hahn and Maxwell~\cite{hahn1951chemical} as well as McCall, Slichter and Gutowski~\cite{gutowsky1951coupling}.  While initially discovered in liquids, Slichter~\cite{slichter1996principles} has presented a theory for the solid state.  For 3D Bloch wavefunctions in a solid the case of the Fermi contact interaction is discussed in Ref.~\cite{slichter1996principles} whereas the case of the dipole-dipole interaction is discussed in Bloembergen and Rowland~\cite{bloembergen1955nuclear}.  These theories, however, do not incorporate in any way the effects of chirality.
We propose instead to investigate the following two pathways:
$$ \mbox{nuclear spin 1} ~ \quad ~  \xlongleftrightarrow{\text{dipole-dipole}}  ~ \quad ~ \stackrel{\text{(wavefunction is enantiospecific)}}{\text{conduction electron spins}}   ~ \quad ~  \xlongleftrightarrow{\text{dipole-dipole}} ~ \quad ~ \mbox{nuclear spin 2} $$
and
$$ \mbox{nuclear spin 1} ~ \quad ~  \xlongleftrightarrow{\text{Fermi contact}}  ~ \quad ~ \stackrel{\text{(wavefunction is enantiospecific)}}{\text{conduction electron spins}}   ~ \quad ~  \xlongleftrightarrow{\text{Fermi contact}} ~ \quad ~ \mbox{nuclear spin 2}. $$
A classic example of chiral molecule is the DNA helix.  DNA is also amenable to simple modeling.  
The hypothetical case of indirect coupling of nuclear spins $\mathbf{I}_1$ and $\mathbf{I}_2$  in a DNA molecule is illustrated in Fig.~\ref{fig:cp}c.

\begin{figure}[h!]
\begin{center}
\includegraphics[width=0.65\textwidth]{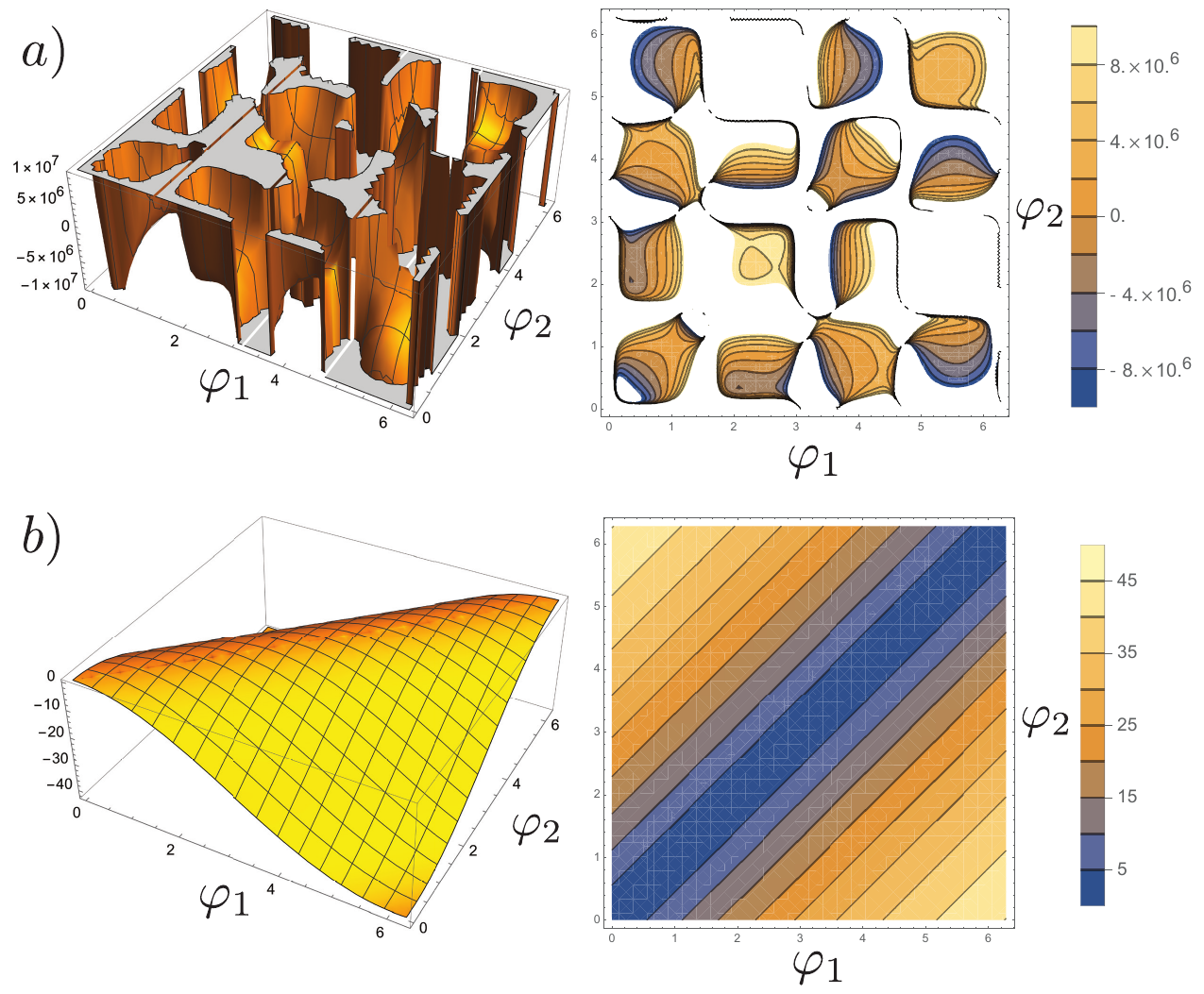}
\end{center}
\caption{Plots of indirect nuclear spin-spin coupling tensor component $\mathtensor{F}_{zz}$
as a function of the position of nucleus 1 and 2  along the DNA chain ($\varphi_1,\varphi_2 \in [0,2\pi]$, with 0 indicating the start and $2\pi$ the end of the helix).  Multiplication by 0.01 gives the coupling strength in Hz.  (a) Contribution from the magnetic dipole interaction.  Peak coupling strengths attain 100 kHz (white regions, right panel).  (b) Contribution from the Fermi contact interaction (values should be multiplied by 0.01 to get units of Hz).}
\label{fig:results}
\end{figure}

The key observation in the present work is that the electronic wavefunction in CISS differs from normal 3D Bloch wavefunctions (e.g.,~\cite{slichter1996principles}) in that it is enantiospecific~\cite{varela2016effective,lopez2022radiation}. Enantiospecificity is related to the SOC interaction and helicity, which takes into account the direction of electron propagation. Another difference is the 1D nature of helical molecules, giving rise to 1D wavefunctions in a band structure model~\cite{varela2016effective,lopez2022radiation}. The physics of one-dimensional systems involves unique mathematical considerations. In Supplementary Text S1 we present a detailed theoretical treatment of the indirect coupling between pairs of nuclear spins in a helical molecule based on spin-dependent mechanisms (electron-nuclear dipole-dipole, Fermi contact). The main result is that both interactions are sufficiently strong to cause observable CP.  The electron-nuclear dipolar contribution to the effective coupling tensor (derived in Supplementary Text S1) depends on chirality. Amplitude estimates are shown in Fig.~\ref{fig:results}a, where coupling strengths between pairs of nuclear spins (assumed to be protons for simplicity) can reach amplitudes that generate observing measurable effects by CP~\cite{bib:ernstJCP} for specific positions of the nuclear spins. The coupling strength depends on the position ($\varphi_1$, $\varphi_2$) of the nuclear spins along the helix.  We remark that this calculation should not be considered quantitative due to the one-dimensional nature of the problem, which leads to the emergence of divergences. This calculation should instead serve to establish the plausibility of the mechanism. As to the Fermi contact interaction, it is generally weaker than dipole-dipole (see Fig.~\ref{fig:results}b), yet sufficiently strong to produce measurable effects~\cite{bib:ernstJCP}. Weak Fermi contact interactions are generally due to low overlap of the electronic wavefunction at the site of the nuclei,  possibly due to a stronger contribution from $p$-wave character of the wavefunction~\cite{varela2016effective,lopez2022radiation} than $s$-wave~\cite{slichter1996principles}. However, as explained in SI for the case of high-field NMR the Fermi contact tensor is not enantioselective. The dipole-dipole term, on the other hand, is. This analysis applies to the DNA toy model only. The situation could be different for real chiral molecules and an independent analysis is warranted on a case-by-case basis.

We sketch the main steps of the derivation presented in SI.  An effective Hamiltonian is derived using second-order perturbation theory:
\begin{align}
\mathcal{H}_{eff} &=  \underbrace{  \left(\frac{2 \mu_0}{3} \right)^2 \gamma_I^2 \gamma_S^2 \hbar^4 \sum_j \mathbf{I}_1 \cdot \frac{ \braket{ 0 | \sum_l \mathbf{S}_l \delta^{(3)}(\mathbf{r}_l - \mathbf{R}_1) |j}  \braket{j| \sum_l \mathbf{S}_l \delta^{(3)}(\mathbf{r}_l - \mathbf{R}_2)|0}  }{ E_0 - E_j } \cdot \mathbf{I}_2 + c.c. }_{ \mathcal{H}_{eff}^{FC} }. \nonumber \\
& + \underbrace{  \left( \frac{\mu_0}{4\pi}\right)^2 \gamma_I^2 \gamma_S^2 \hbar^4 \, p.v. \sum_{\alpha,\alpha'} \sum_{\beta,\beta'}  \sum_j I_1^\alpha  \frac{ \braket{0 | \sum_l \frac{ \delta_{\alpha\beta} - 3 \hat{\tilde{r}}_{1l,\alpha} \hat{\tilde{r}}_{1l,\beta}}{ |\mathbf{R}_1-\mathbf{r}_l|^3} S_l^\beta | j } \braket{ j |\sum_l \frac{ \delta_{\alpha'\beta'} - 3 \hat{\tilde{r}}_{2l,\alpha'} \hat{\tilde{r}}_{2l,\beta'}}{ |\mathbf{R}_2-\mathbf{r}_l|^3}  S_l^{\beta'} |0}}{E_0-E_j} I_2^{\alpha'} + c.c.   }_{ \mathcal{H}_{eff}^{DD} }  \label{eq:heff}
\end{align}

The term on the first line describes the effects of the Fermi contact interaction ($\mathcal{H}_{eff}^{FC}$) whereas term on the second line, the effects of the dipole-dipole interaction ($\mathcal{H}_{eff}^{DD}$).  The Varela spinors \cite{varela2016effective,lopez2022radiation}, which were recently obtained by solving a minimal tight-binding model constructed from valence $s$ and $p$ orbitals of carbon atoms, describe the molecular orbitals of helical electrons in DNA molecules. These spinors can be used to compute the summations by considering them as the electronic states $\ket{j}$:
$$ \boldsymbol \psi_{n,s}^{\nu,\zeta} = \left[ \begin{matrix}
F_A e^{-i \varphi/2} \\
\zeta F_B^* e^{i \varphi/2}
\end{matrix} \right] e^{i \nu \tilde{n} \varphi}, \quad F_A = \frac{ \sqrt{s}}{2} ( s e^{i \theta/2} + e^{-i\theta/2} ), \quad F_B = \frac{\sqrt{s}}{2} ( s e^{ -i\theta/2} - e^{i\theta/2} ), $$
where $\tilde{n}$ is analogous to a wavenumber, $\varphi$ is the angular coordinate along the helix, $\theta$ depends on SOC and is the tilt of the spinor relative to the $z$ axis, $s=\pm 1$ is the electron spin orientation and $\zeta=\pm 1$ labels the enantiomer. By keeping track of $\zeta$ we can determine which contribution(s) depend on enantiomer. 
This leads to the result
\begin{equation}
 \mathcal{H}_{eff}^{DD} = \left( \frac{\mu_0}{4\pi}\right)^2 \gamma_I^2 \gamma_S^2 \sum_{n,n'} \sum_{\alpha,\beta} \sum_{\alpha',\beta'} I_1^\alpha  \frac{
M_{1,\beta}^{\alpha\beta}(\tilde{n}',\tilde{n})M_{2,\beta'}^{\alpha'\beta'}(\tilde{n},\tilde{n}')
}{|T| ( n'-n )    } I_2^{\alpha'} f(\tilde{n})[1-f(\tilde{n}')] + c.c. \label{eq:enan}
\end{equation}
where $\mu_0, \gamma_I, \gamma_S, |T|$ are constants, $f(\tilde{n})$ is a Fermi function, $I_1^\alpha$ are nuclear-spin operators (see SI) and explicit expressions for the matrices $M_{1,\beta}^{\alpha\beta}(\tilde{n}',\tilde{n})$'s are given in Supplementary (SI) section, equations~\ref{eq:Ms1}, ~\ref{eq:Ms2} and \ref{eq:Ms3}.

This expression for the indirect coupling is enantiospecific.  The effective Hamiltonian contains a product $M_{1,\beta}^{\alpha\beta}(\tilde{n}',\tilde{n})M_{2,\beta'}^{\alpha'\beta'}(\tilde{n},\tilde{n}')$.  Explicitly, this term is:
\begin{multline}
\sum_{\beta,\beta'} M_{1,\beta}^{\alpha\beta}(\tilde{n}',\tilde{n})M_{2,\beta'}^{\alpha'\beta'}(\tilde{n},\tilde{n}')=\bigl[ M_{1,x}^{\alpha,1}(\tilde{n}',\tilde{n}) + M_{1,y}^{\alpha,2}(\tilde{n}',\tilde{n}) + M_{1,z}^{\alpha,3}(\tilde{n}',\tilde{n}) \bigr]  \\
 \times  \bigl[  M_{2,x}^{\alpha',1}(\tilde{n},\tilde{n}') + M_{2,y}^{\alpha',2}(\tilde{n},\tilde{n}') + M_{2,z}^{\alpha',3}(\tilde{n},\tilde{n}') \bigr].
\end{multline}
While $M_{i,z}$ is independent of $\zeta$, both $M_{i,x}$ and $M_{i,y}$ depend linearly on $\zeta$ (see equations 2-4 in SI). The term $M_{1,z}^{\alpha,3}(\tilde{n}',\tilde{n}) M_{2,z}^{\alpha',3}(\tilde{n},\tilde{n}')$ does not depend on $\zeta$, since neither factor depend on $\zeta$.   Neither do
$M_{1,x}^{\alpha,1}(\tilde{n}',\tilde{n}) M_{2,x}^{\alpha',1}(\tilde{n},\tilde{n}')$ and $M_{1,y}^{\alpha,2}(\tilde{n}',\tilde{n}) M_{2,y}^{\alpha',2}(\tilde{n},\tilde{n}')$ since $\zeta^2=1$. On the other hand, terms such as $M_{1,z} M_{2,x}$ depend linearly on $\zeta$.  The effect of enantiomer handedness is to flip the sign of this term, leading to a change in the magnitude of the indirect coupling mediated by dipole-dipole interaction.  As explained in SI (and as seen in Fig.~\ref{fig:results}a) the magnitude of this term depends on the exact relative positions of the two nuclei of interest along the helix.

\begin{figure}[h!]
\begin{center}
\includegraphics[width=0.95\textwidth]{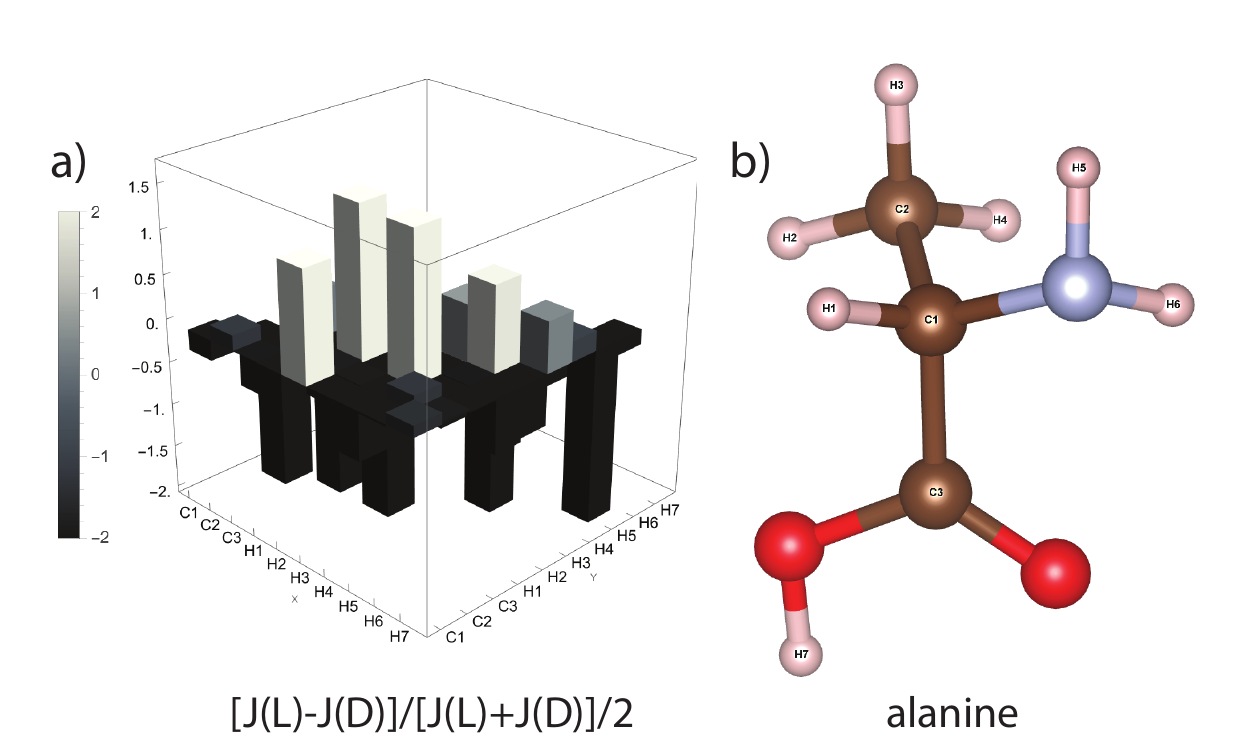}
\end{center}
\caption{Differences in NMR $J$ couplings between ($\textsf{D}, \textsf{L}$) enantiomers can be quantified using the $J$ coupling stereochemical deviation, $[J(\textsf{L})-J(\textsf{D})]/[J(\textsf{L})+J(\textsf{D})]/2$. Nonzero values of this relative difference constitute evidence of chiral selectivity of the scalar coupling.  $J$ couplings between $^1$H and $^{13}$C nuclei were computed by DFT using ORCA for the amino acids: alanine, arginine, aspartic acid, cysteine, glutamic acid, glutamine, glyceraldehyde, methionine, phenylalanine, serine, threonine, tyrosine and valine (see SI for results). The case of alanine is shown here: (a) $J$ coupling stereochemical deviation (b) labeling of atoms in alanine.}
\label{fig:amino}
\end{figure}

As mentioned earlier, the spin-dependent coupling mechanism could be different for real chiral molecules. For the amino acids in Table~\ref{tab:resultssummary}, analytical expressions for the spinors of electronic states, which are essential for the computation of J couplings, are not available to us. We can instead use DFT calculations. In Figure~\ref{fig:amino} we present calculations of J couplings between $^1$H and $^{13}$C nuclei for the two (\textsf{D}, \textsf{L}) enantiomers of alanine.  As seen in the bar plot of Figure~\ref{fig:amino}a, significant relative differences in the J couplings between enantiomers can be observed.   In SI we include DFT results for the remaining amino acids: phenylalanine, arginine, aspartic acid, cysteine, glutamic acid, glutamine, glyceraldehyde (non-amino acid), methionine, serine, threonine, tyrosine and valine.  There, we find that $J$ coupling values depend on the choice of enantiomer for all the molecules.

{\bf Discussion}

We propose a unique CISS-dependent contribution to the conventional indirect nuclear spin-spin coupling mechanism in chiral molecules based on a network of electron-nuclear spin-dependent interactions and enantioselective bond polarization. This finding suggests that NMR-based techniques could be capable of chiral discrimination. The dominant contribution to CP NMR experiments in such cases is, of course, likely due to standard dipole-dipole and non-chiral $J$ couplings. However, our results show that an additive chiral contribution to $J$ could provide observable enantiospecific effects on top of existing conventional mechanisms contributing to CP. For the DNA toy model, the electron-nuclear dipole interaction exhibits dependence on the choice of enantiomer, whereas the Fermi contact interaction does not. However, for real molecules, this mechanism may not always be the same; in which case, DFT calculations provide a more quantitative framework for the analysis of such contributions. Our work may provide a theoretical foundation for experiments that probe chirality by NMR, establishing the first link between the CISS effect and nuclear spins. The CISS effect gives rise to electronic wavefunctions that are enantiospecific owing to the effective SOC interaction, manifesting itself in the Rashba form, in the presence of a chiral structure.  The effective nuclear spin-spin interaction consists of a nuclear spin coupled to delocalized conduction-band electrons, which in turn couple to another nuclear spin.  When averaging this effective interaction over electronic degrees of freedom, we obtain a coupling tensor with  components that are enantiospecific.   We note that the idea of using NMR to probe chirality is not new.  Theoretical studies by Buckingham and co-workers have  proposed the use of electric fields (external, internal) for chiral discrimination by NMR~\cite{garbacz2016chirality,buckingham2014communication,buckingham2006direct,buckingham2004chirality,walls2008measuring}.  A paper by Harris and Jameson ~\cite{harris2006note} explains that the spin-spin ($J$) coupling has a chiral component: $E=J \mathbf{I}_1 \cdot \mathbf{I}_2 + J_{chiral} \mathbf{E} \cdot \mathbf{I}_1 \times \mathbf{I}_2$, where $J_{chiral}$ is a pseudoscalar that changes sign with chirality and $\mathbf{E}$ is an electric field. Different symmetry considerations are involved here, as we do not form a scalar interaction from the three vectors $\mathbf{E}$, $\mathbf{B}$ (magnetic field) and a single spin $\mathbf{I}$.  Instead, we have a tensorial interaction originating from an effective magnetic coupling between two spins $\mathbf{I}_1$ and $\mathbf{I}_2$ where the effective scalar Hamiltonian involves a rank-2 tensor $\mathtensor{F}$, that arises from the averaging of the electron spin-nuclear spin dipolar couplings over the electronic spinor wavefunction.  The effective tensor interaction $\mathbf{I}_1 \cdot \mathtensor{F} \cdot \mathbf{I}_2$ can be decomposed as the sum of three energies, $F \mathbf{I}_1 \cdot \mathbf{I}_2 + \frac{1}{2} \sum_{ij} F_{ij} (I_1^i I_2^j + I_1^j I_2^i) + \mathbf{f} \cdot (\mathbf{I}_1 \times \mathbf{I}_2)$, also known in the field of magnetism as the {\it isotropic} exchange, {\it symmetric} and {\it antisymmetric} parts of the anisotropic exchange, respectively. 
According to the discussion in Supplementary Text S1 (see section ``Enantiospecific NMR Response''), all three coefficients $F$, $F_{ij}$ and $\mathbf{f}$ are enantiospecific for the DNA toy model. This is in contrast with the symmetry of the conventional $J$ coupling tensor in NMR. For real molecules, this situation could be different from the helical structure toy model.

These results have at least four potential implications:  1) They establish NMR as a tool for probing chirality, at least as far as CP experiments are concerned, these two parameters could form building blocks of analytical pulse sequences that sense chirality.  In terms of spectroscopy, solid powders have wide anisotropic lines making it difficult or impossible to observe splittings of resonances.  Thus, chirality effects may be difficult to observe as coherent effects in standard NMR spectra.
On the other hand, a wide range of specialized solid-state NMR experiments were developed by Emsley and co-workers for spectral editing or to probe fine structure in powder spectra~\cite{duma2004principles,mifsud2006assigning,duma2003spin,lesage2003experimental,sakellariou2001spectral,lesage2001through,salager2009enhanced}. These advanced spin control methods are building blocks that could be adapted for chiral discrimination purposes.
2) This establishes chiral molecules as potential components of  quantum information systems through their ability to couple distant nuclear-spin qubits.   3) Our theory, while applied to nuclear spins, could also be extended to localized electronic spins, such as those found in transition metal ions, rare-earth ions and molecular magnets.   The spin-dependent interaction mechanism still holds, and its strength would be 6 orders of magnitude larger due to the higher moment of the Bohr magneton compared to the nuclear magneton.   4) Control of electronic spins could lead to control of nuclear spin states and vice-versa.   It was remarked in a recent paper by Paltiel and co-workers~\cite{bib:paltielnuclearspin} that the control of nuclear spins may lead to control of chemical reactions, which would be remarkable.
Control of electronic spins, of course, would generally not be possible if relaxation times were exceedingly short.   But in the context of CISS, the electron spin is locked to its momentum, giving rise to new possibilities for quantum logic.  For this example, the role of CISS is to create spin-polarized electronic conduction channels that can lead to indirect nuclear spin-spin coupling and associated benefits such as transport of quantum information.  Finally, we note that the application of an electric field for chiral discrimination, as was previously suggested ~\cite{garbacz2016chirality,buckingham2014communication,buckingham2006direct,buckingham2004chirality,walls2008measuring}, is not needed here, as the CISS effect alone generates an observable response.  In particular, an applied current is not required because of a nonzero quantum mechanical probability current (see Supplementary Text S1, Section ``Is an Applied Current Needed to Drive this Effective Interaction?'').  As pointed out by Rossini and co-workers~\cite{bib:rossini} impurities or crystallite size differences could potentially affect the CP process if they give rise to differences in spin-lattice relaxation.  Such differences can be minimized by further purification and recrystallization. Independently from this, however, spin-lattice relaxation rates during the cross-polarization transfer step depend on the choice of enantiomer ($\zeta$).  This is because spin-lattice relaxation can be modulated by the scalar coupling (see Supplementary Text S2), which itself depends on $\zeta$. \hl{In other words, our analysis suggests that both $T_1$ and spin-spin couplings are intrinsically connected through the choice of enantiomer.}

Finally, a word about potential applications.  Chirality is a fundamental determinant of molecular behavior and interaction, notably in biological systems where it influences the specificity of biochemical reactions. Current enantiomeric differentiation techniques, which often necessitate chiral modifiers, can interfere with the native state of biological samples, thus obscuring intrinsic molecular dynamics. The advancement of a non-perturbative method, as described in this work, utilizing the CISS effect for direct chiral recognition, may represent an important leap forward. This approach would not only preserve the pristine condition of the sample but also offer insights into the chiral-driven phenomena at the molecular level. For example, this method could enhance our understanding of enzyme-catalyzed reactions and protein-ligand interactions, elucidating the role of chirality in fundamental biological processes and potentially advancing the design of more selective pharmaceuticals.  Such experiments are crucial for investigating the core aspects of CISS as they bypass the effects of molecule-substrate couplings, instead directly probing the interactions among nuclei within chiral molecules through the electronic response governed by CISS principles.

{\bf Methods}

The theoretical investigations into the NMR $J$ couplings chiral amino acids as well as glyceraldehyde were conducted utilizing DFT as implemented in the  ORCA quantum chemistry package~\cite{neese2020orca} that includes routines for computing  NMR parameters  based on Ref.~\cite{bib:cremer2007}. The amino acids selected for this study were optimized at the B3LYP exchange-correlation~\cite{PhysRevB.37.785} functional with the split-valence Pople basis set, 6-31G(d,p) \cite{safi2022dft} to ensure accurate geometries. Following geometry optimization, the NMR $J$ couplings were calculated using the Gauge-Including Atomic Orbital (GIAO) method combined with the BP86 functional and TZVP basis set ~\cite{jensen2015segmented}. BP86, a widely utilized Generalized Gradient Approximation (GGA) functional, combines the Becke 1988 exchange functional with the triple zeta valence polarization basis set. To account for solvent effects, the geometry optimization and NMR calculations were performed using the Conductor-like Polarizable Continuum Model (CPCM) with water as the solvent. This inclusion aims to more closely simulate the natural aqueous environment of molecules by modeling the solvent as a dielectric polarizable continuum medium. \\

{\bf Acknowledgements:} L.-S.B. acknowledges partial support from NSF CHE-2002313 and would like to thank Alexej Jerschow, Jeffrey Yarger, Daniel Finkelstein-Shapiro and Thomas Fay for helpful discussions (discussions do not imply endorsements). S.V. acknowledges the support given by the Eleonore-Trefftz-Programm and the Dresden Junior Fellowship Programme by the Chair of Materials Science and Nanotechnology at the Dresden University of Technology. G.C. acknowledge the support of the German Research Foundation (DFG) within the project Theoretical Studies on Chirality-Induced Spin Selectivity (CU 44/55-1), and by the transCampus Research Award Disentangling the Design Principles of
Chiral-Induced Spin Selectivity (CISS) at the Molecule-Electrode Interface for Practical Spintronic Applications (Grant No. tCRA 2020-01), and Programme trans-Campus Interplay between vibrations and spin polarization in the CISS effect of helical molecules (Grant No. tC2023-03). V.M acknowledges the support of Ikerbasque, the Basque Foundation for Science, the German Research Foundation for a Mercator Fellowship within the project Theoretical Studies on Chirality-Induced Spin Selectivity (CU 44/55-1), and the W.M. Keck Foundation through the grant ``Chirality, spin coherence and entanglement in quantum biology." \\

{\bf Author Contributions:}  T.G. worked on certain aspects of the theory and performed all DFT computations. J.L.P. helped T.G. with DFT computations of NMR parameters, with inputs from V.M. and L.-S.B. V.M. and S.V. provided critical advice on the CISS effect in the context of the CP experiment.  L.-S.B. developed the analytical theory for the helical model and wrote the first draft. R.N.S. provided critical inputs on the electron spin mechanism. V.S.G., M.G., L.B., S.V., V.M. and G.C. reviewed the manuscript. All authors conceived the project and contributed to the design of the study.  All authors have critically examined the results and helped write and/or revise the manuscript. \\

{\bf Competing Interests:} The authors declare no competing interests.\\

\section*{Supplementary Information}


{\bf TABLE OF CONTENTS}\par

\vspace*{1\baselineskip}
{\bf Supplementary Text}\par
{\bf Text S1.} Theory of Indirect Coupling in DNA Single Helix Model. \par
{\bf Text S2.} Enantiospecificity in Cross-Polarization. \par
{\bf Text S3.} J Coupling  Stereochemical Deviations  (DFT) for Amino Acids. \par
{\bf Text S4.} Raw Data for J Couplings (DFT) in Amino Acids. \par

\section*{Text S1. Theory of Indirect Coupling in DNA Single Helix Model}

Our theoretical model to derive the effective Hamiltonian is based on second-order perturbation theory.   Effective Hamiltonians can be obtained by partitioning the Hamiltonian matrix into low energy (0) and high-energy blocks (1):
$$H=\left[ \begin{matrix} H_{00} & T_{01}\\
T_{10} & H_{11} 
\end{matrix} \right]. $$
To this matrix corresponds a resolvent (Green's function):
$$ G(\epsilon) = \frac{1}{\epsilon-H} = \left[ \begin{matrix}
\epsilon - H_{00} & T_{01} \\
T_{10} & \epsilon - H_{11}
\end{matrix} \right]^{-1}. $$
Inversion of the block 2$\times$2 matrix gives:
$$G_{00}(\epsilon) = \left( \epsilon - H_{00} -  T_{01}(\epsilon - H_{11})^{-1} T_{10}  \right)^{-1},$$
which has the form of a resolvent with effective Hamiltonian:
$$ G(\epsilon) = \frac{1}{\epsilon-H_{eff}}, \qquad H_{eff} = H_{00} + T_{01}(\epsilon - H_{11})^{-1} T_{10}. $$
Because $H_{eff}$ itself depends on the unknown energy $\epsilon$,
the Schr\"odinger equation is then solved self-consistently.  This method was introduced by L\"owdin~\cite{lowdin1951note} and is nowadays known as downfolding.  
If instead we replace the (unknown) energy $\epsilon$ by a known energy of the low-energy states, $\epsilon_0$:
$$ \approx H_{00} + T_{01}(\epsilon_0 - H_{11})^{-1} T_{10}, $$
taking the matrix element yields the familiar expression for second-order perturbation theory:
$$ E_0  = \epsilon_0 + \sum_{j,j'}  \braket{0 | T_{01} |j} \braket{ j | (\epsilon_0 - H_{11})^{-1} | j' } \braket{ j' | T_{10} | 0 } = \epsilon_0 + \sum_{j}  \frac{  | \braket{0 | \hat{V} | j } |^2}{ \epsilon_0 - \epsilon_j }  $$
  where $\hat{V} \equiv T_{01}$, $T_{10}^\dagger=T_{01}$ and $H_{11}$ is assumed diagonalized by the unperturbed basis $\{ \ket{j} \}$.   A pedagogical description of the perturbation theory approach to deriving effective spin-Hamiltonians can be found in chapter 4 of Slichter~\cite{slichter1996principles}. 
Electronic wavefunctions, which encode chirality, were obtained from the following papers by Varela and co-workers~\cite{varela2016effective,lopez2022radiation}.
In the work of Santos and San Sebastian~\cite{santos2018chirality,san2020enantiospecific} a mechanism 
for indirect nuclear spin-spin coupling was proposed based on spin-orbit coupling (SOC).   SOC is indeed expected to play a crucial role.  The exact mechanism for the indirect coupling, however, remains unclear.   In our calculation and proposed mechanism, SOC enters through the use of the Varela wavefunctions~\cite{varela2016effective,lopez2022radiation} as part of the perturbation theoretic calculation.  We find that the magnetic interaction that can explain the existence of the effective spin-Hamiltonian is dipolar in nature.   Nuclear spin 1 couples magnetically to delocalized conduction band elections, which then couple magnetically to nuclear spin 2.

\subsection*{CISS \& Indirect Spin-Spin Coupling}

The magnetic dipole interaction between two spins $\mathbf{I}_1$ and $\mathbf{I}_2$ is of the form of a scalar product of irreducible rank-2 tensors, $[\mathbf{I}_1 \otimes \mathbf{I}_2]^{(2)}$ and $[\nabla \otimes \nabla ]^{(2)}\frac{1}{r}$:
$$ \mathcal{H} =  - 3 \sqrt{5} \frac{\mu_0}{4\pi} \hbar^2  \gamma_I \gamma_S  \left[ [\mathbf{I}_1 \otimes \mathbf{I}_2]^{(2)} \otimes  [\nabla \otimes \nabla ]^{(2)} (1/r) \right]_0^{(0)}  $$
where $[\mathtensor{A} \otimes\mathtensor{B}]_0^{(0)}$ denotes the scalar tensor product of two irreducible tensors $\mathtensor{A}$ and $\mathtensor{B}$, with the convention that all vectors are expressed as rank-1 spherical tensors, i.e. for $\mathbf{r}=(x,y,z)$ we have $\mathbf{r}_{(1),\pm 1} = \mp (1/\sqrt{2})(x \pm i y)$, $r_0^{(1)}=z$.  The irreducible components of the product of two tensors $\mathtensor{T}_{(k_1)}$ and $\mathtensor{T}_{(k_2)}$ are obtained as the linear combinations:
$$  \left[ \mathtensor{T}_{(k_1)} \otimes \mathtensor{T}_{(k_2)} \right]_{\kappa_3}^{(k_3)} = \sum_{\kappa_1,\kappa_2} \left( k_1 \kappa_1 k_2 \kappa_2 | k_1 k_2 k_3 \kappa_3 \right) \mathtensor{T}_{(k_1),\kappa_1} \mathtensor{T}_{(k_2),\kappa_2}, $$
where $\left( k_1 \kappa_1 k_2 \kappa_2 | k_1 k_2 k_3 \kappa_3 \right)$ are Clebsch-Gordan coefficients.  This formula can be used to describe any number of pairwise magnetic interactions, whose contributions to the energy are additive, provided that extra care is taken at the point $r=0$, by adding a Dirac delta function.   The latter is a result from distribution theory (see below).  Two localized spins $\mathbf{I}_1$ and $\mathbf{I}_2$ centered at $\mathbf{R}_1$ and $\mathbf{R}_2$, respectively, 
interact with a set of electron spins $\{ \mathbf{S}_l\} $ each centered at $\{ \mathbf{r}_l\}$
according to the sum of Fermi contact and dipole-dipole interactions (energy units):
$$ \mathcal{H}(\{\mathbf{R}_i\},\{\mathbf{r}_l\}) = \frac{\mu_0}{4\pi} \hbar^2 \Biggl[  \underbrace{ \frac{ 8\pi}{3} \gamma_I \gamma_S \mathbf{I}_1 \cdot \sum_l \mathbf{S}_l \delta^{(3)}(\mathbf{R}_1-\mathbf{r}_l) + \frac{ 8\pi}{3}   \gamma_I \gamma_S \mathbf{I}_2 \cdot \sum_l \mathbf{S}_l \delta^{(3)}(\mathbf{R}_2-\mathbf{r}_l) }_{\mbox{Fermi contact},  \mathcal{H}_{F_c}(\{\mathbf{R}_i\},\{\mathbf{r}_l\}) } $$
$$ +  \underbrace{ \gamma_I \gamma_S \, p.v. \sum_l \frac{ \delta_{\alpha\beta} - 3 \hat{\tilde{r}}_{1l,\alpha} \hat{\tilde{r}}_{1l,\beta}}{ |\mathbf{R}_1-\mathbf{r}_l|^3} I_1^\alpha S_l^\beta + \gamma_I \gamma_S \, p.v. \sum_l \frac{ \delta_{\alpha\beta} - 3 \hat{\tilde{r}}_{2l,\alpha} \hat{\tilde{r}}_{2l,\beta}}{ |\mathbf{R}_2-\mathbf{r}_l|^3} I_2^\alpha S_l^\beta }_{ \mbox{dipole-dipole},  \mathcal{H}_{DD}(\{\mathbf{R}_i\},\{\mathbf{r}_l\})  }  \Biggr] $$
where $p.v.$ denotes Cauchy principal value, and the  component of unit vector $\hat{\tilde{r}}_{il,\alpha}$ are:
 $$\hat{\tilde{r}}_{il,\alpha} =\frac{R_{i,\alpha}-r_{l,\alpha}}{\mathbf{R}_i-\mathbf{r}_l}.  $$
Since $\mathbf{I}_1$ and $\mathbf{I}_2$ are assumed to be localized spins, $\mathbf{R}_1$ and $\mathbf{R}_2$ are treated as fixed parameters.

An effective Hamiltonian can be obtained through second-order perturbation theory:
\begin{align}
\mathcal{H}_{eff} &=  \underbrace{  \left(\frac{2 \mu_0}{3} \right)^2 \gamma_I^2 \gamma_S^2 \hbar^4 \sum_j \mathbf{I}_1 \cdot \frac{ \braket{ 0 | \sum_l \mathbf{S}_l \delta^{(3)}(\mathbf{r}_l - \mathbf{R}_1) |j}  \braket{j| \sum_l \mathbf{S}_l \delta^{(3)}(\mathbf{r}_l - \mathbf{R}_2)|0}  }{ E_0 - E_j } \cdot \mathbf{I}_2 + c.c. }_{ \mathcal{H}_{eff}^{FC} }. \nonumber \\
& + \underbrace{  \left( \frac{\mu_0}{4\pi}\right)^2 \gamma_I^2 \gamma_S^2 \hbar^4 \, p.v. \sum_{\alpha,\alpha'} \sum_{\beta,\beta'}  \sum_j I_1^\alpha  \frac{ \braket{0 | \sum_l \frac{ \delta_{\alpha\beta} - 3 \hat{\tilde{r}}_{1l,\alpha} \hat{\tilde{r}}_{1l,\beta}}{ |\mathbf{R}_1-\mathbf{r}_l|^3} S_l^\beta | j } \braket{ j |\sum_l \frac{ \delta_{\alpha'\beta'} - 3 \hat{\tilde{r}}_{2l,\alpha'} \hat{\tilde{r}}_{2l,\beta'}}{ |\mathbf{R}_2-\mathbf{r}_l|^3}  S_l^{\beta'} |0}}{E_0-E_j} I_2^{\alpha'} + c.c.   }_{ \mathcal{H}_{eff}^{DD} }  \label{eq:heff}
\end{align}
It will be convenient for us to express the dipole interaction tensor in terms of the Hessian of $1/\tilde{r}$:
$$ \nabla \nabla \frac{1}{|\mathbf{r-r'}|}  =p.v.\left( - \frac{ \delta_{ij}}{ |\mathbf{r-r'}|^3 }  + \frac{3 (r_i-r_i') (r_j-r_j')}{ |\mathbf{r-r'}|^5 }  \right) \hat{\mathbf{e}}_i \hat{\mathbf{e}}_j  - \frac{4\pi}{3} \delta^{(3)}( \mathbf{r-r'} ) I $$
where $p.v.$ denotes the Cauchy principal value and $I$ is the identity tensor.     The Dirac delta term is an artifact from distribution theory originating from the point $\mathbf{r}=\mathbf{r'}$.   The Dirac term is of no consequence here, as we show (see Section ``Fermi Contact Term'') that Fermi contact terms do not contribute significantly to the effective interaction in the context of CISS.

\subsubsection*{Matrix Elements of Single-Body Operators} 

In the above expression, we have matrix elements of $\sum_l \mathbf{S}_l \delta^{(3)}(\mathbf{r}_l - \mathbf{R})$, which is a sum of single-body operators.  The states $\ket{0}$ and $\ket{j}$ are Slater determinants that are products of Bloch functions.  In the language of second-quantization a single-body operator $\mathbf{H}$ has the form
$$ \mathbf{H} = \underbrace{ \sum_{i=1}^N \mathbf{h}(i) }_{\mbox{first quantization}} =\underbrace{  \sum_{m,n} \braket{ m | \mathbf{H} | n } a_m^\dagger a_{n},}_{\mbox{second quantization}} $$
where $a_m^\dagger$ and $a_n$ are fermionic creation and annihilation operators, respectively.   The equal signs here are taken loosely in the sense that the different $\mathbf{H}$'s involved operate on different spaces (e.g. Hilbert vs Fock space).  The $N$-body wavefunctions are Slater determinants:
$$ \ket{0} = a_{j_1}^\dagger a_{j_2}^\dagger \dots a_{j_N}^\dagger  \ket{vac} $$
$$ \ket{j} = a_{j_{1'} }^\dagger a_{j_{2'} }^\dagger \dots a_{j_{N'} }^\dagger  \ket{vac}, $$
where $\ket{vac}$ denotes the vacuum state.
The matrix elements of $\mathbf{H}$ are:
\begin{align*}
\braket{j | \mathbf{H} | 0 } &= \bra{vac}  a_{j_{N'} } \dots a_{j_{2'} } a_{j_{1'} }  \cdot \sum_{m,n} \braket{ m | \mathbf{H} | n } a_m^\dagger a_{n}  \cdot a_{j_1}^\dagger a_{j_2}^\dagger \dots a_{j_N}^\dagger  \ket{vac}  \\
  &=  \sum_{m,n} \braket{ m | \mathbf{H} | n }  \bra{vac}  a_{j_{N'} } \dots a_{j_{2'} } a_{j_{1'} }  a_m^\dagger a_{n} a_{j_1}^\dagger a_{j_2}^\dagger \dots a_{j_N}^\dagger  \ket{vac}  
 \end{align*}
Using anticommutation relations for the fermionic operators:
$$ \{ a_j^\dagger, a_{j'} \} = \delta_{jj'}, \qquad \{ a_j,a_{j'} \} = 0, \qquad \{ a_j^\dagger, a_{j'}^\dagger \}=0,  $$
we see that moving the $a_n$ operator to the right will annihilate the vacuum state to give zero unless $n \in \{ j_1, j_2, \dots, n_N \}$.    If $n$ belongs to that set, say $n=j_k$, then:
\begin{align*}
 a_n a_{j_1}^\dagger \dots a_{j_k}^\dagger \dots a_{j_N}^\dagger  \ket{vac} =&  (-1)^{k-1} a_{j_1}^\dagger \dots \underbrace{  (a_n  a_{j_k}^\dagger) }_{1 -  a_{j_k}^\dagger a_n   } \dots a_{j_N}^\dagger  \ket{vac} \\
  =&  (-1)^{k-1} a_{j_1}^\dagger \dots  (1 -  a_{j_k}^\dagger a_n) \dots a_{j_N}^\dagger  \ket{vac} \\
  =&  (-1)^{k-1} a_{j_1}^\dagger \dots  (1) \dots a_{j_N}^\dagger  \ket{vac} \\
   & -  (-1)^{k-1} a_{j_1}^\dagger \dots  a_{j_k}^\dagger \dots a_{j_N}^\dagger a_n (-1)^{N-k} \ket{vac}
 \end{align*}
The second term is zero because $a_n$ acts on the vacuum.  Likewise, moving $a_m^\dagger$ to the left, we get (say $m=j_{d'}$):
\begin{align*}
\bra{vac}  a_{j_{N'} } \dots a_{j_{d'}} \dots a_{j_{2'} } a_{j_{1'} }  a_m^\dagger =& \bra{vac} a_{j_N'} \dots \underbrace{ (a_{j_{d'}} a_m^\dagger ) }_{ 1 - a_m^\dagger a_{j_{d'}}  } \dots  a_{j_2'} a_{j_1'} (-1)^{d'-1} \\
=& \bra{vac} a_{j_{N'} } \dots  (1 - a_m^\dagger a_{j_{d'}}) \dots  a_{j_{2'} } a_{j_{1'} } (-1)^{d'-1} \\
 =& \bra{vac} a_{j_{N'} } \dots (1) \dots  a_{j_{2'} } a_{j_{1'} } (-1)^{d'-1} \\
   & - \bra{vac} a_m^\dagger  a_{j_{N'} } \dots a_{j_{d'}}  \dots  a_{j_2'} a_{j_{1'} } (-1)^{d'-1} (-1)^{N-d'}
\end{align*} 
The second term is zero as $a_m^\dagger$ acts on $\bra{vac}$ to its left.
Thus, we are left with:
\begin{multline*}
\bra{vac}  a_{j_{N'} } \dots a_{j_{d'}} \dots a_{j_{2'} } a_{j_{1'} }  (a_m^\dagger  a_n) a_{j_1}^\dagger \dots a_{j_k}^\dagger \dots a_{j_N}^\dagger  \ket{vac}  \\
= \bra{vac} a_{j_{N'} } \dots (1) \dots  a_{j_{2'} } a_{j_{1'} }  a_{j_1}^\dagger \dots  (1) \dots a_{j_N}^\dagger  \ket{vac} (-1)^{d'+k}
\end{multline*}
which is zero unless $\{ j_1, \dots, \hat{j_k}, \dots, j_N \} = \{ j_{1'}, \dots, \hat{j}_{d'}, \dots, j_{N'} \}$ (where the hat indicates omission of a term).   Thus, the average $\braket{j | a_m^\dagger a_n | 0}$ is zero unless $n \in \{ j_1, \dots, j_N \}$, $d' \in \{ j_{1'}, \dots, j_{N'} \}$ and the remaining indices are all identical $\{ j_1, \dots, \hat{j_k}, \dots, j_N \} = \{ j_{1'}, \dots, \hat{j}_{d'}, \dots, j_{N'} \}$.  If the remaining indices are identical (but not necessarily in the same order), then
\begin{multline*}
 \bra{vac} a_{j_{N'} } \dots (1) \dots  a_{j_{2'} } a_{j_{1'} }  a_{j_1}^\dagger \dots  (1) \dots a_{j_N}^\dagger  \ket{vac}  \\
= (-1)^P  \bra{vac} a_{j_{N} } \dots (1) \dots  a_{j_{2} } a_{j_{1} }  a_{j_1}^\dagger \dots  (1) \dots a_{j_N}^\dagger \ket{vac} 
=(-1)^P, 
\end{multline*}
through repeated application of the anticommutator $a_m a_m^\dagger = 1 - a_m^\dagger a_m$ in the center term (and dropping $a_m^\dagger a_m$ because it vanishes when moving $a_m$ all the way to the right).  $P$ is the permutation needed to bring them to the same order.

We have just proved that:
\begin{align*}
 \braket{j | \mathbf{H} | 0 } &= \bra{vac} a_{j_{N'} } \dots  a_{j_{2'} } a_{j_{1'} }  \mathbf{H}
   a_{j_1}^\dagger a_{j_2}^\dagger \dots a_{j_N}^\dagger  \ket{vac}  \\
  &= \bra{vac} a_{j_{N'} } \dots  a_{j_{2'} } a_{j_{1'} }  \left(  \sum_{m,n} \braket{ m | \mathbf{H} | n } a_m^\dagger a_{n} \right) a_{j_1}^\dagger a_{j_2}^\dagger \dots a_{j_N}^\dagger  \ket{vac}  \\
  &=  \sum_{m \in \bra{j} \atop n \in \ket{0} } \braket{ m | \mathbf{H} | n }  \bra{vac} a_{j_{N'} } \dots  a_{j_{2'} } a_{j_{1'} }  ( a_m^\dagger a_{n}) a_{j_1}^\dagger a_{j_2}^\dagger \dots a_{j_N}^\dagger  \ket{vac}  \\
 &= \begin{cases}
  \sum_{m \in \bra{j} \atop n \in \ket{0} } \bra{ vac } a^\dagger_m \mathbf{H}  a_n \ket{vac} (-1)^{d'+k+P} & \mbox{if~} \{ j_1, \dots, \hat{j_k}, \dots, j_N \} = \{ j_{1'}, \dots, \hat{j}_{d'}, \dots, j_{N'} \}\\
  0 & \mbox{otherwise}
  \end{cases}   
 \end{align*}
 where the summation runs over all
 $$ m \in \bra{j} = \{ j_{1'}, j_{2'}, \dots, j_{N'} \}, \qquad \mbox{and} \qquad n \in \ket{0} = \{ j_1, j_2, \dots, j_N \}. $$
The states $a_n \ket{vac}$ are Bloch functions, leading to $\sum_{\mathbf{kk}'ss'} \braket{ \mathbf{k}'s' | \mathbf{H}   | \mathbf{k}s}  (-1)^{d'+k+P} $. The sign $(-1)^{d'+k+P}$ disappears when forming the product: $\sum_{\mathbf{kk}'ss'} \braket{ \mathbf{k}'s' | \mathbf{H}  | \mathbf{k}s }  \braket{ \mathbf{k}s | \mathbf{H}  | \mathbf{k}'s' }$.

\subsection*{Dipole-Dipole Interaction Term}

To compute the matrix elements 
$$ \langle \mathbf{k}s | \frac{ \delta_{\alpha\beta} - 3 \hat{\tilde{r}}_{il,\alpha} \hat{\tilde{r}}_{il,\beta}}{ |\mathbf{R}_i-\mathbf{r}_l|^3} S_l^\beta   | \mathbf{k}' s \rangle 
$$
we use the spinor-valued Bloch wavefunctions derived by Varela~\cite{varela2016effective,lopez2022radiation}, which we shall abbreviate as:
$$ \boldsymbol\psi_{n,s}^{\nu,\zeta}  = \frac{\sqrt{s}}{2} \left[ \begin{matrix}
(s e^{i\theta/2} + e^{-i\theta/2}) e^{-i \varphi/2} \\
\zeta(s e^{i\theta/2} - e^{-i\theta/2}) e^{i\varphi/2} 
\end{matrix}
\right] e^{i \nu \tilde{n} \varphi}= \left[ \begin{matrix}
F_A e^{-i \varphi/2} \\
\zeta F_B^* e^{i \varphi/2} 
\end{matrix} \right] e^{i \nu \tilde{n} \varphi}, $$
$$ F_A = \frac{ \sqrt{s}}{2} ( s e^{i \theta/2} + e^{-i\theta/2} ), \qquad F_B = \frac{\sqrt{s}}{2} ( s e^{ -i\theta/2} - e^{i\theta/2} ). $$
The symbols we will need are:
\begin{itemize}
\item $s=\pm 1$ is the spin label
\item $\nu=\pm 1$ is the electron propagation direction
\item $\zeta=\pm 1$ labels the right- and left-handedness of the helix
\item $\tilde{n}=n/2M\mathcal{N}$ labels the subbands corresponding to the discrete modes due to longitudinal confinement within a helix
\item $n$ is an integer in the interval $0 < n < M$
\item $M$ is the number of bases per turn.
\item $\varphi$ ranges from 0 to $2\pi \mathcal{N}$
\item $\mathcal{N}$ is the number of turns in the helix.
\end{itemize} 

\begin{figure}[h]
\includegraphics[height=0.28\textwidth]{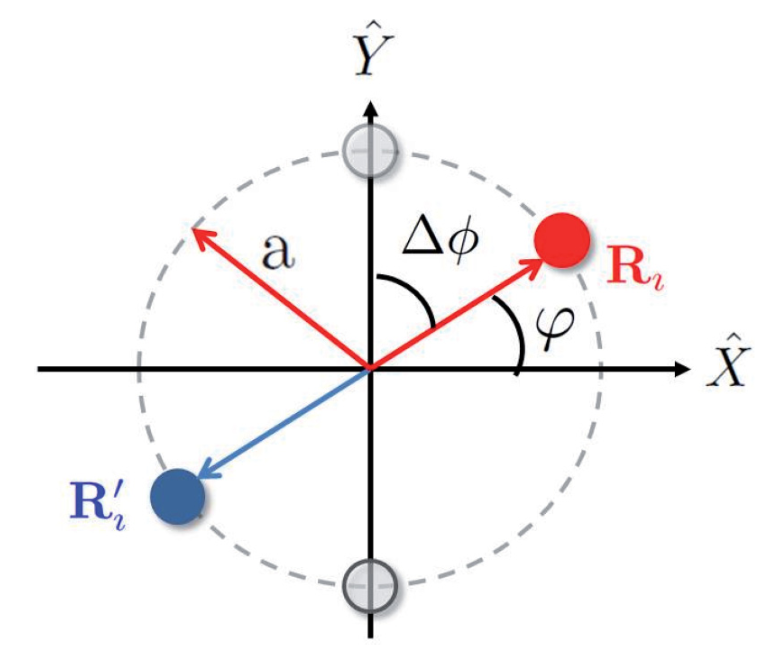} 
\includegraphics[height=0.28\textwidth]{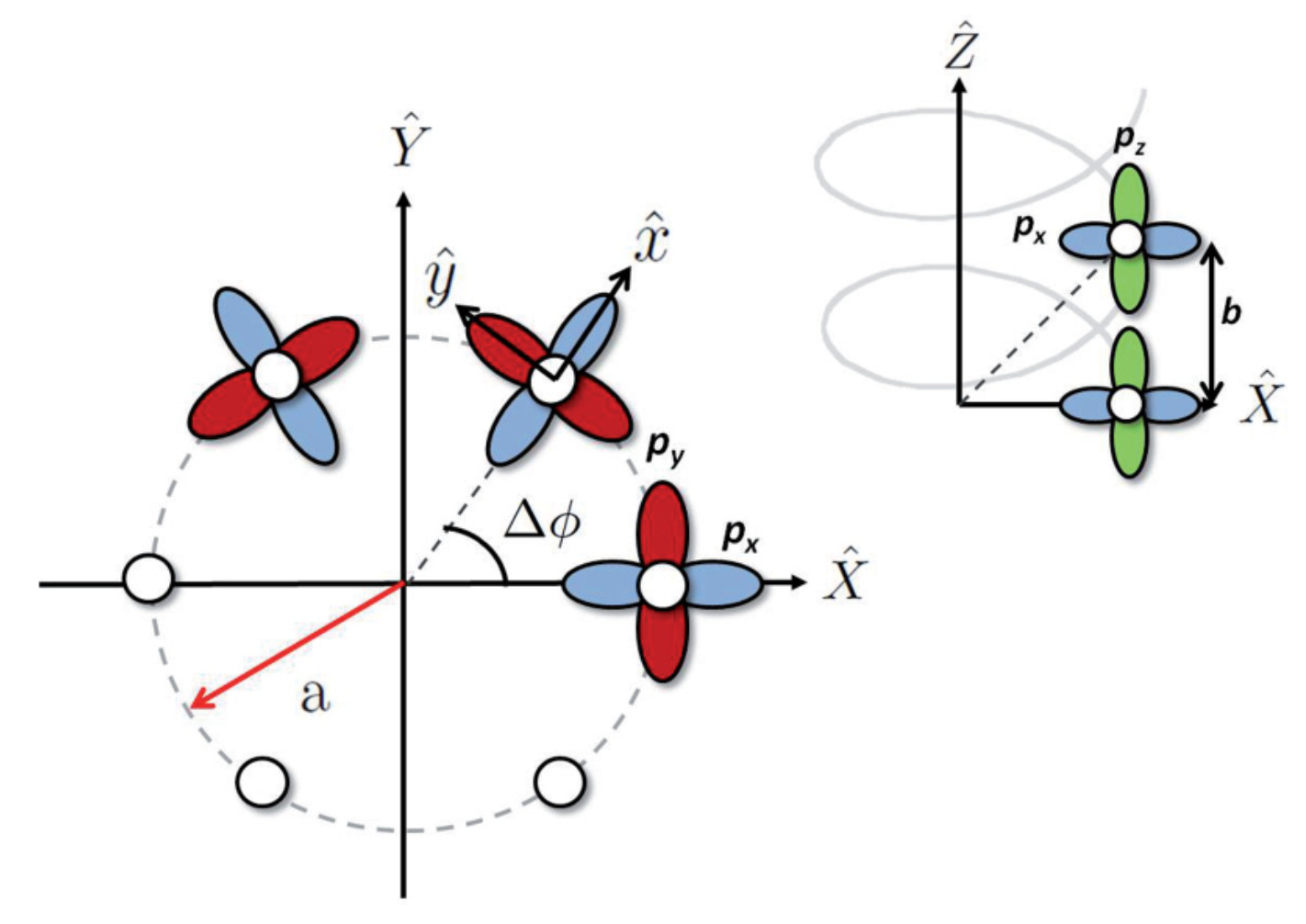} 
\includegraphics[height=0.28\textwidth]{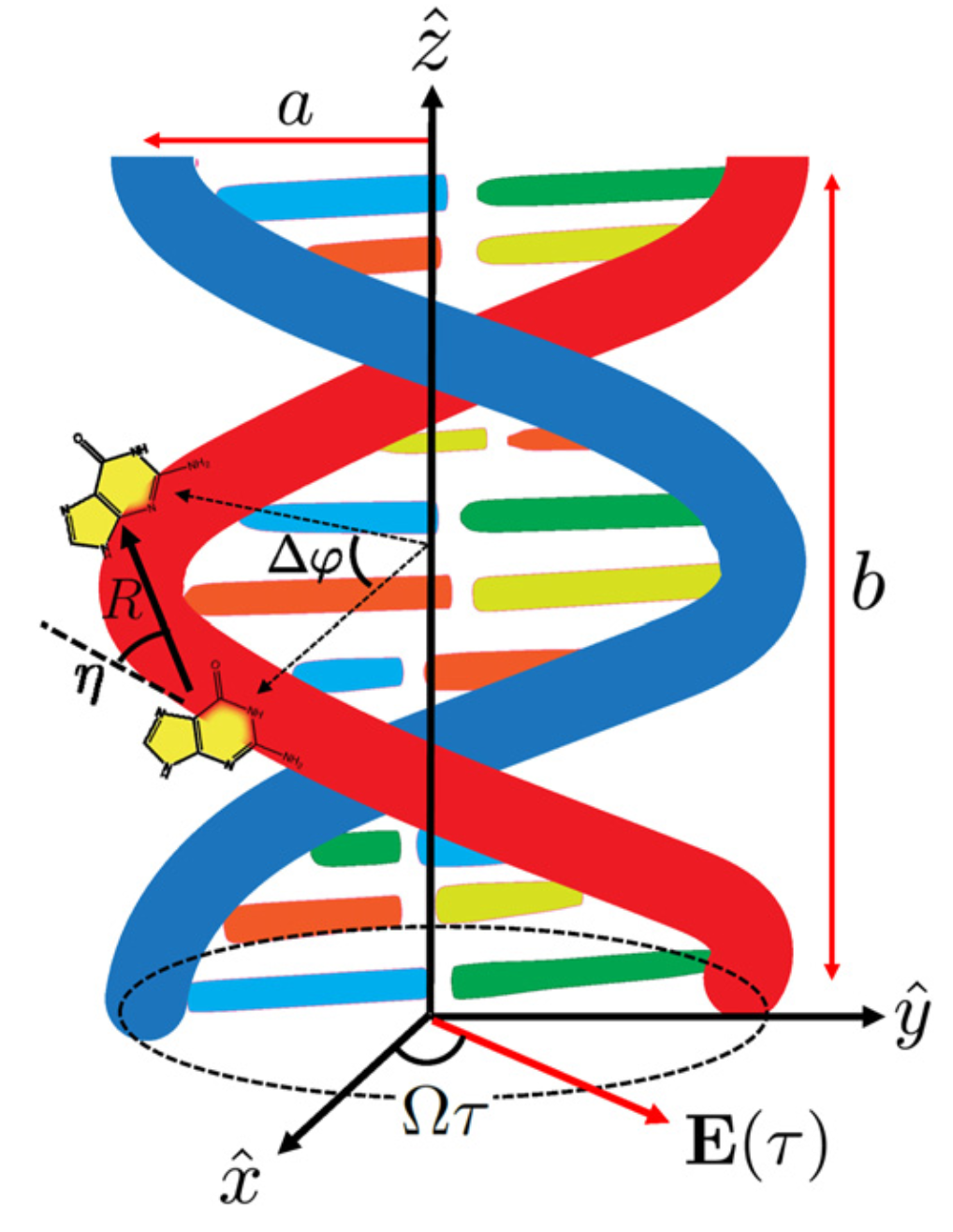}  
\caption{Geometry of the helical DNA-like model used for computation of the electronic band structure for the CISS effect.  Illustration reproduced from the work of Varela and L\'opez~\cite{varela2016effective,lopez2022radiation}.  }
\label{fig:geometry}
\end{figure}

\noindent Conversion from helicoidal to Cartesian coordinates is done using the mapping:
$$ \varphi \mapsto (a\cos \varphi, a\sin \varphi, b \varphi), \varphi\in [0,T]. $$
This leads to the substitutions:
$$ \mathbf{R}_1 = (a \cos \varphi_1, a \sin \varphi_1, b \varphi_1) $$
$$ \mathbf{R}_2 = (a \cos \varphi_2, a \sin \varphi_2, b \varphi_2) $$
$$ \mathbf{r}_l = (a \cos \varphi_l, a \sin \varphi_l, b \varphi_l). $$
Note: $\varphi_l = (r_l)_z/b=z_l/b$.
In helicoidal coordinates, integration over the spatial coordinates involves integrals of the type:
$$ \frac{1}{2\pi} \int_0^{2\pi}  e^{i [\nu(\tilde{n}'-\tilde{n}) + \xi] \varphi_l } \left[ (a\cos\varphi_1 - a\cos\varphi_l)^2 + (a\sin\varphi_1 - a\sin\varphi_l)^2 + (b\varphi_1-b\varphi_l)^2  \right]^{-3/2} d\varphi_l  $$
where $\xi=-1,0,1$. According to Varela and Lop\'ez~\cite{varela2016effective,lopez2022radiation},  $a=2$ nm, $b=3.38$ nm, $\Delta \varphi=36^\circ$, $R=0.70$ nm (interbase distance).    The geometry is shown in Fig.~\ref{fig:geometry}.

The last term is the largest for two reasons:  $b>a$ so $b^2 > a^2$ (by a factor of almost 3); also $\varphi$ ranges from 0 to $2\pi$ whereas $\cos$ and $\sin$ range from -1 to 1, when $\mathcal{N}=1$.
That's a factor $((4\pi)/4)^2 \approx 9.9$.  Therefore, it can be larger by almost 30$\times$.
We note that the integral of the DD tensor can be computed as:
$$ \nabla_{\mathbf{R}_i} \nabla_{\mathbf{R}_i }  \frac{1}{2\pi} \int_0^{2\pi} d\varphi_l \frac{ e^{i  [\nu(\tilde{n}'-\tilde{n}) + \xi] \varphi_l } }{ |\mathbf{R}_i-\mathbf{r}_l| }. $$
We therefore expand it:
\begin{multline*}
\frac{1}{\sqrt{  (a\cos\varphi_1  - a\cos\varphi_l)^2 + (a\sin\varphi_1 - a\sin\varphi_l)^2 + (b\varphi_1-b\varphi_l)^2  }} \\
\approx \frac{1}{b|\varphi_1-\varphi_l|}-\frac{a^2}{2 b^3|\varphi_1-\varphi_l|^{3}}+\frac{3 a^4}{8 b^5|\varphi_1-\varphi_l|^5}+O\left(a^6\right).  
\end{multline*}
We shall take $M=10$ and $\mathcal{N}=1$. This gives:
$$ \nabla_{\mathbf{R}_i} \nabla_{\mathbf{R}_i}  \frac{1}{2\pi} \int_0^{2\pi} d\varphi_l  e^{i [ \nu(\tilde{n}'-\tilde{n}) + \xi] \varphi_l } \left[ 
\frac{1}{b|\varphi_1-\varphi_l|}-\frac{a^2}{2 b^3|\varphi_1-\varphi_l|^{3}}+\frac{3 a^4}{8 b^5|\varphi_1-\varphi_l|^5}+O\left(a^6\right) 
 \right].  
$$ 
Integrating the first term:
$$ \lim_{\epsilon\rightarrow 0} \int_0^{2\pi} d\varphi_l  
 \frac{e^{i [ \nu(\tilde{n}'-\tilde{n}) + \xi ] \varphi_l } }{|\varphi_1-\varphi_l| - i\epsilon} = p.v. \int_0^{2\pi} d\varphi_l  
 \frac{e^{i [ \nu(\tilde{n}'-\tilde{n}) + \xi ] \varphi_l } }{|\varphi_1-\varphi_l| } + i \pi e^{i [ \nu(\tilde{n}'-\tilde{n}) + \xi ] \varphi_1 } $$
where $0 \le \varphi_1 \le 2\pi$.  Where we made use of the Plemelj-Sokhotski theorem:
$$ \lim_{\epsilon\rightarrow 0} \frac{1}{x \pm i \epsilon} = p.v. \frac{1}{x} \mp i \pi \delta(x). $$
Furthermore making use of the Cauchy principal value
$$ p.v. \int_{-\infty}^\infty \frac{ e^{ix}}{x} dx = i \pi,\qquad  p.v. \int_{-\infty}^\infty \frac{ e^{i(x-x_0)}}{x} dx = i \pi e^{i x_0}, $$
we conclude that
$$ \int_0^{2\pi} d\varphi_l  
 \frac{e^{i  [\nu(\tilde{n}'-\tilde{n}) + \xi ] \varphi_l } }{|\varphi_1-\varphi_l  | - i0 } 
 =  i \pi e^{i [\nu(\tilde{n}'-\tilde{n}) + \xi] \varphi_1 }  + i \pi e^{i [\nu(\tilde{n}'-\tilde{n}) + \xi] \varphi_1 } = 2 i \pi e^{i [ \nu(\tilde{n}'-\tilde{n}) + \xi] \varphi_1 }. $$ 
We then take the gradient (twice). But we must make use of the chain rule: $\partial_{R_j} = \frac{ \partial \varphi_1}{ \partial R_j } \frac{ \partial}{ \partial \varphi_1}
$, where
$$ \frac{ \partial  x}{ \partial \varphi_1 } = - a \sin\varphi, \quad \frac{ \partial y}{  \partial \varphi_1 } =  a \cos\varphi, \quad \frac{ \partial z }{  \partial \varphi_1 } = b.  $$
Inverting
$$ \frac{ \partial  \varphi_1 }{ \partial x} = - \frac{1}{ a \sin\varphi}, \quad \frac{ \partial  \varphi_1 }{ \partial y  } = \frac{1}{a \cos\varphi}, \quad \frac{ \partial  \varphi_1 }{ \partial z } = \frac{1}{b}.  $$
The transformation is not 1-1.  Therefore,
$$ \nabla_{\mathbf{R}_1} = \left( - \frac{1}{a\sin\varphi_1 } \partial_{\varphi_1} , \frac{1}{a\cos\varphi_1} \partial_{\varphi_1}, \frac{1}{b} \partial_{\varphi_1} \right). $$
The gradient is defined on $[0,2\pi] \setminus \{ 0, \pi/2, \pi, 3\pi/2, 2\pi \}$. Care will be needed around singularities. 
Taking the gradient once:
\begin{multline*}
\nabla_{\mathbf{R}_1} \int_0^{2\pi} d\varphi_l  
 \frac{e^{i [ \nu(\tilde{n}'-\tilde{n}) + \xi] \varphi_l } }{ b |\varphi_1-\varphi_l|} 
= 2 i \pi b^{-1} \nabla_{\mathbf{R}_1}  e^{i  [ \nu(\tilde{n}'-\tilde{n}) + \xi ] \varphi_1 } \\
=\frac{2\pi i}{b} \left( - \frac{1}{a\sin\varphi_1}, \frac{1}{a\cos\varphi_1} , \frac{1}{b} \right) i [ \nu(\tilde{n}'-\tilde{n}) + \xi]  e^{i  [ \nu(\tilde{n}'-\tilde{n}) + \xi ] \varphi_1 }.
\end{multline*}
Taking another gradient we obtain the matrix:
\begin{align*}
\nabla_{\mathbf{R}_1} & \nabla_{\mathbf{R}_1} 
 \int_0^{2\pi} d\varphi_l  
 \frac{e^{i [ \nu(\tilde{n}'-\tilde{n}) + \xi ] \varphi_l } }{b |\varphi_1-\varphi_l|} = -2 \pi  [ \nu(\tilde{n}'-\tilde{n}) + \xi ]  e^{i [ \nu(\tilde{n}'-\tilde{n}) + \xi]  \varphi_1 }  b^{-1} \\
 & \times  \left[ 
 \begin{smallmatrix}
 i \frac{1}{a^2 \sin\varphi_1 } \csc (\varphi_1 ) ( [ \nu(\tilde{n}'-\tilde{n}) + \xi] +i \cot (\varphi_1 )) & - \frac{1}{a^2 \sin\varphi_1 }  \sec (\varphi_1 ) (\tan (\varphi_1 )+i [ \nu(\tilde{n}'-\tilde{n}) + \xi]  ) & i  \frac{ [ \nu(\tilde{n}'-\tilde{n}) + \xi ] }{a^2\sin\varphi_1 } \\
- i \frac{1}{a^2 \cos \varphi_1 } \csc (\varphi_1 ) ( [ \nu(\tilde{n}'-\tilde{n}) + \xi] +i \cot (\varphi_1 )) & \frac{1}{a^2 \cos\varphi_1 }  \sec (\varphi_1 ) (\tan (\varphi_1 )+i [ \nu(\tilde{n}'-\tilde{n}) + \xi ]  ) & - i  \frac{ [ \nu(\tilde{n}'-\tilde{n}) + \xi ] }{a^2\cos\varphi_1 } \\
-i \frac{1}{b^2 } \csc (\varphi_1 ) ( [ \nu(\tilde{n}'-\tilde{n}) + \xi ]  -i \cot (\varphi_1 )) & - \frac{1}{b^2}  \sec (\varphi_1 ) (\tan (\varphi_1 )+i  [ \nu(\tilde{n}'-\tilde{n}) + \xi ]  )& -i  \frac{ [ \nu(\tilde{n}'-\tilde{n}) + \xi ] }{b^2}
 \end{smallmatrix}
 \right].   
\end{align*}
In fact, we need matrix elements of the form 
$$\nabla_{\mathbf{R}_i} \nabla_{\mathbf{R}_i}  \frac{1}{2\pi} \int_0^{2\pi} d\varphi_l  \frac{ e^{-i [ \nu(\tilde{n}'-\tilde{n}) + \xi ] \varphi_l } }{ |\mathbf{R}_i-\mathbf{r}_l| }
 \braket{\tilde{n}' |  S_l^{\beta} | \tilde{n}},$$ 
 where $\beta=x,y,z$. 
Using the matrix elements of Pauli matrices $\braket{ \boldsymbol\psi_{n',s}^{\nu,\zeta} | \sigma_i |  \boldsymbol\psi_{n,s}^{\nu,\zeta} }$, $i =x,y,z$, we obtain:
\begin{align}
M_{1,x}(\tilde{n}',\tilde{n}) &\equiv  2 \pi b^{-1} \zeta F_A^* F_B^*  [\nu(\tilde{n}'-\tilde{n})+1]  e^{i [\nu(\tilde{n}'-\tilde{n})+1]  \varphi_1 }   \nonumber  \\
 & \times  \left[ 
 \begin{smallmatrix}
 i \frac{1}{a^2 \sin\varphi_1 } \csc (\varphi_1 ) ([\nu(\tilde{n}'-\tilde{n})+1] +i \cot (\varphi_1 )) & - \frac{1}{a^2 \sin\varphi_1 }  \sec (\varphi_1 ) (\tan (\varphi_1 )+i [\nu(\tilde{n}'-\tilde{n})+1]  ) & i  \frac{[\nu(\tilde{n}'-\tilde{n})+1] }{a^2\sin\varphi_1 } \\
- i \frac{1}{a^2 \cos \varphi_1 } \csc (\varphi_1 ) ( [\nu(\tilde{n}'-\tilde{n})+1]  +i \cot (\varphi_1 )) & \frac{1}{a^2 \cos\varphi_1 }  \sec (\varphi_1 ) (\tan (\varphi_1 )+i [\nu(\tilde{n}'-\tilde{n})+1]  ) & - i  \frac{[\nu(\tilde{n}'-\tilde{n})+1] }{a^2\cos\varphi_1 } \\
i \frac{1}{a^2 } \csc (\varphi_1 ) ([\nu(\tilde{n}'-\tilde{n})+1]  +i \cot (\varphi_1 )) & - \frac{1}{a^2}  \sec (\varphi_1 ) (\tan (\varphi_1 )+i [\nu(\tilde{n}'-\tilde{n})+1]  )& i  \frac{[\nu(\tilde{n}'-\tilde{n})+1] }{a^2}
 \end{smallmatrix}
 \right]   \nonumber  \\
& + 2 \pi \zeta b^{-1} F_A F_B  [\nu(\tilde{n}'-\tilde{n})-1]  e^{i [\nu(\tilde{n}'-\tilde{n})-1] \varphi_1 }  \nonumber  \\
 & \times  \left[ 
 \begin{smallmatrix}
 i \frac{1}{a^2 \sin\varphi_1 } \csc (\varphi_1 ) ([\nu(\tilde{n}'-\tilde{n})-1] +i \cot (\varphi_1 )) & - \frac{1}{a^2 \sin\varphi_1 }  \sec (\varphi_1 ) (\tan (\varphi_1 )+i [\nu(\tilde{n}'-\tilde{n})-1]  ) & i  \frac{[\nu(\tilde{n}'-\tilde{n})-1] }{a^2\sin\varphi_1 } \\
- i \frac{1}{a^2 \cos \varphi_1 } \csc (\varphi_1 ) ( [\nu(\tilde{n}'-\tilde{n})-1]  +i \cot (\varphi_1 )) & \frac{1}{a^2 \cos\varphi_1 }  \sec (\varphi_1 ) (\tan (\varphi_l )+i [\nu(\tilde{n}'-\tilde{n})-1]  ) & - i  \frac{[\nu(\tilde{n}'-\tilde{n})-1] }{a^2\cos\varphi_1 } \\
-i \frac{1}{b^2 } \csc (\varphi_1 ) ([\nu(\tilde{n}'-\tilde{n})-1]  +i \cot (\varphi_1 )) & \frac{1}{b^2}  \sec (\varphi_1 ) (\tan (\varphi_1 )+i [\nu(\tilde{n}'-\tilde{n})-1]  )& -i  \frac{[\nu(\tilde{n}'-\tilde{n})-1] }{b^2} 
 \end{smallmatrix}
 \right]  \label{eq:Ms1}
\end{align}

\begin{align}
M_{1,y}(\tilde{n}',\tilde{n}) &\equiv  - 2 \pi i \zeta b^{-1} F_A^* F_B^*  [\nu(\tilde{n}'-\tilde{n})+1]  e^{i [\nu(\tilde{n}'-\tilde{n})+1]  \varphi_1 }  \nonumber  \\
 & \times  \left[ 
 \begin{smallmatrix}
 i \frac{1}{a^2 \sin\varphi_1 } \csc (\varphi_1 ) ([\nu(\tilde{n}'-\tilde{n})+1] +i \cot (\varphi_1 )) & - \frac{1}{a^2 \sin\varphi_1 }  \sec (\varphi_1 ) (\tan (\varphi_1 )+i [\nu(\tilde{n}'-\tilde{n})+1]  ) & i  \frac{[\nu(\tilde{n}'-\tilde{n})+1] }{a^2\sin\varphi_1 } \\
- i \frac{1}{a^2 \cos \varphi_1 } \csc (\varphi_1 ) ( [\nu(\tilde{n}'-\tilde{n})+1]  +i \cot (\varphi_1 )) & \frac{1}{a^2 \cos\varphi_1 }  \sec (\varphi_1 ) (\tan (\varphi_1 )+i [\nu(\tilde{n}'-\tilde{n})+1]  ) & - i  \frac{[\nu(\tilde{n}'-\tilde{n})+1] }{a^2\cos\varphi_1 } \\
i \frac{1}{a^2 } \csc (\varphi_1 ) ([\nu(\tilde{n}'-\tilde{n})+1]  +i \cot (\varphi_1 )) & - \frac{1}{a^2}  \sec (\varphi_1 ) (\tan (\varphi_1 )+i [\nu(\tilde{n}'-\tilde{n})+1]  )& i  \frac{[\nu(\tilde{n}'-\tilde{n})+1] }{a^2}
 \end{smallmatrix}
 \right]   \nonumber \\
& + 2 \pi i \zeta b^{-1} F_A F_B  [\nu(\tilde{n}'-\tilde{n})-1]  e^{i [\nu(\tilde{n}'-\tilde{n})-1] \varphi_1 }  \nonumber   \\
 & \times  \left[ 
 \begin{smallmatrix}
 i \frac{1}{a^2 \sin\varphi_1 } \csc (\varphi_1 ) ([\nu(\tilde{n}'-\tilde{n})-1] +i \cot (\varphi_1 )) & - \frac{1}{a^2 \sin\varphi_1 }  \sec (\varphi_1 ) (\tan (\varphi_1 )+i [\nu(\tilde{n}'-\tilde{n})-1]  ) & i  \frac{[\nu(\tilde{n}'-\tilde{n})-1] }{a^2\sin\varphi_1 } \\
- i \frac{1}{a^2 \cos \varphi_1 } \csc (\varphi_1 ) ( [\nu(\tilde{n}'-\tilde{n})-1]  +i \cot (\varphi_1 )) & \frac{1}{a^2 \cos\varphi_1 }  \sec (\varphi_1 ) (\tan (\varphi_1 )+i [\nu(\tilde{n}'-\tilde{n})-1]  ) & - i  \frac{[\nu(\tilde{n}'-\tilde{n})-1] }{a^2\cos\varphi_1 } \\
-i \frac{1}{b^2 } \csc (\varphi_1 ) ([\nu(\tilde{n}'-\tilde{n})-1]  -i \cot (\varphi_1 )) & - \frac{1}{b^2}  \sec (\varphi_1 ) (\tan (\varphi_1 )+i [\nu(\tilde{n}'-\tilde{n})-1]  )& -i  \frac{[\nu(\tilde{n}'-\tilde{n})-1] }{b^2}
 \end{smallmatrix}
 \right]   \label{eq:Ms2}
\end{align}

\begin{align}
M_{1,z}(\tilde{n}',\tilde{n}) &\equiv 2 \pi b^{-1} ( |F_A|^2 - \zeta^2 |F_B|^2)  \nu(\tilde{n}-\tilde{n}')  e^{i \nu(\tilde{n}-\tilde{n}')  \varphi_1 }   \nonumber  \\
 & \times  \left[ 
 \begin{smallmatrix}
 i \frac{1}{a^2 \sin\varphi_1 } \csc (\varphi_1 ) (\nu(\tilde{n}-\tilde{n}') +i \cot (\varphi_1 )) & - \frac{1}{a^2 \sin\varphi_1 }  \sec (\varphi_1 ) (\tan (\varphi_1 )+i \nu(\tilde{n}-\tilde{n}')  ) & i  \frac{\nu(\tilde{n}-\tilde{n}') }{a^2\sin\varphi_1 } \\
- i \frac{1}{a^2 \cos \varphi_1 } \csc (\varphi_1 ) ( \nu(\tilde{n}-\tilde{n}')  +i \cot (\varphi_1 )) & \frac{1}{a^2 \cos\varphi_1 }  \sec (\varphi_1 ) (\tan (\varphi_1 )+i \nu(\tilde{n}-\tilde{n}')  ) & - i  \frac{\nu(\tilde{n}-\tilde{n}') }{a^2\cos\varphi_1 } \\
-i \frac{1}{b^2 } \csc (\varphi_1 ) (\nu(\tilde{n}-\tilde{n}') + i \cot (\varphi_1 )) & - \frac{1}{b^2}  \sec (\varphi_1 ) (\tan (\varphi_1 )+i \nu(\tilde{n}-\tilde{n}
) )& -i  \frac{\nu(\tilde{n}-\tilde{n}') }{b^2}
 \end{smallmatrix}
 \right]  \label{eq:Ms3}
\end{align}
and similarly for $M_{2,x}$, $M_{2,y}$
 and $M_{2,z}$. For the energy denominator of our perturbation expression, we start from the linear energy dispersion relation (see Eq.~109, Ref.~\cite{varela2016effective}):
$$ E_{n,s}^{\nu,\zeta} = |T| \tilde{n} - s\nu \frac{ \sqrt{ T^2 + (4\lambda_{SO}^{in})^2 } }{ 2} $$
with $\tilde{n} = n / 2M \mathcal{N}$ according to Ref.~\cite{varela2016effective}. In their follow-up work, Ref.~\cite{lopez2022radiation}, the quantity that multiplies $|T|$ is $n$, not $\tilde{n}$ (see Eq.~13 in Ref.~\cite{lopez2022radiation}).    The spin dependence of the denominator introduces degeneracies in the system, which manifest themselves as singularities in the non-degenerate perturbation expansion.  To circumvent this, one may justify the approximation $E_{n,s}^{\nu,\zeta} \approx |T| n$ at large $n$.   Secondly, we may invoke degenerate perturbation theory, which excludes singular terms from the perturbation summations.  On a practical level, singularities can be excluded from the denominator by taking $s=s'$, which could be justified physically based on the spin-momentum locking property of electron transport in CISS, which ultimately originates from the spin-orbit coupling interaction. This leads to:
\begin{equation}
 \mathcal{H}_{eff}^{DD} = \left( \frac{\mu_0}{4\pi}\right)^2 \gamma_I^2 \gamma_S^2 \sum_{n,n'} \sum_{\alpha,\beta} \sum_{\alpha',\beta'} I_1^\alpha  \frac{ 
M_{1,\beta}^{\alpha\beta}(\tilde{n}',\tilde{n})M_{2,\beta'}^{\alpha'\beta'}(\tilde{n},\tilde{n}')
}{|T| ( n'-n )    } I_2^{\alpha'} f(\tilde{n})[1-f(\tilde{n}')] + c.c. \label{eq:enan}
\end{equation}
We will take the limit of zero temperature. There are lots of terms.  All terms can be obtained analytically in closed form and pose no difficulty.  We will therefore not work out all terms except to illustrate the fate of different functional forms encountered.  Setting $\xi=0$ for simplicity, we find terms of the form
$$  \sum_{n,n'=0}^M (\tilde{n}-\tilde{n}') e^{i \nu (\tilde{n}-\tilde{n}') \varphi_1} e^{i \nu (\tilde{n}-\tilde{n}') \varphi_2}  f(\tilde{n})[1-f(\tilde{n}')] \frac{1}{\tilde{n}-\tilde{n}'},$$
which, in the low temperature limit, becomes
$$=  \sum_{\tilde{n}=0}^{\tilde{n}_F} \sum_{\tilde{n}'=\tilde{n}_F}^M e^{i \nu (\tilde{n}-\tilde{n}') (\varphi_1-\varphi_2) } = \frac{e^{-\frac{i (\tilde{n}_F-1) \nu (\varphi_1-\varphi_2) }{2 M \mathcal{N}}} \left( e^{\frac{i (\tilde{n}_F+1) \nu (\varphi_1-\varphi_2) }{2 M \mathcal{N}}} - 1 \right) \left( e^{\frac{i \nu (\varphi_1-\varphi_2) (\tilde{n}_F-M-1)}{2 M \mathcal{N} }} - 1 \right)}{\left( e^{\frac{i \nu  (\varphi_1-\varphi_2) }{2 M \mathcal{N} }} - 1 \right)^2}. $$
This term does not become singular unless $\varphi_1=\varphi_2$ (two localized spins at the same exact position), which goes against the idea of coupling two distinct spins localized in different positions.

Also, we have terms of the type:
$$  \sum_{n,n'=0}^M (\tilde{n}-\tilde{n}')^2 e^{i \nu (\tilde{n}-\tilde{n}') \varphi_1} e^{i \nu (\tilde{n}-\tilde{n}') \varphi_2}  f(\tilde{n})[1-f(\tilde{n}')] \frac{1}{\tilde{n}-\tilde{n}'}=  \sum_{\tilde{n}=0}^{\tilde{n}_F} \sum_{\tilde{n}'=\tilde{n}_F}^M(\tilde{n}-\tilde{n}') e^{i \nu (\tilde{n}-\tilde{n}') (\varphi_1-\varphi_2) }  $$
which can be obtained by differentiating the previous expression with respect to $(\varphi_1-\varphi_2)$ and dividing by $i\nu$. The result is:
$$ = \frac{ e^{-\frac{i (\tilde{n}_F-1) \nu (\varphi_1-\varphi_2) }{2 M \mathcal{N} }}}{2 M \mathcal{N} \left(e^{\frac{i \nu (\varphi_1-\varphi_2) }{2 M \mathcal{N}}} - 1 \right)^3} \bigl(-2 e^{\frac{i (\tilde{n}_F+1) \nu  (\varphi_1-\varphi_2) }{2 M \mathcal{N} }}+(\tilde{n}_F+1) e^{\frac{i \nu (\varphi_1-\varphi_2) }{2 M \mathcal{N} }}+(\tilde{n}_F-M+1) e^{\frac{i \nu (\varphi_1-\varphi_2) (2 \tilde{n}_F-M)}{2 M \mathcal{N} }} $$
$$+M e^{\frac{i \nu (\varphi_1-\varphi_2) (\tilde{n}_F-M-1)}{2 M \mathcal{N} }}-(M+2) e^{\frac{i \nu (\varphi_1-\varphi_2) (\tilde{n}_F-M)}{2 M \mathcal{N} }}+(M+1-\tilde{n}_F) e^{\frac{i \nu (\varphi_1-\varphi_2) (2 \tilde{n}_F-M+1)}{2 M \mathcal{N} }}-\tilde{n}_F+1\bigr). $$

\subsubsection*{Enantiospecific NMR Response\label{eq:enansp}}

It should now be clear that the effective dipole-dipole coupling tensor is enantiospecific upon inspection of Eq.~(\ref{eq:enan}), which contains a product $M_{1,\beta}^{\alpha\beta}(\tilde{n}',\tilde{n})M_{2,\beta'}^{\alpha'\beta'}(\tilde{n},\tilde{n}')$.  Explicitly, this term is:
\begin{multline*}
\sum_{\beta,\beta'} M_{1,\beta}^{\alpha\beta}(\tilde{n}',\tilde{n})M_{2,\beta'}^{\alpha'\beta'}(\tilde{n},\tilde{n}')=\bigl[ M_{1,x}^{\alpha,1}(\tilde{n}',\tilde{n}) + M_{1,y}^{\alpha,2}(\tilde{n}',\tilde{n}) + M_{1,z}^{\alpha,3}(\tilde{n}',\tilde{n}) \bigr]  \\
 \times  \bigl[  M_{2,x}^{\alpha',1}(\tilde{n},\tilde{n}') + M_{2,y}^{\alpha',2}(\tilde{n},\tilde{n}') + M_{2,z}^{\alpha',3}(\tilde{n},\tilde{n}') \bigr].
\end{multline*}
While $M_{i,z}$ is independent of $\zeta$, both $M_{i,x}$ and $M_{i,y}$ depend linearly on $\zeta$.  The tensor (indices $\alpha,\alpha'$)
contains terms such as $M_{1,z} M_{2,x}$, which depend linearly on $\zeta$.  The effect of enantiomer handedness is to flip the sign of this term, leading to a change in the magnitude of the dipole-dipole interaction.  The term $M_{1,z}^{\alpha,3}(\tilde{n}',\tilde{n}) M_{2,z}^{\alpha',3}(\tilde{n},\tilde{n}')$ does not depend on $\zeta$, since neither factor depend on $\zeta$.   Neither do 
$M_{1,x}^{\alpha,1}(\tilde{n}',\tilde{n}) M_{2,x}^{\alpha',1}(\tilde{n},\tilde{n}')$ and $M_{1,y}^{\alpha,2}(\tilde{n}',\tilde{n}) M_{2,y}^{\alpha',2}(\tilde{n},\tilde{n}')$ since $\zeta^2=1$.

\subsubsection*{Order of magnitude estimate}

From this equation:
$$ \mathcal{H}_{eff}^{DD} = \left( \frac{\mu_0}{4\pi}\right)^2 \gamma_I^2 \gamma_S^2 \hbar^4 \sum_{n,n'} \sum_{\alpha,\beta} \sum_{\alpha',\beta'} I_1^\alpha  \frac{ 
M_{1,\beta}^{\alpha\beta}(\tilde{n}',\tilde{n})M_{2,\beta'}^{\alpha'\beta'}(\tilde{n},\tilde{n}')
}{|T| (n'-n) } I_2^{\alpha'} f(\tilde{n})[1-f(\tilde{n}')] + c.c. $$
Taking a typical term from the matrix $M$ the magnitude is 
$$ \left|\mathcal{H}_{eff}^{DD}\right| \sim  \left( \frac{\mu_0}{4\pi}\right)^2 \gamma_I^2 \gamma_S^2  \hbar^4 \sum_{\alpha,\beta} \sum_{\alpha',\beta'}  
\left|\frac{ 
(2 \pi ( |F_A|^2 - \zeta^2 |F_B|^2)  )^2
}{|T|  } \right|  \frac{1}{b^2 a^4} $$
where $|F_A|, |F_B| \le 1$. So,
$$ \left|\mathcal{H}_{eff}^{DD}\right| \sim  \left( \frac{\mu_0}{4\pi}\right)^2 \gamma_I^2 \gamma_S^2  \hbar^4 16 
\left|\frac{ 
(4 \pi   )^2
}{T (2M\mathcal{N}) } 
 \right|  \frac{1}{b^2 a^4} \sim  \left( \mu_0\right)^2 \gamma_I^2 \gamma_S^2 \hbar^4  16 
\left|\frac{ 
  1
}{T (2M\mathcal{N}) } \right|  \frac{1}{b^2 a^4}  \sim   
\frac{ 
 2 \mu_0^2 \gamma_I^2 \gamma_S^2  \hbar^4 
}{|T|(2M\mathcal{N}) }  \frac{1}{b^2 a^4}. $$
According to Varela, $T=2 t_{in}$, $t_{in}=-10$ meV = 1.6 $\times 10^{-21}$ J.   Plugging numbers:
$$ \left|\mathcal{H}_{eff}^{DD}\right|  \sim   
\frac{ 
 2 \mu_0^2 (\gamma_I\hbar)^2 (\gamma_S\hbar)^2 
}{|T|(2M\mathcal{N})}  \frac{1}{b^2 a^4} $$
$$\approx \frac{ 2 (1.257 \times 10^{-6} \mathrm{N\cdot A^{-2})^2} (\mathrm {-1.76\times 10^{11}\,rad{\cdot }s^{-1}{\cdot }T^{-1}} )^2 (\mathrm {267.5\times 10^{6}\,rad{\cdot }s^{-1}{\cdot }T^{-1}} )^2 }{ (1.6\times 10^{-21} ~\mbox{J}) (2 \times 10^{-9}~\mbox{m})^2 (3.38 \times 10^{-9}~\mbox{m})^4 (2M\mathcal{N})  } \cdot (1.05\times 10^{-34}~\mbox{J.s})^4 $$ 
$$\approx \frac{1.02\times 10^{-36}~\mbox{J}}{ 2M\mathcal{N} } \approx \frac{0.01~\mbox{Hz} }{2M\mathcal{N} }. $$
The reader can verify that the units are Joules.
But 0.01 Hz/$(2M\mathcal{N}) $ is not the full story.  We also have multiplicative factors of the form:
$$\frac{e^{-\frac{i (\tilde{n}_F-1) \nu (\varphi_1-\varphi_2) }{2 M \mathcal{N}}} \left( e^{\frac{i (\tilde{n}_F+1) \nu (\varphi_1-\varphi_2) }{2 M \mathcal{N}}} - 1 \right) \left( e^{\frac{i \nu (\varphi_1-\varphi_2) (\tilde{n}_F-M-1)}{2 M \mathcal{N} }} - 1 \right)}{\left( e^{\frac{i \nu  (\varphi_1-\varphi_2) }{2 M \mathcal{N} }} - 1 \right)^2}, $$
which blow up when
$|\varphi_1-\varphi_2|/(2 M \mathcal{N})$ is an integer multiple of $2\pi$, i.e. when $|\varphi_1-\varphi_2|=4\pi M \mathcal{N}m$, $m\in\mathbb{N}$.  We note that when the denominator blows up the numerator does not, because it depends on the Fermi energy $\tilde{n}_F$.
As to the denominator, we mentioned earlier that it blows up when  $\varphi_1=\varphi_2$, a situation that is not allowed (or worst, when $M\mathcal{N}\rightarrow\infty$).

Finally, we also have factors of tan$(\varphi_1)$, csc$(\varphi_1)$, cot$(\varphi_1)$ and sec$(\varphi_1)$, all of which can ``blow up'' depending on the argument to the function ($\varphi_1$ or $\varphi_2$ here).  For example, tan and sec blow up at $(n + \frac{1}{2})\pi$, $n\in \mathbb{Z}$ whereas csc and cot blow up at $n\pi$.  Recall that csc$(\varphi_1)=1/\sin(\varphi_1)$ and sec$(\varphi_1)=1/\cos(\varphi_1)$. Some of the terms are:  

1) sec$(\varphi_1)\tan(\varphi_1)/\sin(\varphi_1)=$sec$^2(\varphi_1)$. Similar to sec, this functions blows up at the points $(n + \frac{1}{2})\pi$. 

2) csc$(\varphi_1)/\sin(\varphi_1)=$csc$^2(\varphi_1)$, which blows up at $n\pi/2$. 

3) sec$(\varphi_1)/\sin(\varphi_1)=$csc$(\varphi_1)$sec$(\varphi_1)$, which blows up at 
$(n + \frac{1}{2})\pi$. 

4) csc$(\varphi_1)$cot$(\varphi_1)/\sin(\varphi_1)=$csc$^2(\varphi_1)$cot$(\varphi_1)$, which blows up at the points $n\pi$.  We rule out this scenario, as the two nuclear spins are not likely to be located at the center of the molecule. \\

\noindent Let's plot one such matrix element. Let's take the upper left hand corner element. Recall that $\tilde{n} = n / 2M \mathcal{N}$ and $0 \le n \le M$.  Let's choose Fermi level to be in the middle ($n=M/2$ so that $\tilde{n} = 1/4\mathcal{N}$, so that the levels  are half-filled). We will stick with $\mathcal{N}=1$, $M=10$ so that $\tilde{n}_F = 1/40$.  The amplitude of the $I_1^z \mathtensor{F}_{zz}  I_2^z$ term is plotted as a function of $\varphi_1,\varphi_2 \in [0,2\pi]$ in Fig. 2a of the main text. 

\subsection{Fermi Contact Term}

The Fermi contact term:
$$ \mathcal{H}_{eff}^{FC} =  \left(\frac{2 \mu_0}{3} \right)^2 \gamma_I^2 \gamma_S^2 \hbar^4 \sum_{\mathbf{k},s} \sum_{\mathbf{k}',s'}  \mathbf{I}_1 \cdot \frac{ \braket{ \mathbf{k}s | \sum_l \mathbf{S}_l \delta^{(3)}(\mathbf{r}_l - \mathbf{R}_1) | \mathbf{k}'s' }  \braket{ \mathbf{k}'s' | \sum_l \mathbf{S}_l \delta^{(3)}(\mathbf{r}_l - \mathbf{R}_2)| \mathbf{k}s }  }{ E_{\mathbf{k}s} - E_{\mathbf{k}'s' } } \cdot \mathbf{I}_2 + c.c.$$
involves computing integrals of the form
$$ \braket{ \tilde{n}' s | \sum_l \mathbf{S}_l \delta^{(3)}(\mathbf{r}_l - \mathbf{R}_1) | \tilde{n} s }   \equiv  \braket{ \boldsymbol \psi_{n',s}^{\nu,\zeta}  | \sum_l \mathbf{S}_l \delta^{(3)}(\mathbf{r}_l - \mathbf{R}_1) | \boldsymbol \psi_{n,s}^{\nu,\zeta}  }  $$
using the Varela spinors
$$ \boldsymbol \psi_{n,s}^{\nu,\zeta} = \left[ \begin{matrix}
F_A e^{-i \varphi/2} \\
\zeta F_B^* e^{i \varphi/2} 
\end{matrix} \right] e^{i \nu \tilde{n} \varphi}, \quad F_A = \frac{ \sqrt{s}}{2} ( s e^{i \theta/2} + e^{-i\theta/2} ), \quad F_B = \frac{\sqrt{s}}{2} ( s e^{ -i\theta/2} - e^{i\theta/2} ). $$
There are 3 such terms, one for $S_x$, $S_y$ and $S_z$:
$$ \braket{ \boldsymbol \psi_{n',s}^{\nu,\zeta}  | \sum_l \frac{\sigma_x}{2} \delta^{(3)}(\mathbf{r}_l - \mathbf{R}_1) | \boldsymbol \psi_{n,s}^{\nu,\zeta}  }, \,
\braket{ \boldsymbol \psi_{n',s}^{\nu,\zeta}  | \sum_l \frac{\sigma_y}{2} \delta^{(3)}(\mathbf{r}_l - \mathbf{R}_1) | \boldsymbol \psi_{n,s}^{\nu,\zeta} ) } \\
   \braket{ \boldsymbol \psi_{n',s}^{\nu,\zeta}  | \sum_l \frac{\sigma_z}{2} \delta^{(3)}(\mathbf{r}_l - \mathbf{R}_1) | \boldsymbol \psi_{n,s}^{\nu,\zeta}  }.   $$
We may temporarily denote these terms as $s_x$, $s_y$ and $s_z$, respectively. See Section ``Matrix Elements of the 3D Dirac Delta Function'' for a derivation.    They give rise to an effective spin-spin coupling tensor:
$$ \left[ \begin{matrix}
s_x s_x & s_x s_y & s_x s_z \\
s_y s_x & s_y s_y & s_y s_z \\
s_z s_x & s_z s_y & s_z s_z 
\end{matrix} \right]  = \dots  $$
Taking care of ensuring that the $s$ on the left involves a braket of the form $\braket{n'|\cdot |n}$ while the $s$ on the right is of the form $\braket{n|\cdot |n'}$.   Because the Fermi contact term does not contain any singularities we expect that its magnitude may be too weak for cross-polarization.   For indirect electron-electron coupling, we gain a factor $10^6$ from the stronger gyromagnetic ratios.

Our Fermi contact contribution becomes
\begin{equation} \mathcal{H}_{eff}^{FC} = \left( \frac{ 2\mu_0}{3} \right)^2  \frac{ 4^2 \gamma_I^2 \gamma_S^2 \hbar^4}{ (2\pi s) a^4 b^2 |\varphi_1| |\varphi_2| } \sum_{n,n'} \mathbf{I}_1 \cdot \frac{ \boldsymbol{\mathsf{F}}_c(\tilde{n}',\tilde{n})
  }{ |T|(n'-n) } \cdot \mathbf{I}_2 f(\tilde{n}) [1-f(\tilde{n}')] + c.c.
\label{eq:fcc}
\end{equation}
where the Fermi-contact coupling tensor $\boldsymbol{\mathsf{F}}_c(\tilde{n}',\tilde{n})$ is given by the outer product (dyadic):
$$ \boldsymbol{\mathsf{F}}_c(\tilde{n}',\tilde{n}) =   \left[ \begin{matrix}
 \zeta (F_A^* F_B^* e^{i\varphi_1} + F_A F_B e^{- i\varphi_1})  e^{ - i \nu (\tilde{n}'-\tilde{n}) \varphi_1} \\
 i \zeta (F_A F_B e^{-i\varphi_1}  - F_A^* F_B^* e^{i\varphi_1} ) e^{ - i \nu (\tilde{n}'-\tilde{n}) \varphi_1}  \\
 (|F_A|^2 -|F_B|^2) e^{-i\nu (\tilde{n}'-\tilde{n})\varphi_1}  
\end{matrix} \right] \otimes  \left[ \begin{matrix}
 \zeta (F_A^* F_B^* e^{i\varphi_2} + F_A F_B e^{- i\varphi_2} )  e^{ - i  \nu (\tilde{n}-\tilde{n}') \varphi_2} \\
 i \zeta (F_A F_B e^{- i\varphi_2}  - F_A^* F_B^* e^{i\varphi_2} ) e^{ - i  \nu (\tilde{n}-\tilde{n}') \varphi_2}  \\
 (|F_A|^2 -|F_B|^2) e^{-i\nu (\tilde{n}-\tilde{n}')\varphi_2}  
\end{matrix} \right]^T.  $$
Summation over $n,n'$, i.e. $\boldsymbol{\mathsf{F}} \equiv \sum_{n,n'}  \boldsymbol{\mathsf{F}}_c(\tilde{n}',\tilde{n}) f(\tilde{n}) [1-f(\tilde{n}')] $, is straightforward, as explained previously. This yields the tensor components $\boldsymbol{\mathsf{F}}_{xx}$, $\boldsymbol{\mathsf{F}}_{xy}$, $\boldsymbol{\mathsf{F}}_{xz}$, $\boldsymbol{\mathsf{F}}_{yx}$, $\boldsymbol{\mathsf{F}}_{yy}$, $\boldsymbol{\mathsf{F}}_{yz}$, $\boldsymbol{\mathsf{F}}_{zx}$, $\boldsymbol{\mathsf{F}}_{zy}$ and $\boldsymbol{\mathsf{F}}_{zz}$.

In high-field NMR the  coefficient of the $I_z^1I_z^2$ term, $\boldsymbol{\mathsf{F}}_{zz}$ matters most. Here, $\boldsymbol{\mathsf{F}}_{zz}$ does not depend on $\zeta$. Consequently, in high-field NMR we do not expect the Fermi contact contribution to the indirect coupling to be enantioselective.  We note that along the line $\varphi_1=\varphi_2$ we have 0 times $\infty$, which is undefined.  This is not a problem since $\varphi_1 \ne \varphi_2$ is a requirement. The interaction strength is plotted as a function of $\varphi_1,\varphi_2 \in [0,2\pi]$ in Fig. 2b of the main text. Outside of the $\varphi_1=\varphi_2$ region ($\varphi_1 \ne \varphi_2$) the magnitude is at most 10, which is a rather weak mechanism for CP.  We note that the factor $\frac{1}{|\varphi_1| |\varphi_2|}$ in Eq.~(\ref{eq:fcc}) could in principle ``blow up'' if $|\varphi_1|=0$,  $|\varphi_2|=0$ or both.   Firstly, we cannot have both nuclear spins in the same location ($|\varphi_1|= |\varphi_2|=0$).  Also, even if one spin is close to the edge of the molecule, it is unlikely to be exactly at the position $|\varphi_1|=0$ since the electronic wavefunction always extends beyond the nucleus.  The distance of closest approach would be bounded from below by (at least) the Bohr radius ($\sim 0.5 \times 10^{-10}$ m).  For a helix pitch of 3.38 nm, this would correspond to a lower bound of $\sim 0.1$ radians, which is still at least 6 orders of magnitude weaker than the dipole interaction contribution.

\subsection{Is an Applied Current Needed to Drive this Effective Interaction?\label{sec:probcurrent}}

Intuitively, a current is needed to mediate the coupling between two localized spins.  One may ask whether or not a voltage must be applied across the molecule.  Fortunately, such a current naturally arises in quantum mechanics through the concept of probability current.  Indeed, an electron in state $\psi(x)$ gives rise to a current density
$$ \vec{j}(x) = - \frac{ e\hbar }{ 2 m_e i } ( \psi^*(x) \nabla \psi(x) - \psi(x) \nabla \psi^*(x) ). $$
For spinors, the expression is  
$$ \vec{j}(x) = - \frac{ e\hbar }{ 2 m_e i } (  \boldsymbol\psi^\dagger  \nabla \boldsymbol\psi  - 
  (\nabla \boldsymbol \psi^\dagger ) \boldsymbol \psi  ). $$
In order to compute this current, we need the derivative operator along the arc length of the curve.
The metric tensor in local coordinates for the parametrized curve $ \varphi \mapsto (a\cos \varphi, a\sin \varphi, b \varphi), \varphi\in [0,T] $ is:
$$ g_{\varphi,\varphi} = \langle \frac{ \partial }{ \partial \varphi} , \frac{ \partial }{ \partial \varphi}  \rangle = (-a \sin\varphi, a\cos\varphi , b) \cdot (-a \sin\varphi, a\cos\varphi , b) = a^2 \sin^2\varphi + a^2 \cos^2\varphi + b^2 = a^2+b^2. $$
Denoting the length by $\ell$, the element of length is:
$$ d\ell^2 = g_{\varphi,\varphi}  d\varphi \otimes d\varphi = (a^2+b^2) d\varphi^2, \qquad  d\ell = \sqrt{ a^2+b^2} d\varphi. $$
We will also need to make a slight adjustment to the Varela spinors, which are defined over the interval $[0,2\pi]$, and do not have units of inverse square root length.  Normally, quantum mechanical wavefunctions $|\psi(x)|$ have dimensions of $1/\sqrt{x}$ and we have the normalization $ \int_0^L  |\psi(x)|^2 dx=1$ whereas here we have $ \frac{1}{2\pi} \int_0^{2\pi}  |\psi(\varphi)|^2 d\varphi = 1$. Consider the following mapping and associated change of variables:
$$ \varphi \mapsto \ell=\sqrt{a^2+b^2} \varphi, \qquad \varphi = \frac{\ell }{ \sqrt{a^2+b^2} }, \qquad d\varphi =  \frac{d\ell}{ \sqrt{a^2+b^2} }. $$
The integral becomes:
$$ \frac{1}{2\pi} \int_0^{2\pi \sqrt{a^2+b^2} } \frac{1 }{ \sqrt{a^2+b^2} }   |\psi( \frac{\ell }{ \sqrt{a^2+b^2} } )|^2 d\ell  = 1. $$
or (with $L=2\pi \sqrt{a^2+b^2}$):
$$  \int_0^{2\pi \sqrt{a^2+b^2} } \left( \frac{ \left|\psi \left( \frac{ \ell }{ \sqrt{a^2+b^2} } \right) \right|^2}{ 2\pi \sqrt{a^2+b^2} } \right) d\ell  = 1. $$
This means that the wavefunction with proper units is obtained by rescaling:
$$ \psi \mapsto \frac{ \psi \left( \frac{ \ell }{ \sqrt{a^2+b^2} } \right) }{ \sqrt{2\pi} (a^2+b^2)^{1/4} }.  $$
Taking the derivative of the spinor:
$$ \boldsymbol \psi_{n,s}^{\nu,\zeta}  = \left[ \begin{matrix}
F_A e^{-i \varphi/2} \\
\zeta F_B^* e^{i \varphi/2} 
\end{matrix} \right] e^{i \nu \tilde{n} \varphi}, $$
we get:
$$\nabla \boldsymbol \psi   =  \frac{ d}{ d\ell } \frac{ \boldsymbol \psi( \tfrac{ \ell }{\sqrt{a^2+b^2}} ) }{(a^2+b^2)^{1/4}}  =   \frac{ i }{\sqrt{ a^2+b^2} (a^2+b^2)^{1/4} } \left[ \begin{matrix}
F_A e^{-i \varphi/2}  (\nu \tilde{n} -\frac{1}{2})\\
\zeta F_B^* e^{i \varphi/2} (\nu \tilde{n} +\frac{1}{2})
\end{matrix} \right] e^{i \nu \tilde{n} \varphi} $$
and
$$ (\nabla \boldsymbol \psi^\dagger )  = \frac{-i }{ \sqrt{ a^2+b^2} (a^2+b^2)^{1/4} }  \left[ \begin{matrix}
F_A^* e^{i \varphi/2} (\nu \tilde{n} -\frac{1}{2}) &
\zeta F_B e^{-i \varphi/2} (\nu \tilde{n} +\frac{1}{2})
\end{matrix} \right] e^{- i \nu \tilde{n} \varphi}. $$
Computing the products of spinors:
$$ \boldsymbol \psi^\dagger  \nabla \boldsymbol \psi  = \left[ \begin{matrix}
F_A^* e^{i \varphi/2} &
\zeta F_B e^{- i \varphi/2} 
\end{matrix} \right] e^{ - i \nu \tilde{n} \varphi}  \frac{i }{ \sqrt{ a^2+b^2} (a^2+b^2)^{1/2} } \left[ \begin{matrix}
F_A e^{-i \varphi/2} (\nu \tilde{n} -\frac{1}{2}) \\
\zeta F_B^* e^{i \varphi/2} (\nu \tilde{n} +\frac{1}{2})
\end{matrix} \right] e^{i \nu \tilde{n} \varphi}  $$
$$= \frac{i }{ \sqrt{ a^2+b^2} (a^2+b^2)^{1/2} } (|F_A|^2 (\nu \tilde{n} -\tfrac{1}{2}) + |F_B|^2 (\nu \tilde{n} +\tfrac{1}{2}) ). $$
Similarly,
$$  (\nabla \boldsymbol \psi^\dagger ) \boldsymbol \psi =  \frac{- i  }{ \sqrt{ a^2+b^2} (a^2+b^2)^{1/2} } (|F_A|^2 (\nu \tilde{n} - \tfrac{1}{2}) + |F_B|^2 (\nu \tilde{n} +\tfrac{1}{2}) ). $$
Noting\footnote{$$ |F_A|^2 = F_A^* F_A = \frac{|s|}{4} ( s e^{-i \theta/2} + e^{i\theta/2} ) ( s e^{i \theta/2} + e^{-i\theta/2} ) = \frac{1}{4} (s^2 + s(e^{-i \theta} + e^{i\theta}) + 1 ) =  \frac{1}{2} (1 + s \cos\theta) $$
$$ |F_B|^2 = F_B^* F_B = \frac{|s|}{4} ( s e^{i \theta/2} - e^{- i\theta/2} ) ( s e^{- i \theta/2} - e^{i\theta/2} ) = \frac{1}{4} (s^2 - s(e^{-i \theta} + e^{i\theta}) + 1 ) =  \frac{1}{2} (1 - s \cos\theta) $$
$$ |F_A|^2 +  |F_B|^2 = 1,\qquad  |F_A|^2 -  |F_B|^2 = s \cos\theta $$} that $ |F_A|^2 +  |F_B|^2 = 1$, and
$$ |F_A|^2 (\nu \tilde{n} -\tfrac{1}{2}) + |F_B|^2 (\nu \tilde{n} +\tfrac{1}{2}) = \nu\tilde{n} (|F_A|^2 + |F_B|^2) + \tfrac{1}{2}(|F_B|^2 - |F_A|^2) = \nu \tilde{n} - \tfrac{s}{2} \cos\theta $$
we obtain the formula
$$ j =  \frac{ 2 e\hbar  }{  m_e(a^2+b^2)} (\nu \tilde{n} - \tfrac{s}{2} \cos\theta). $$
This is because
$$ \psi^\dagger \frac{d}{d \ell } \psi    \rightarrow \frac{ \psi^\dagger}{ (a^2+b^2)^{1/4} } \frac{ d}{ d \ell } \left( \frac{ \psi}{ (a^2+b^2)^{1/4} }  \right). $$
Taking the derivative $d/d\ell$ brings down a factor $1/\sqrt{a^2+b^2}$.  The overall length scaling factor is:
$$ \frac{ 1 }{ (a^2+b^2)^{1/4} }     \frac{ 1 }{ \sqrt{a^2+b^2} }     \frac{ 1 }{ (a^2+b^2)^{1/4} }  = \frac{ 1 }{ a^2+b^2 }. $$

\noindent Plugging some numbers in the multiplicative coefficient:
$$ j = \frac{ (2)(1.6 \times 10^{-19}~\mbox{C})(1.05\times 10^{-34}~\mbox{J.s}) }{ (9.1\times 10^{-31} ~\mbox{kg}) [(2\times 10^{-9}~\mbox{m})^2 + (3.38 \times 10^{-9}~\mbox{m})^2] } \left( \frac{n}{2M\mathcal{N}} + \tfrac{s}{2}\cos\theta \right), $$
or
$$ j =  2 \left( \frac{n}{2M\mathcal{N}} + \tfrac{s}{2}\cos\theta \right) (1.2 \times 10^{-6}) ~\mbox{A}. $$
We get the correct units (Amp\`{e}re) after having rescaled the wavefunction.  Let's check the magnitude. 
For $\theta=\frac{\pi}{2}$, $n=M=10$ and  $\mathcal{N}=1$, the current $j$ is 2.4 $\mu$A.  
Since $b=3.38$ nm, the length of one helix is 33.8 nm. (Similarly, $\mathcal{N}=3$, $n=M$ gives a current of 800 nA whereas  $n=1$, $M=10$, $\mathcal{N}=1$, gives 80 nA.)  For comparison, the I-V curves presented in the paper by Wang~\cite{wang2018dna} (see Fig. 4) show currents up to 220 nA.

\subsubsection*{Embedding of 1D Spinor Wavefunction in 3D}

The normalized Varela spinors (see Section ``Normalization of Spinors'')
$$ \boldsymbol \psi (\varphi)  =   \frac{1}{\sqrt{2\pi s}}  \left[ \begin{matrix}
F_A e^{-i \varphi/2} \\
\zeta F_B^* e^{i \varphi/2} 
\end{matrix} \right] e^{i \nu \tilde{n} \varphi} \chi_{[0,2\pi]}(\varphi) $$
one-parameter mappings $\boldsymbol \psi: \varphi \in [0,2\pi] \mapsto \mathbb{C}^2$.   However, computation of the Fermi contact interaction requires three-dimensional wavefunctions because the Dirac delta distribution in the Hamiltonian, $ \delta^{(3)}(\mathbf{r}-\mathbf{r}_0)$, must be integrated over three dimensional space.   It is unclear how to embed a 1D function into a 3D space because there are an infinite number of possibilities.

We can limit the options by invoking a simple argument from probability theory. Let's say we have a random variable $X$ and its associated probability density function, $p_X(x)$.  By definition, it must be normalized:
$$ \int_{-\infty}^\infty  p_X(x) dx =1. $$
If we also want it to be a function of $y$ and $z$ in order to describe the behavior of two additional random variables, $Y$ and $Z$.   If $X$, $Y$ and $Z$ are statistically independent, the joint probability density is a product.  It must be normalized as well:
$$ \int_{-\infty}^\infty \int_{-\infty}^\infty \int_{-\infty}^\infty p_X(x) p_Y(y) p_Z(z) \, dx \, dy \, dz  =1. $$
We must then have:
$$ \int_{-\infty}^\infty  p_Y(y) dy =1, \qquad \int_{-\infty}^\infty  p_Z(z) dz =1. $$
What are the requirements on $p_Y$ and $p_Z$?  In the context of quantum mechanics, the norm of the wavefunction must have dimension of inverse length since $\alpha$ and $\beta$ have dimensions of length.  Let's call these functions $f$ and $g$:
$$ \int_{-\infty}^\infty f(\alpha) d\alpha=1, \qquad \int_{-\infty}^\infty g(\beta) d\beta=1. $$
When averaging a distance, we would like to fix the values of the averages, i.e. $\langle \alpha \rangle=a$, $\langle \beta \rangle=a$, 
$$ \langle \alpha \rangle = \int_{-\infty}^\infty \alpha f(\alpha) d\alpha := a, \qquad  \langle b \rangle = \int_{-\infty}^\infty \beta g(\beta) d\beta := b. $$
The choice of distribution is not so important\footnote{Choosing a different distribution will affect the overall scaling slightly, depending on the moments of the distribution.  For example, if a random variable $X$ follows a Rayleigh distribution with PDF $f(x) = \frac{x}{ \sigma^2} e^{-x^2/(2\sigma^2)}$, $x\ge 0$ we would fulfill the requirements of a continuous wavefunction and first derivative across space.  However, 
the mean is $\langle X \rangle = \sigma \sqrt{\pi/2}$. One may argue that the extra factor $\sqrt{\pi/2}$ does significantly alter the result.  We could also impose additional conditions on the moments if needed.}
as long as the correct averages are obtained when computing the moments of interest.  We choose the uniform distribution for simplicity:
$$ f(\alpha) = \frac{2}{a} ~\mbox{for~} 0 \le \alpha \le a. $$
In the context of a quantum mechanical wavefunction, we would take a one-dimensional wavefunction $\psi(\varphi)$ and construct a three-dimensional function
$$ \Psi(\alpha,\beta,\varphi) = \psi(\varphi) \sqrt{ \frac{2}{a} } \chi_{[0,a]}(\alpha) \sqrt{ \frac{2}{b} }  \chi_{[0,b]}(\beta) $$
such that the probability gives us the correct average position:
\begin{multline*}
\langle \alpha \rangle =  \int_{-\infty}^\infty  \int_{-\infty}^\infty  \int_{-\infty}^\infty  | \Psi(\alpha,\beta,\varphi)|^2 \alpha \, d\beta \, d\alpha \, d\varphi  \\
=\int_{-\infty}^\infty  \int_{-\infty}^\infty  \int_{-\infty}^\infty  | \psi(\varphi) |^2 \frac{2}{a} \chi_{[0,a]}(\alpha) \frac{2}{b} \chi_{[0,b]}(\beta)  \alpha \, d\beta \, d\alpha \, d\varphi =  a
\end{multline*}
\begin{multline*}
\langle \beta \rangle =  \int_{-\infty}^\infty  \int_{-\infty}^\infty  \int_{-\infty}^\infty  | \Psi( \alpha, \beta, \varphi)|^2 \beta \, d\beta \, d\alpha \, d\varphi \\
=  \int_{-\infty}^\infty  \int_{-\infty}^\infty  \int_{-\infty}^\infty  | \psi(\varphi) |^2 \frac{2}{a} \chi_{[0,a]}(\alpha) \frac{2}{b} \chi_{[0,b]}(\beta)  \beta \, d\beta \, d\alpha \, d\varphi =  b.
\end{multline*}
We therefore redefine the Varela spinors as follows:
$$ \boldsymbol \psi(\alpha,\beta,\varphi)   =   \frac{1}{\sqrt{2\pi s}}  \left[ \begin{matrix}
F_A e^{-i \varphi/2} \\
\zeta F_B^* e^{i \varphi/2} 
\end{matrix} \right] e^{i \nu \tilde{n} \varphi} \chi_{[0,2\pi]}(\varphi) \sqrt{ \frac{2}{a} } \chi_{[0,a]}(\alpha) \sqrt{ \frac{2}{b} }  \chi_{[0,b]}(\beta). $$
In this manner, integration over $\mathbb{R}^3$ can be performed.    When computing averages of $\alpha$ and $\beta$, integration over all space of the norm of this wavefunction will return $a$ and $b$, the correct radius and pitch of the helix.

\subsubsection*{Matrix Elements of the 3D Dirac Delta Function\label{sec:avgdir}}

Let us return to the Fermi contact interaction and start by checking that sensible results are obtained upon integration.   In helicoidal coordinates the 3D Dirac distribution is (see Section ``Pull-back of the Delta Function in Curvilinear Coordinates''):
$$ \delta^{(3)}(\mathbf{r}-\mathbf{r}_0) = \frac{1}{|a \varphi |} \delta(a-a_0) \delta(b-b_0) \delta(\varphi-\varphi_0). $$
The three averages needed are ($a>0$, $b>0$, $\varphi>0$ so the absolute values are unimportant):
\begin{multline*}
\braket{ \boldsymbol \psi_{n',s}^{\nu,\zeta}  |  \sigma_x \delta^{(3)}(\mathbf{r} - \mathbf{R}_1) | \boldsymbol \psi_{n,s}^{\nu,\zeta}  } \\
= \lim_{\epsilon \rightarrow 0} \int_{-\infty}^\infty  \int_{-\infty}^\infty  \int_{-\infty}^\infty \frac{1}{ 2\pi s}  (F_A^* F_B^* e^{ i\varphi/2} \zeta  e^{ i\varphi/2} + \zeta F_A F_B  e^{- i\varphi/2}  e^{ - i\varphi/2} ) e^{ i\nu (\tilde{n}-\tilde{n}')\varphi} \\
\times  \frac{1}{|\alpha \varphi |} \delta(\alpha - a + \epsilon) \delta(\beta-b + \epsilon) \delta(\varphi-\varphi_1 + \epsilon)   \frac{2}{a} \chi_{[0,a]}(\alpha) \frac{2}{b} \chi_{[0,b]}(\beta) \chi_{[0,2\pi]}(\varphi) \, d\beta \, d\alpha \, d\varphi \\
=\frac{2}{\pi s}  \frac{1}{a^2b} \zeta ( F_A F_B e^{-i\varphi_1 } + c.c.)  \frac{ e^{  i \nu (\tilde{n}-\tilde{n}')  \varphi_1} }{ |\varphi_1| }
\end{multline*}
where the limiting procedure $\lim_{\epsilon \rightarrow 0}$ was introduced to ensure that the Dirac delta function is concentrated inside the domain of integration, as defined by the indicator functions.  Next, 
\begin{multline*}
\braket{ \boldsymbol \psi_{n',s}^{\nu,\zeta}  | \sigma_y \delta^{(3)}(\mathbf{r} - \mathbf{R}_1) | \boldsymbol \psi_{n,s}^{\nu,\zeta}  }  \\
= \lim_{\epsilon \rightarrow 0} \int_{-\infty}^\infty  \int_{-\infty}^\infty  \int_{-\infty}^\infty \frac{i}{ 2\pi s}(F_A F_B e^{-i\varphi/2} \zeta  e^{-i\varphi/2} - \zeta F_A^* F_B^*  e^{i\varphi/2}  e^{i\varphi/2} ) e^{i\nu (\tilde{n}-\tilde{n}')\varphi} \\ \times  \frac{1}{|\alpha \varphi |} \delta(\alpha - a + \epsilon) \delta(\beta-b + \epsilon) \delta(\varphi-\varphi_1 + \epsilon)   \frac{2}{a} \chi_{[0,a]}(\alpha) \frac{2}{b} \chi_{[0,b]}(\beta) \chi_{[0,2\pi]}(\varphi) \, d\beta \, d\alpha \, d\varphi
\end{multline*}
$$ =\frac{2 i }{ \pi s}  \frac{1}{a^2b} \zeta  (F_A F_B e^{- i\varphi_1}  - c.c. )   \frac{ e^{ i \nu (\tilde{n}-\tilde{n}')  \varphi_1} }{ |\varphi_1| }. $$
Finally,
$$  \braket{ \boldsymbol \psi_{n',s}^{\nu,\zeta}  |  \sigma_z \delta^{(3)}(\mathbf{r} - \mathbf{R}_1) | \boldsymbol \psi_{n,s}^{\nu,\zeta}  }  =  \lim_{\epsilon \rightarrow 0} 
 \int_{-\infty}^\infty  \int_{-\infty}^\infty  \int_{-\infty}^\infty  (|F_A|^2 -|F_B|^2) e^{i\nu (\tilde{n}-\tilde{n}')\varphi}   $$
$$ \qquad \qquad \qquad \qquad   \times \frac{1}{|\alpha \varphi |} \delta(\alpha - a + \epsilon) \delta(\beta-b + \epsilon) \delta(\varphi-\varphi_1 + \epsilon)   \frac{2}{a} \chi_{[0,a]}(\alpha) \frac{2}{b} \chi_{[0,b]}(\beta) \chi_{[0,2\pi]}(\varphi) \, d\beta \, d\alpha \, d\varphi    $$
$$ =\frac{2  }{ \pi s}  \frac{1}{a^2b}  ( |F_A|^2 -|F_B|^2)   \frac{ e^{  i  \nu (\tilde{n}-\tilde{n}')  \varphi_1} }{ |\varphi_1| }. $$

\subsubsection*{Pull-back of the Delta Function in Curvilinear Coordinates}

Consider the mapping:
$$ (a,b,\varphi) \mapsto (x,y,z)=(a \cos \varphi , a \sin \varphi , b \varphi ). $$
The Jacobian of this transformation is:
$$ \frac{ \partial (x,y,z) }{ \partial (a,b,\varphi) } =  \frac{ \partial (a \cos \varphi, a \sin \varphi  , b \varphi) }{ \partial (a,b,\varphi) } = \left[ 
\begin{matrix}
\cos\varphi & 0 & -a\sin\varphi  \\
\sin\varphi & 0 & a\cos\varphi \\
0 & \varphi & b 
\end{matrix} \right]. $$
The determinant of which is $-a \varphi \sin^2\varphi - a\varphi\cos^2 \varphi = -a\varphi$.
Therefore, the 3D Dirac delta function is
$$ \delta^{(3)}(\mathbf{r}-\mathbf{r}_0) = \frac{1}{|a \varphi |} \delta(a-a_0) \delta(b-b_0) \delta(\varphi-\varphi_0). $$
Let $\Phi: A \mapsto B$.  This expression has the following meaning:
$$ \int_{B} f(\mathbf{r}) \boxed{  \delta^{(3)}(\mathbf{r}-\mathbf{r}_0) } d^3\mathbf{r} =  f(\mathbf{r}_0) = \int_{A} (f \circ \Phi)(a,b,\varphi) \boxed{  \frac{1}{|a \varphi |} \delta(a-a_0) \delta(b-b_0) \delta(\varphi-\varphi_0) } |a\varphi| \,  da \, db \, d\varphi $$
$$ = \int_{A} (f \circ \Phi)(a,b,\varphi) \delta(a-a_0) \delta(b-b_0) \delta(\varphi-\varphi_0)  \, da \, db \, d\varphi = (f \circ \Phi)(a_0, b_0, \varphi_0), $$
where $(f \circ \Phi)(a,b,\varphi)$ denotes the pull-back of the scalar-valued function $f$, i.e. $(\Phi^* f)(a,b,\varphi)$.

\subsubsection*{Normalization of Spinors\label{eq:norspin}}

In the 2016 paper the Varela spinors are defined as:
$$ \boldsymbol \psi_{n,s}^{\nu,\zeta}(\varphi)  = \left[ \begin{matrix}
F_A e^{-i \varphi/2} \\
\zeta F_B^* e^{i \varphi/2} 
\end{matrix} \right] e^{i \nu \tilde{n} \varphi}. $$
For convenience, we abbreviate $\boldsymbol \psi_{n,s}^{\nu,\zeta}(\varphi)$ as $\boldsymbol \psi (\varphi)$ or simply, $\boldsymbol \psi$.
The inner product is:
$$ \boldsymbol \psi^\dagger   \boldsymbol \psi  = \left[ \begin{matrix}
F_A^* e^{i \varphi/2} &
\zeta F_B e^{- i \varphi/2} 
\end{matrix} \right] e^{ - i \nu \tilde{n} \varphi}  \left[ \begin{matrix}
F_A e^{-i \varphi/2} \\
\zeta F_B^* e^{i \varphi/2} 
\end{matrix} \right] e^{i \nu \tilde{n} \varphi} = |F_A|^2 + \zeta^2 |F_B|^2 = |F_A|^2 + |F_B|^2=s$$
since $|F_A|^2 +  |F_B|^2 = s$.  When normalizing, we integrate over all space:
$$ \braket{  \boldsymbol \psi^\dagger  |  \boldsymbol \psi  } =  \int_0^{2\pi} \boldsymbol \psi^\dagger(\varphi) \boldsymbol \psi (\varphi) d\varphi = 2 \pi s. $$
Therefore,  proper normalization of the probability amplitudes requires a factor of $\frac{1}{\sqrt{2\pi s}}$.  Also, for convenience we will multiply them with an indicator function:
$$ \chi_{[0,2\pi]}(\varphi) = \begin{cases}
1 & \varphi \in [0,2\pi] \\
0 & \mbox{otherwise}
\end{cases}. $$
This enables us to extend the integration limits to infinity. We define them as:
$$ \boldsymbol \psi_{n,s}^{\nu,\zeta} (\varphi)  = \frac{1}{\sqrt{2\pi s}} \left[ \begin{matrix}
F_A e^{-i \varphi/2} \\
\zeta F_B^* e^{i \varphi/2} 
\end{matrix} \right] e^{i \nu \tilde{n} \varphi} \chi_{[0,2\pi]}(\varphi). $$
Then, the factor of $2\pi$ is not needed.  We can also integrate over all space:
$$ \braket{  \boldsymbol \psi^\dagger  |  \boldsymbol \psi  } =  \int_{-\infty}^\infty \boldsymbol \psi^\dagger (\varphi) \boldsymbol \psi (\varphi) \chi_{[0,2\pi]}(\varphi)  d\varphi = 1. $$

\section*{Text S2. Enantiospecificity in Cross-Polarization}

The dependence of the bilinear coupling $\mathbf{I}_1 \cdot \mathtensor{F} \cdot \mathbf{I}_2$ on chirality implies that spin-relaxation may also depend on chirality.  Because the kinetics of the cross-polarization experiment depend on spin relaxation times, we look at the impact of the indirect coupling on spin-lattice relaxation in the rotating frame. Consider a general Hamiltonian of the form $\mathcal{H} = \mathcal{H}_Z + \mathcal{H}_{SB} + \mathcal{H}_B$, where $\mathcal{H}_Z$ is the unperturbed spin part (e.g. Zeeman interaction), $\mathcal{H}_{SB}$ is the spin-bath interaction including any spin-spin interactions and $\mathcal{H}_B$ is the bath Hamiltonian.  We denote the corresponding Liouvillian superoperator as $\mathtensor{L}=\mathtensor{L}_Z + \mathtensor{L}_{SB} + \mathtensor{L}_B$.   The Liouvillian superoperator is defined as the commutation superoperator, $\mathtensor{L} \rho \equiv i [\mathcal{H}, \rho]$. Consider a set of operators $\{ F_k\}_{k=1}^m$ and an inner product $(A,B)=\mbox{Tr}(A^\dagger B)$, often denoted $\langle\langle A|B \rangle\rangle$. The operators $F_k$ are orthogonal, $\langle\langle F_i|F_j \rangle\rangle=0$ for $i \ne j$.  A projection superoperator can be defined as $\pi = \sum_{k=1}^m \frac{ | F_k \rangle\rangle  \langle\langle F_k | }{ \langle\langle F_k | F_k \rangle\rangle }$. It is customary to transform the Liouvillian to the interaction representation (see Section ``Appendix: Interaction Representation for Spin-Lock Experiment (Review)'') generated by $\mathtensor{L}_Z$ and write the resulting Liouvillian as $\mathtensor{L}^*(t) \equiv \mathtensor{L}_B + \mathtensor{L}_{SB}(t)$, where $*$ denotes the interaction representation and $\mathtensor{L}_{SB}(t) \equiv e^{ \mathtensor{L}_Z t} \mathtensor{L}_{SB}$.  It will be convenient to study irreversible processes using  projection operator methods, which yield a quantum master equation for an observable $F_k$:
\begin{align*}
\frac{d}{dt} \langle\langle F_k|\rho(t) \rangle\rangle =& -i \langle\langle F_k | \mathtensor{L}^*(t) \pi | \rho(t) \rangle\rangle - i \langle\langle F_k| \mathtensor{L}^*(t) \vec{T} e^{-i\int_{0}^t d\tau (1-\pi) \mathtensor{L}^*(\tau) }
 (1-\pi) |\rho(0) \rangle\rangle \\
 & - \sum_{j=1}^m \int_0^t dt' \frac{  \langle\langle F_k| \mathtensor{L}^*(t) \vec{T} e^{-i\int_{t'}^t d\tau (1-\pi) \mathtensor{L}^*(\tau) }
 (1-\pi) \mathtensor{L}^*(t') | F_j \rangle\rangle }{ \langle\langle F_j | F_j \rangle\rangle} \langle\langle F_j|\rho(t') \rangle\rangle
\end{align*}
where $\vec{T}$ denotes Dyson time-ordering of the exponential~\cite{HaeberlenBook}. Insertion of the expression for the projection operator $\pi$ into the first term shows that it causes coherent evolution (Bloch term). The second term can be made to vanish with a proper choice of initial conditions.  The last term is the dissipative term we are interested in. Under some assumptions (weak collision limit, short correlation time limit, average Liouvillian, stationarity of the memory kernel), $\vec{T} e^{-i\int_{t'}^t d\tau (1-\pi) \mathtensor{L}^*(\tau) }$ is approximated by $e^{-i(t-t')(1-\pi)\mathtensor{L}^* }$, where $\mathtensor{L}^*$ is a time-averaged Liouvillian in the interaction representation.  It is customary to drop the Zeeman and spin-bath parts ($\| \mathtensor{L}_B \| \gg \|\mathtensor{L}_Z + \mathtensor{L}_{SB}\|$), leaving only the bath part, i.e. $\mathtensor{L}^* \approx \mathtensor{L}_B$, in the exponential. Since there is no spin part left, the projector term is dropped leaving $e^{-i(t-t')\mathtensor{L}_B}$.  Finally, the term $(1-\pi) \mathtensor{L}^*(t') | F_j \rangle\rangle$ gives two terms: $\mathtensor{L}^*(t') | F_j \rangle\rangle$ and $-\pi \mathtensor{L}^*(t') | F_j \rangle\rangle$.  The second term is not of interest in the derivation of dissipation rates because it describes a product of average frequencies; we are instead interested in deviations from  averages. We therefore consider the first term.  The dissipation term in this wide-sense stationary Redfield limit reads:
\begin{equation}
 \sum_{j=1}^m \int_0^\infty d\tau \frac{ \langle\langle F_k| \mathtensor{L}^*(0) e^{i\tau \mathtensor{L}_B}
\mathtensor{L}^*(\tau) | F_j \rangle\rangle }{ \langle\langle F_j | F_j \rangle\rangle } \langle\langle F_j|\rho(t) \rangle\rangle  \equiv \sum_{j=1}^m \langle\langle F_j|\rho(t) \rangle\rangle \int_0^\infty \mathcal{K}_{kj}(\tau) d\tau,
\label{eq:FjFk}
\end{equation}
where $W(F_k) \equiv  \int_0^\infty \mathcal{K}_{kj}(\tau) d\tau$ is a dissipation rate for the state $F_k$.   In the study of cross-polarization we shall be interested in the quantity $T_{1\rho}$, the spin-lattice relaxation time in the rotating frame.  For $\{ F_k \}$ we will take the set $\{ F_x, F_y, F_z \}$, where $F_j \equiv I_j^u$, $u=1,2$.

\subsection{Homonuclear Case}

We begin with the homonuclear case.  While homonuclear spin systems are not commonplace in cross-polarization, such experiments are possible, involving different interacting subensembles. Two interacting nuclear spins of the same type (homonuclear case), $\mathbf{I}_1$ and $\mathbf{I}_2$ are coupled to a bath of electron spins $\{ \mathbf{S}_l \}$. The spin-bath interaction is taken to be a sum of pairwise couplings.  In a spin-lock experiment where $T_{1\rho}$ is normally measured, the Zeeman Hamiltonian $\mathcal{H}_Z$ is the sum of two components, the Zeeman and RF parts (in frequency units) are: $\mathcal{H}_Z = \omega_0(I_{z}^1 + I_z^2) + \omega_S \sum_l S_{z}^l + \omega_1  (I_x^1+I_x^2)\cos(\omega_0 t)$.  For simplicity: 1) the two nuclear spins are assumed to be chemically and magnetically equivalent; 2) we also neglect off-resonance effects for simplicity.  (Those assumptions can easily be dropped, posing no difficulties, but here we'd like to keep our expressions simple.)  The Hermitian operator $\omega_0(I_{z}^1 + I_z^2) + \omega_S \sum_l S_{z}^l$ can be used to effect a transformation to the ``doubly rotating frame''.  In the secular (rotating-wave) approximation this leaves $\mathcal{H}_Z^* = \omega_1  (I_x^1+I_x^2)$.   The nuclear-nuclear term of the spin-bath Hamiltonian in the rotating frame, $\mathcal{H}_{SB}^*$, acquires a phase factor $e^{- i \omega_0 q t}$:
\begin{align*}
 \mathcal{H}_{SB}^*(t) =& \sum_{q=-2}^2 (-1)^q \underbrace{ A_{2,q}( \mathbf{I}^1, \mathbf{I}^2) T_{2,-q}(\mathbf{I}^1,\mathbf{I}^2) e^{- i \omega_0 q t}   }_{\mbox{nuclear-nuclear}} \\
& + \sum_{j=1}^m \underbrace{ \left[ A_{2,0}( \mathbf{I}^1, \mathbf{S}^j) T_{2,0}^*(\mathbf{I}^1,\mathbf{S}^j) + A_{2,0}( \mathbf{I}^2, \mathbf{S}^j) T_{2,0}^*(\mathbf{I}^2,\mathbf{S}^j) \right] }_{\mbox{electron-nuclear}}
\end{align*}
whereas the electron-nuclear term does not.  It is also truncated to longitudinal order.  
Instead of describing an ``effective'' interaction, the components $A_{2,0}( \mathbf{I}^2, \mathbf{S}^j)$ of the electron-nuclear tensor would typically be the components of a direct hyperfine interaction.   On the other hand, the $k=2$ components of the spherical tensor $A$ coupling nuclear spins $\mathbf{I}^1$ to $\mathbf{I}^2$ given in terms of the Cartesian components of the indirect coupling tensor $\mathtensor{F}$ are:
\begin{align*}
  A_{2,0} ( \mathbf{I}^1, \mathbf{I}^2) =& \frac{ - \mathtensor{F}_{xx} - \mathtensor{F}_{yy} + 2 \mathtensor{F}_{zz} }{\sqrt{6}} \\
  A_{2,\pm 1} ( \mathbf{I}^1, \mathbf{I}^2) =& \mp \frac{ \mathtensor{F}_{zx} + \mathtensor{F}_{xz} \pm i ( \mathtensor{F}_{zy} + \mathtensor{F}_{yz}) }{2} \\
  A_{2,\pm 2} ( \mathbf{I}^1, \mathbf{I}^2) =& \frac{ \mathtensor{F}_{xx} - \mathtensor{F}_{yy} \pm i ( \mathtensor{F}_{xy} + \mathtensor{F}_{yx}) }{2}.
\end{align*}
For the case of the dipole interaction, the tensor components can be inferred from Eq.~(\ref{eq:enan}). Explicitly, they are:
\begin{equation}
\mathtensor{F}_{\alpha \alpha'}  = \left( \frac{\mu_0}{4\pi}\right)^2 \gamma_I^2 \gamma_S^2 \sum_{n,n'} \sum_{\beta,\beta'}  \frac{ 
M_{1,\beta}^{\alpha\beta}(\tilde{n}',\tilde{n})M_{2,\beta'}^{\alpha'\beta'}(\tilde{n},\tilde{n}')
}{|T| (\tilde{n}'-\tilde{n})    } f(\tilde{n})[1-f(\tilde{n}')] + c.c.
\label{eq:Ftensor}
\end{equation}
where the $M$ matrices are given by Eqs.~(\ref{eq:Ms1}), (\ref{eq:Ms2}) and~(\ref{eq:Ms3}).  As discussed in Section~\ref{eq:enansp} the tensor $\sum_{\beta,\beta'} M_{1,\beta}^{\alpha\beta}(\tilde{n}',\tilde{n})M_{2,\beta'}^{\alpha'\beta'}(\tilde{n},\tilde{n}')$ is enantiospecific. Therefore, the tensor components $\mathtensor{F}_{\alpha\alpha'}$ will be enantiospecific.  And as we will see later, the relaxation rate $(T_{1\rho})^{-1}$ will be enantiospecific. 
The irreducible tensors for the homonuclear interaction are:
$$
T_{2,0} ( \mathbf{I}^1, \mathbf{I}^2) = \frac{1}{\sqrt{6}}
 (3I_{z}^1 I_{z}^2 - \mathbf{I}^1 \cdot \mathbf{I}^2 ) \quad
 T_{2,\pm 1}  ( \mathbf{I}^1, \mathbf{I}^2) = \mp \frac{1}{2} ( I_{z}^1
 I_{\pm}^2 + I_{\pm}^1 I_{z}^2 ) \quad
T_{2,\pm 2} ( \mathbf{I}^1, \mathbf{I}^2) =\frac{1}{2}
 I_{\pm}^1 I_{\pm}^2
$$
whereas tensor components for the electron-nuclear interactions with $q=\pm 1, \pm 2$ average to zero except for $q=0$, which becomes 
$$T_{2,0}^*(\mathbf{I}^1,\mathbf{S}^j)=\lim_{T\rightarrow \infty} \frac{1}{T} \int_0^T e^{-i \omega_0 (I_z^1+I_z^2) t } T_{2,0}(\mathbf{I}^1,\mathbf{S}^j) e^{i \omega_0 (I_z^1+I_z^2) t } dt = \sqrt{2/3} \, I_z S_z.$$  
Next, to eliminate the remaining Zeeman term, $\mathcal{H}_Z^* = \omega_1  (I_x^1+I_x^2)$ from the Hamiltonian we effect another transformation to a second interaction frame using this residual Zeeman term (see Section ``Appendix: Interaction Representation for Spin-Lock Experiment (Review)'').  This involves rotating the spin-bath Hamiltonian $\mathcal{H}_{SB}^*(\tau)$ by an angle $\omega_1 t$ about the $X$ axis.  This can be accomplished with the help of Wigner matrices. The tensors $T_{2,-q}(\mathbf{I}^1,\mathbf{I}^2)$ transform as:
$$ T_{2,-q}(\mathbf{I}^1,\mathbf{I}^2)  \rightarrow \mathtensor{R}(\Omega) T_{k,-q}  = \sum_{q'=-2}^2 T_{2,-q'}(\mathbf{I}^1,\mathbf{I}^2)  \mathscr{D}_{q',-q}^{(2)}(\Omega) $$
where $\Omega = \alpha \, \beta \, \gamma$ is a set of Euler angles, $\mathtensor{R}(\Omega)$ is a rotation operator and $\mathscr{D}_{m' m}^{(j)} (\alpha \, \beta \, \gamma) = e^{-i(m' \alpha + m\gamma)} d_{m' m}^{(j)} (\beta)$.   From the definition of Euler rotation, $\mathtensor{R}(\alpha \, \beta \, \gamma) = \mathtensor{R}_z(\alpha) \mathtensor{R}_y(\beta) \mathtensor{R}_z(\gamma)$, a rotation about the $X$ axis by an angle $\beta$ can be effected using the sequence $\mathtensor{R}(\tfrac{\pi}{2} , \beta , -\tfrac{\pi}{2}) = \mathtensor{R}_z(\tfrac{\pi}{2}) \mathtensor{R}_y(\beta) \mathtensor{R}_z(-\tfrac{\pi}{2})$.  The elements of the Wigner matrix are therefore $$\mathscr{D}_{m' m}^{(j)} (\tfrac{\pi}{2} , \beta , -\tfrac{\pi}{2}) = e^{-i(m' \pi/2 - m\pi/2)} d_{m' m}^{(j)} (\beta),$$ where $\beta = \omega_1 t$. Thus, $\mathscr{D}_{q',-q}^{(2)} (\tfrac{\pi}{2} , \omega_1 t , -\tfrac{\pi}{2})= e^{-i(q' \pi/2 +q\pi/2)} d_{q',-q}^{(2)} (\omega_1 t)$.

Finally, the electron-nuclear tensor component $T_{2,0}^*(\mathbf{I}^1,\mathbf{S}^j)$ transforms according to the rule:
$$ T_{2,0}^*(\mathbf{I}^1,\mathbf{S}^j)  \rightarrow \sqrt{2/3} S_z^j \left( I_z^1 \cos(\omega_1 t) - I_y^1 \sin(\omega_1 t) \right). $$
The spin-bath Hamiltonian in this second interaction representation is therefore:
\begin{multline*}
 \mathcal{H}_{SB}^{**}(t) = \sum_{q'=-2}^2 \sum_{q=-2}^2 (-1)^q  A_{2,q}( \mathbf{I}^1, \mathbf{I}^2) T_{2,-q}(\mathbf{I}^1,\mathbf{I}^2) e^{- i \omega_0 q t}  e^{-i(q' \pi/2 +q\pi/2)} d_{q',-q}^{(2)} (\omega_1 t)    \\ 
 + \sqrt{2/3}  \sum_{j=1}^m   \Bigl[ A_{2,0}( \mathbf{I}^1, \mathbf{S}^j) S_z^j \left( I_z^1 \cos(\omega_1 t) - I_y^1 \sin(\omega_1 t) \right) \\
  \qquad + A_{2,0}( \mathbf{I}^2, \mathbf{S}^j) S_z^j \left( I_z^2 \cos(\omega_1 t) - I_y^2 \sin(\omega_1 t) \right) \Bigr]. 
\end{multline*}
We could of course carry on with the calculation including all terms.  While this poses no extra difficulties other than additional bookkeeping, it carries no pedagogical value.   Since, our goal is to provide an intuitive understanding of the link between chirality and nuclear spin relaxation, we simplify the expression by dropping all terms where $q \ne 0$.  This is equivalent to assuming that motion is slower than the Larmor period ($2\pi/\omega_1$).   Using the middle column ($q=0$) of the reduced Wigner matrix
\begin{equation}
d^{(2)}(\beta) = \left( \begin{smallmatrix}
 \cos^4(\beta/2)  & -\frac{1}{2} \sin\beta(1+\cos\beta) &  \sqrt{\frac{3}{8}}\sin^2\beta  & \frac{1}{2}\sin\beta(\cos\beta-1)  &  \sin^4(\beta/2) \\
 \frac{1}{2}\sin\beta(1+\cos\beta) &  \frac{1}{2} (2\cos\beta-1)(\cos\beta+1)  &  -\sqrt{\frac{3}{2}}\sin\beta\cos\beta  & \frac{1}{2}(2\cos\beta+1)(1-\cos\beta) &  \frac{1}{2}\sin\beta(\cos\beta-1)  \\
  \sqrt{\frac{3}{8}}\sin^2\beta  & \sqrt{\frac{3}{2}}\sin\beta\cos\beta &  \frac{1}{2}(3\cos^2\beta-1) & -\sqrt{\frac{3}{2}}\sin\beta\cos\beta  &  \sqrt{\frac{3}{8}}\sin^2\beta  \\
 -\frac{1}{2}\sin\beta(\cos\beta-1)  & \frac{1}{2}(2\cos\beta+1)(1-\cos\beta) &   \sqrt{\frac{3}{2}}\sin\beta\cos\beta  & \frac{1}{2}(2\cos\beta-1)(\cos\beta+1)  &  -\frac{1}{2}\sin\beta(1+\cos\beta)  \\
  \sin^4(\beta/2)  &  -\frac{1}{2}\sin\beta(\cos\beta-1) &  \sqrt{\frac{3}{8}}\sin^2\beta  &  \frac{1}{2} \sin\beta(1+\cos\beta)  & \cos^4(\beta/2) 
\end{smallmatrix} \right)
\label{eq:reducedd2}
\end{equation}
the sum over $q'$ is:
\begin{multline*}
\sum_{q'=-2}^2 e^{-i q' \pi/2} d_{q',0}^{(2)} (\omega_1 t) = - \sqrt{\frac{3}{8}}\sin^2(\omega_1t)  +  i  \sqrt{\frac{3}{2}}\sin(\omega_1t)\cos(\omega_1t) + \frac{1}{2}(3\cos^2(\omega_1t)-1)  \\
   + (-i) \sqrt{\frac{3}{2}}\sin\beta\cos(\omega_1t)  - \sqrt{\frac{3}{8}}\sin^2(\omega_1t) \\
 =\frac{1}{8} \left(3 e^{-2 i \omega_1t}+\left(2 \sqrt{6}+3\right) e^{2 i \omega_1t }-2 \sqrt{6}+2\right)
\end{multline*}
The spin-bath Hamiltonian becomes:
\begin{multline*}
 \mathcal{H}_{SB}^{**}(t) = A_{2,0}( \mathbf{I}^1, \mathbf{I}^2) T_{2,0}(\mathbf{I}^1,\mathbf{I}^2)  \frac{1}{8} \left(3 e^{-2 i \omega_1t}+\left(2 \sqrt{6}+3\right) e^{2 i \omega_1t }-2 \sqrt{6}+2\right)  \\ 
 + \sqrt{2/3}  \sum_{u=1,2} \sum_{j=1}^m A_{2,0}( \mathbf{I}^u, \mathbf{S}^j) S_z^j \left( I_z^u \cos(\omega_1 t) - I_y^u \sin(\omega_1 t) \right) 
\end{multline*}

We are interested in how nuclear spin 1 (or 2) relaxes under conditions of spin-lock.  Denote by $u$ the spin of interest  ($u=1,2$).  Write down the operator corresponding to its state in terms of the basis 
$$ |F_u \rangle\rangle = \mathscr{Y}^{(1)q}(\mathbf{I}^u) $$
where $q=-1,0,1$. We now compute the numerator of the memory function, 
$$\langle\langle F_k| \mathtensor{L}^{**}(0) e^{i\tau \mathtensor{L}_B}
\mathtensor{L}^{**}(\tau) | F_j \rangle\rangle.$$
According to Eq.~(\ref{eq:FjFk}), when a state is prepared along $X$, spin-locking is concerned with the evolution of the operator $F_x$.  Therefore, three terms will contribute to the rate of thermalization of $F_x$: 
$$ \frac{ \langle\langle F_x | \mathtensor{L}^*(0) e^{i\tau \mathtensor{L}_B}
\mathtensor{L}^*(\tau) | F_x \rangle\rangle }{ \langle\langle F_x | F_x \rangle\rangle }, \quad
\frac{ \langle\langle F_x | \mathtensor{L}^*(0) e^{i\tau \mathtensor{L}_B}
\mathtensor{L}^*(\tau) | F_y \rangle\rangle }{ \langle\langle F_y | F_y \rangle\rangle }, \quad
\frac{ \langle\langle F_x | \mathtensor{L}^*(0) e^{i\tau \mathtensor{L}_B}
\mathtensor{L}^*(\tau) | F_z \rangle\rangle }{ \langle\langle F_z | F_z \rangle\rangle } $$
In order to illustrate the rate dependence on chirality, it will be sufficient to compute the first term only, with $F_x=I_x^u$, $u=1,2$. Starting from the commutator $
\mathtensor{L}^{**}(\tau) F_u  \equiv [\mathcal{H}_{SB}^{**}(\tau), I_x^u ]$.  The nuclear-nuclear term has a commutator $ [ T_{2,0}(\mathbf{I}^1,\mathbf{I}^2) , I_x^u)  ]$.
Using the relation $[I_x,I_y]=iI_z$ (and cyclic permutations), we find
$$  [T_{2,0} ( \mathbf{I}^1, \mathbf{I}^2), I_x^1] = \frac{1}{\sqrt{6}} [
 3I_{z}^1 I_{z}^2 - \mathbf{I}^1 \cdot \mathbf{I}^2, I_x^1]  = \frac{i}{\sqrt{6}} (2  I_y^1 I_z^2 +  I_z^1 I_y^2) $$
and a similar expression exists for the commutator with $I_x^2$ (by flipping 1 and 2). The electron-nuclear term has the commutator
$$ [ S_z^j ( I_z^1 \cos(\omega_1t) - I_y^1 \sin(\omega_1t), I_x^1] =  i S_z^j  ( I_y^1 \cos(\omega_1 t)  + I_z^1 \sin(\omega_1t)), $$
Acting on $\mathtensor{L}^*(\tau)F$ with the bath propagator $e^{ \mathtensor{L}_B \tau}$ introduces a time dependence $A_{kq}( \mathbf{I}^1, \mathbf{I}^2)  \rightarrow A_{kq}( \mathbf{I}^1, \mathbf{I}^2)(\tau)$.  The final expression is:
\begin{multline*}
e^{\mathtensor{L}_B \tau } \mathtensor{L}_{SB}^{**}(\tau) F_u  = A_{2,0}( \mathbf{I}^1, \mathbf{I}^2)(\tau) \frac{i}{8\sqrt{6}}  \left[ \delta_{u,1} (2  I_y^1 I_z^2 +  I_z^1 I_y^2) + \delta_{u,2} (2  I_y^2 I_z^1 +  I_z^2 I_y^1) \right]  \\
 \times  \left(3 e^{-2 i \omega_1\tau}+\left(2 \sqrt{6}+3\right) e^{2 i \omega_1\tau }-2 \sqrt{6}+2\right)  \\ 
 + i \sqrt{2/3} ( I_y^u \cos(\omega_{1} \tau )  + I_z^u \sin(\omega_{1} \tau )) \sum_{j=1}^m  A_{2,0}( \mathbf{I}^u, \mathbf{S}^j)(\tau) S_z^j. 
\end{multline*}
Next, we compute the inner product:
$$ \frac{ \langle \langle \mathtensor{L}_{SB}^{**}(0) F_u  |  e^{ \mathtensor{L}_B \tau } \mathtensor{L}_{SB}^{**}(\tau) F_u \rangle\rangle }{ \langle\langle F_u | F_u \rangle\rangle } \equiv \frac{ \mbox{Tr} \bigl[ (\mathtensor{L}_{SB}^{**}(0) F_u )^\dagger e^{\mathtensor{L}_B \tau} \mathtensor{L}_{SB}^{**}(\tau) F_u \bigr] }{ \mbox{Tr}[ F_u^\dagger F_u ] }, $$
for $F_u = I_x^u$. Upon multiplying the ``bra'' $\langle\langle \mathtensor{L}_{SB}^{**}(0) I_x^u |$ and the ``ket'' $|e^{i\mathtensor{L}_B \tau} \mathtensor{L}_{SB}^{**}(\tau) I_x^u \rangle\rangle$, which involves taking a trace over all space, we get:
\begin{multline*}
=  C \cdot \langle  A_{2,0}^\dagger( \mathbf{I}^1, \mathbf{I}^2)(0) A_{2,0}( \mathbf{I}^1, \mathbf{I}^2)(\tau) \rangle  \left(3 e^{-2 i \omega_1 \tau }+\left(2 \sqrt{6}+3\right) e^{2 i \omega_1 \tau }-2 \sqrt{6}+2\right)  \\
+ \frac{2}{3}  C' \cdot \cos(\omega_{1} \tau )  \sum_{j=1}^m  \langle A_{2,0}^\dagger( \mathbf{I}^u, \mathbf{S}^j)(0) A_{2,0}( \mathbf{I}^u, \mathbf{S}^j)(\tau) \rangle_B.  
\end{multline*}
where $C=\tfrac{1}{192} \frac{ I^1(I^1+1)(2I^1+1) I^2(I^2+1)(2I^2+1) }{ I^u(I^u+1)(2I^u+1) (2I^{\tilde{u}}+1) } = \frac{1}{192} I^{\tilde{u}}(I^{\tilde{u}}+1)$ and $C' = \frac{ \mathrm{Tr} [(I_y^u)^2] (S_z^j)^2] }{ \mathrm{Tr}[(I_x^u)^2]} = \frac{ 1}{3} S^j(S^j+1)(2S^j+1)=\frac{1}{2}$ since $S^j=S=\frac{1}{2}$.  We have used $\mbox{Tr} \left[ (I_y^1 I_z^2 +  I_z^1 I_y^2)^2 \right] =2 \cdot \frac{1}{3} I^1(I^1+1)(2I^1+1) \cdot \frac{1}{3} I^2(I^2+1)(2I^2+1)$ and $\mbox{Tr}_{I_1,I_2}[(I_x^1)^\dagger I_x^1]= \mbox{Tr}_{I_1}[I_x^\dagger I_x] \cdot \mbox{Tr}_{I_2}[\mathbf{1}]=\frac{1}{3} I^1(I^1+1) (2 I^1+1)  \cdot (2 I^2+1)$.
According to Abragam~\cite{abragam1961principles} the bath autocorrelation functions, $\langle A_{2,0}^\dagger( \mathbf{I}^1, \mathbf{I}^2)(0) A_{2,0}( \mathbf{I}^1, \mathbf{I}^2)(\tau) \rangle $ are computed as thermal averages, i.e. 
$$\langle AA(t) \rangle_B \equiv \mbox{tr}[\rho AA(t) ], \quad \rho(\mathcal{H}_B)=\frac{ \exp(-\beta_L \mathcal{H}_B) }{ \mbox{Tr}[\exp(-\beta_L \mathcal{H}_B)] }, $$ 
which is the Boltzmann density matrix and $\mbox{tr}$ denotes the partial trace over the bath degrees of freedom.  $\beta_L$ is the (inverse) temperature of the bath (lattice).

The spin-lock relaxation rate, $ \frac{1}{T_{1\rho}}$,  is therefore equal to:
\begin{multline*}
 \frac{1}{T_{1\rho}} = \sum_j \int_0^\infty \frac{ \langle\langle F_k| \mathtensor{L}^*(0) e^{i\tau \mathtensor{L}_B}
\mathtensor{L}^*(\tau) | F_j \rangle\rangle }{ \langle\langle F_j | F_j \rangle\rangle } d\tau  \\
= \underbrace{ 3 C  \left[ 3 J_{II}(-2\omega_1) + (2\sqrt{6}+3) J_{II}(2\omega_1) + 2(1- \sqrt{6})J_{II}(0)  \right] + \frac{3}{2} C' J^c_{IS}(\omega_1) }_{XX}  + XY + XZ.
\end{multline*}
where $XY$ corresponds to the $\langle\langle F_x^u |  \cdot  | F_y^u \rangle\rangle$ term and $XZ$, to the $\langle\langle F_x^u | \cdot  | F_z^u \rangle\rangle$ term.
The spectral density functions are:
$$ J_{II}(\omega) = \int_0^\infty \langle  A_{2,0}^\dagger( \mathbf{I}^1, \mathbf{I}^2)(0) A_{2,0}( \mathbf{I}^1, \mathbf{I}^2)(\tau) \rangle_B \, e^{i \omega \tau } \, d\tau, $$
$$ J^c_{IS}(\omega) = \int_0^\infty \langle  A_{2,0}^\dagger( \mathbf{I}^u, \mathbf{S}^j)(0) A_{2,0}( \mathbf{I}^u, \mathbf{S}^j)(\tau) \rangle_B \, \cos(\omega \tau) \, d\tau,  $$
which was assumed, for simplicity, to be independent of $u$ and $j$. The homonuclear spectral density function $J_{II}(\omega)$, and by extension, $T_{1\rho}$, is enantiospecific, since the $A_{2,0}( \mathbf{I}^1, \mathbf{I}^2)(\tau)$'s depend on the tensor components $\sum_{\beta,\beta'} M_{1,\beta}^{\alpha\beta}(\tilde{n}',\tilde{n})M_{2,\beta'}^{\alpha'\beta'}(\tilde{n},\tilde{n}')$, which are themselves enantiospecific.

\subsection{Heteronuclear Case}

The heteronuclear case is widely used in NMR to transfer polarization from high-gamma abundant nuclei to low-gamma dilute spins. The spin operators $\mathbf{I}_1$ and $\mathbf{I}_2$ now refer to different nuclei (e.g. $^1$H and $^{13}$C). The Zeeman Hamiltonian is now $\mathcal{H}_Z = \omega_{0,1} I_{z}^1 + \omega_{0,2} I_z^2 + \omega_S \sum_l S_z^l +  \omega_{1,1}  I_x^1 \cos(\omega_{0,1} t) +  \omega_{1,2}  I_x^2 \cos(\omega_{0,2} t)$.   The Hartmann-Hahn condition corresponds to $\omega_{1,1} = \omega_{1,2}$, where $\omega_{1,1} = \gamma_{I_1} B_{1,1}$ and $\omega_{1,2} = \gamma_{I_2} B_{1,2}$. The operator $ \omega_{0,1} I_{z}^1 + \omega_{0,2} I_z^2 + \omega_S \sum_l S_z^l$ is used to effect a transformation to the ``interaction representation''.  In the secular (rotating-wave) approximation this leaves $\mathcal{H}_Z^* = \omega_{1,1}  I_x^1+ \omega_{1,2} I_x^2$.  We will limit our discussion to the heteronuclear $I-I$ term.  The spin-bath Hamiltonian in the rotating frame, $\mathcal{H}_{SB}^*$ is:
\begin{align*}
 \mathcal{H}_{SB}^*(t) =& A_{2,0}( \mathbf{I}^1, \mathbf{I}^2) T_{2,0}^*(\mathbf{I}^1,\mathbf{I}^2)  
\end{align*}
where $T_{2,0}^*(\mathbf{I}^1,\mathbf{I}^2)=\sqrt{2/3} I_z^1 I_z^2$.  Terms oscillating at $\omega_{0,1}-\omega_{0,2}$ or higher frequency have been dropped.

To eliminate the remaining Zeeman term, 
$\mathcal{H}_Z^* = \omega_{1,1}  I_x^1+ \omega_{1,2} I_x^2$
from the Hamiltonian we effect another transformation to a second interaction frame using this residual Zeeman term (see Section ``Appendix: Interaction Representation for Spin-Lock Experiment (Review)'').  Using the rule
$$ T_{2,0}^*(\mathbf{I}^1,\mathbf{I}^2)  \rightarrow \sqrt{2/3} \left( I_z^1 \cos(\omega_{1,1} \tau ) - I_y^1 \sin(\omega_{1,1}  \tau ) \right)  \left( I_z^2 \cos(\omega_{1,2} \tau ) - I_y^2 \sin(\omega_{1,2} \tau ) \right), $$
the spin-bath Hamiltonian in this double interaction representation is therefore:
\begin{multline*}
 \mathcal{H}_{SB}^{**}( \tau ) =  \sqrt{2/3} A_{2,0}( \mathbf{I}^1, \mathbf{I}^2) \left( I_z^1 \cos(\omega_{1,1} \tau ) - I_y^1 \sin(\omega_{1,1} \tau ) \right)  \left( I_z^2 \cos(\omega_{1,2} \tau ) - I_y^2 \sin(\omega_{1,2} \tau ) \right).
\end{multline*}
Commuting this with $I_x^u$, $u=1,2$:
\begin{multline*}
\mathtensor{L}_{SB}^{**}(\tau) I_x^u =   \sqrt{2/3} A_{2,0}( \mathbf{I}^1, \mathbf{I}^2)(\tau) [ \left( I_z^1 \cos(\omega_{1,1} \tau ) - I_y^1 \sin(\omega_{1,1} \tau ) \right)  \left( I_z^2 \cos(\omega_{1,2} \tau ) - I_y^2 \sin(\omega_{1,2} \tau ) \right), I_x^u]  \\
 = i  \sqrt{2/3} A_{2,0}( \mathbf{I}^1, \mathbf{I}^2)(\tau) \left( I_y^u \cos(\omega_{1,u} \tau ) + I_z^u \sin(\omega_{1,u} \tau )  \right)  \left( I_z^{\tilde{u}} \cos(\omega_{1,\tilde{u}} \tau ) - I_y^{\tilde{u}} \sin(\omega_{1,\tilde{u}} \tau ) \right).  
\end{multline*}
where $\tilde{u}=3-u$ (i.e. $u=1 \rightarrow \tilde{u}=2$ and $u=2 \rightarrow \tilde{u}=1$).  Acting on $\mathtensor{L}_{SB}^{**}(\tau) I_x^u$ with the bath propagator $e^{ \mathtensor{L}_B \tau}$ introduces a time dependence $A_{kq}( \mathbf{I}^1, \mathbf{I}^2)  \rightarrow A_{kq}( \mathbf{I}^1, \mathbf{I}^2)(\tau)$.  The final expression is:
\begin{multline*}
e^{\mathtensor{L}_B \tau } \mathtensor{L}_{SB}^{**}(\tau) I_x^u  \\
=  i \sqrt{2/3} A_{2,0}( \mathbf{I}^1, \mathbf{I}^2)(\tau) \left( I_y^u \cos(\omega_{1,u} \tau ) + I_z^u \sin(\omega_{1,u} \tau )  \right)  \left( I_z^{\tilde{u}} \cos(\omega_{1,\tilde{u}} \tau ) - I_y^{\tilde{u}} \sin(\omega_{1,\tilde{u}} \tau ) \right).
\end{multline*}
Next, we compute the inner product by multiplying the ``bra'' $\langle\langle \mathtensor{L}_{SB}^{**}(0) I_x^u  )^\dagger|$ and the ``ket'' $|e^{i\mathtensor{L}_B \tau} \mathtensor{L}_{SB}^{**}(\tau) I_x^u  \rangle\rangle$, 
\begin{multline*}
\frac{ \langle\langle \mathtensor{L}_{SB}^{**}(0) I_x^u ,  e^{ \mathtensor{L}_B \tau } \mathtensor{L}_{SB}^{**}(\tau) I_x^u \rangle\rangle  }{ \langle\langle I_x^u |  I_x^u \rangle\rangle }  =  \frac{  \mbox{Tr} \bigl[ (\mathtensor{L}_{SB}^{**}(0) I_x^u )^\dagger e^{\mathtensor{L}_B \tau} \mathtensor{L}_{SB}^{**}(\tau) I_x^u \bigr]}{ \mbox{Tr} [ (I_x^u)^\dagger U_x^u]} \\
 = \frac{2}{3} \langle A_{2,0}( \mathbf{I}^1, \mathbf{I}^2)(0) A_{2,0}( \mathbf{I}^1, \mathbf{I}^2)(\tau) \rangle_B   \cos(\omega_{1,u} \tau ) \cos(\omega_{1,\tilde{u}} \tau ) \cdot \frac{1}{3} I^{\tilde{u}}(I^{\tilde{u}}+1) \\
 = \frac{1}{3} \langle A_{2,0}( \mathbf{I}^1, \mathbf{I}^2)(0) A_{2,0}( \mathbf{I}^1, \mathbf{I}^2)(\tau) \rangle_B [1 + \cos(2\omega_{1} \tau )] \cdot \frac{1}{3} I^{\tilde{u}}(I^{\tilde{u}}+1),
\end{multline*}
where we used $\frac{ \mathrm{Tr}[(I_y^u)^2 (I_z^{\tilde{u}})^2] }{ \mathrm{Tr}[ (I_x^u)^2] }=\frac{ \frac{1}{3} I^u(I^u+1)(2I^u+1) \cdot \frac{1}{3} I^{\tilde{u}}(I^{\tilde{u}}+1)(2I^{\tilde{u}}+1)  }{ \frac{1}{3} I^u(I^u+1)(2I^u+1) \cdot (2I^{\tilde{u}}+1) } = \frac{1}{3} I^{\tilde{u}}(I^{\tilde{u}}+1)$. The last line follows from the Hartmann-Hahn condition.
Commuting with $I_y^u$, $u=1,2$:
\begin{multline*}
\mathtensor{L}_{SB}^{**}(\tau) I_y^u =   [ \left( I_z^1 \cos(\omega_{1,1} \tau ) - I_y^1 \sin(\omega_{1,1} \tau ) \right)  \left( I_z^2 \cos(\omega_{1,2} \tau ) - I_y^2 \sin(\omega_{1,2} \tau ) \right),  I_y^u]  \\
 =  - i  I_x^u \cos(\omega_{1,u} \tau )  \left( I_z^{\tilde{u}} \cos(\omega_{1,\tilde{u}} \tau ) - I_y^{\tilde{u}} \sin(\omega_{1,\tilde{u}} \tau ) \right).  
\end{multline*}
Similarly, we find:
$$ \frac{ \langle\langle \mathtensor{L}_{SB}^{**}(0) I_y^u ,  e^{ \mathtensor{L}_B \tau } \mathtensor{L}_{SB}^{**}(\tau) I_x^u \rangle\rangle  }{ \langle\langle I_y^u |  I_y^u \rangle\rangle }  
 = \frac{1}{3} \langle A_{2,0}( \mathbf{I}^1, \mathbf{I}^2)(0) A_{2,0}( \mathbf{I}^1, \mathbf{I}^2)(\tau) \rangle_B [1 + \cos(2\omega_{1} \tau )] \cdot \frac{1}{3} I^{\tilde{u}}(I^{\tilde{u}}+1). $$
Finally, commutation with $I_z^u$, $u=1,2$:
\begin{multline*}
\mathtensor{L}_{SB}^{**}(\tau) I_z^u =   [ \left( I_z^1 \cos(\omega_{1,1} \tau ) - I_y^1 \sin(\omega_{1,1} \tau ) \right)  \left( I_z^2 \cos(\omega_{1,2} \tau ) - I_y^2 \sin(\omega_{1,2} \tau ) \right),  I_z^u]  \\
 =  - i  I_x^u \sin(\omega_{1,u} \tau )  \left( I_z^{\tilde{u}} \cos(\omega_{1,\tilde{u}} \tau ) - I_y^{\tilde{u}} \sin(\omega_{1,\tilde{u}} \tau ) \right)
\end{multline*}
whereas
$$ \frac{ \langle\langle \mathtensor{L}_{SB}^{**}(0) I_z^u |  e^{ \mathtensor{L}_B \tau } \mathtensor{L}_{SB}^{**}(\tau) I_x^u \rangle\rangle  }{ \langle\langle I_z^u |  I_z^u \rangle\rangle ) }  
 =0,$$
because $\sin(0)=0$.   The heteronuclear spin-lock relaxation rate for spin $u$ is:
\begin{multline*}
 \frac{1}{T_{1\rho}^u} = \sum_j \int_0^\infty \frac{ \langle\langle F_k| \mathtensor{L}^*(0) e^{i\tau \mathtensor{L}_B}
\mathtensor{L}^*(\tau) | F_j \rangle\rangle }{ \langle\langle F_j | F_j \rangle\rangle } d\tau  \\
= \frac{1}{3} \cdot \frac{2}{3} I^{\tilde{u}}(I^{\tilde{u}}+1) \int_0^\infty \langle A_{2,0}( \mathbf{I}^1, \mathbf{I}^2)(0) A_{2,0}( \mathbf{I}^1, \mathbf{I}^2)(\tau) \rangle_B [1 + \cos(2\omega_{1} \tau )] d\tau \\
= \frac{2}{9}  I^{\tilde{u}}(I^{\tilde{u}}+1) [ J_{II}(0) + J^c_{II}(2\omega_1)],
\end{multline*}
which is enantiospecific through the dependence of spectral density functions on chirality.

\subsection{Relationship to the Cross-Polarization Experiment}

We have demonstrated the existence of an effective nuclear spin-spin indirect coupling of the type $\mathbf{I}^1 \cdot \mathtensor{F} \cdot \mathbf{I}^2$, where the coupling tensor $\mathtensor{F} $ is enantiospecific.  This has consequences for the cross-polarization (CP) experiment.  For a more detailed expos\'e of the theory of CP in solids, the reader can consult Ref.~(\cite{michel1994cross}).  We will limit our discussion to the case of cross-polarization in non-rotating solids.
In the CP experiment, one considers two nuclei $\mathbf{I}^1$ and $\mathbf{I}^2$, each with their inverse spin temperature ($\beta_{I1}, \beta_{I2}$ respectively).  The spin temperatures evolve according to the sum of cross-polarization rate ($T_{I1-I2}^{-1}$) and rotating-frame relaxation ($T_{1\rho}^{-1}$) terms:
$$ \dot{\beta}_{I2} = - \frac{1}{T_{I1-I2}} ( \beta_{I2} - \beta_{I1} ) - \frac{1}{T_{1\rho, I2}} \beta_{I2} $$
$$ \dot{\beta}_{I1} = - \frac{1}{T_{I1-I2}} ( \beta_{I1} - \beta_{I2} ) - \frac{1}{T_{1\rho, I1}} \beta_{I1}. $$
When the initial conditions are $\beta_{I1}(0)=\beta_{I1,0}$ and $\beta_{I2}(0)=0$, the spin systems evolve according to
$$ \beta_{I1}(t) = \beta_{I1,0} \exp(-t/T_{1\rho, I1}) $$
$$ \beta_{I2}(t) = (\beta_{I1,0}/\lambda) \exp(-t/T_{1\rho, I1})  [ 1- \exp(-t/T_{I1-I2}) ] $$
where
$$ \lambda = 1 + \frac{ T_{I^1-I^2}}{ T_{1\rho, I2}} - \frac{ T_{I1-I2} }{ T_{1\rho, I1}}. $$
The $I^2$ spin magnetization reaches a maximum at time $t_m$:
$$ t_m = \frac{ T_{I1-I2} \cdot T_{1\rho, I1} }{ T_{1\rho, I1} - T_{I1-I2} } \log \frac{ T_{1\rho, I1}}{ T_{I1-I2}}. $$
The asymptotic value of $I_2$ spin magnetization the asymptotic value is
$$ \frac{M_{I2}(t \rightarrow \infty) }{ M_{I2,0} } \cong \frac{ \gamma_{I1} }{ \gamma_{I2} }. $$
It can be shown that at the Harmann-Hahn condition the CP rate, $(T_{I1-I2})^{-1}$, is:
$$ \frac{1}{T_{I1-I2}} = \frac{3}{2} M_2^{I1-I2} \left( \frac{ 2\pi }{5 M_2^{I1} } \right)^{1/2} \cong 1.681 \cdot \frac{ M_2^{I1-I2}}{ ( M_2^{I1})^{1/2} } $$
where $M_2^{I1}$ and $M_2^{I1-I2}$ are the Van Vleck second moments (in units of s$^{-2}$) for the $I^1$ and $I^1-I^2$ spin systems, respectively (see Abragam~\cite{abragam1961principles} and Stejskal \& Memory~\cite{stejskal1994high}).    The  $I^1$  spin system generally refers to  $I^1$  spins interacting with the lattice.  For example, it could be due to homonuclear  $I^1$- $I^1$ spin-spin interactions.   The $I^1-I^2$ spin system, on the other hand, refers to the $I^1-I^2$ heteronuclear interaction, which occurs through the tensor $\mathtensor{F}$.  The second moment $M_2^{I1-I2}$ is enantiospecific because it is proportional to the trace, $\mbox{tr}\{ [ \mathbf{I}^1 \cdot \mathtensor{F} \cdot \mathbf{I}^2 {}^{**},I_x]^2 \}$, where $I_x=I_x^1+I_x^2$ and $\mathbf{I}^1 \cdot \mathtensor{F} \cdot \mathbf{I}^2 {}^{**}$ is the indirect nuclear spin-spin interaction in the spin lock frame ($**$) and the components of the tensor $\mathtensor{F}$ are enantiospecific (see Section ``Text S2. Enantiospecificity in Cross-Polarization'').  Therefore, the CP rate $(T_{I^1-I^2})^{-1}$ and time at maximum $t_m$ both depend on chirality through the components of the $\mathtensor{F}$ tensor.

\subsection{Appendix: Interaction Representation for Spin-Lock Experiment (Review)}

This is a review of a well known procedure in quantum mechanics.   The usual transformation to the interaction representation proceeds by decomposing the Liouvillian as:
$\mathtensor{L}(t) = \mathtensor{L}_0(t) + \mathtensor{L}_V(t)$ (with $'0'$ referring to the unperturbed part and $'V'$ to the perturbation), the
Heisenberg equation of motion is
\begin{equation*}
 \frac{dA(t)}{dt} = \bigl[ \mathtensor{L}_0(t) +
\mathtensor{L}_V(t) \bigr] A(t).
\end{equation*}
\noindent The transformed operators are:
$$ \tilde{A} = \overrightarrow{T} \exp \left( - \int_0^t d\tau
 \mathtensor{L}_0(\tau) \right) A, \qquad
 \tilde{\mathtensor{L}} = \overrightarrow{T} \exp \left( - \int_0^t d\tau
 \mathtensor{L}_0(\tau) \right) \mathtensor{L},$$
\noindent where $\overrightarrow{T}$ is the Dyson time-ordering operator~\cite{HaeberlenBook} going from
left to right, with the most recent (in time) Liouvillian to the right. These
transformations are anti-causal. The (total) time-derivative evaluated at the most recent time point is:
\begin{align*}
 \frac{ d \tilde{A} }{dt} =& \frac{d}{dt}  \overrightarrow{T} \exp
 \left( - \int_0^t d\tau \mathtensor{L}_0(\tau) \right) A \nonumber \\
 =& - \overrightarrow{T} \exp\left( - \int_0^t d\tau
 \mathtensor{L}_0(\tau) \right) \mathtensor{L}_0(t) A +
 \overrightarrow{T} \exp\left( -
 \int_0^t d\tau \mathtensor{L}_0(\tau) \right)
 \bigl[ \mathtensor{L}_0(t) A +
 \mathtensor{L}_V(t) A(t) \bigr] \nonumber \\
 =& \tilde{\mathtensor{L}}_V(t) \tilde{A}.
\label{eq:interrp}
\end{align*}
Thus, in a rotating frame of reference, a Liouville equation
\begin{equation*}
\frac{d \tilde{A}}{dt} = \tilde{\mathtensor{L}}_V \tilde{A}
\end{equation*}
\noindent holds, where $\tilde{\mathtensor{L}}_V$ is the perturbation
part of the Liouvillian expressed in the rotating frame.  
It is also possible to effect yet another transformation to a second frame
by decomposing the time derivative $\tilde{\mathtensor{L}}_V$ further:
\begin{equation*}
\tilde{\mathtensor{L}}_V = \tilde{\mathtensor{L}}_1 +
\tilde{\mathtensor{L}}_2,
\end{equation*}
\noindent where $\tilde{\mathtensor{L}}_1$ is the
part corresponding to the RF field and
$\tilde{\mathtensor{L}}_2$ is the remaining ``internal''
part of the Liouvillian (e.g. spin-spin
interactions).
Consider the transformation
\begin{equation*}
 \hat{\tilde{A}} = \overrightarrow{T} \exp \left( - \int_0^t
 \tilde{\mathtensor{L}}_1(t') dt' \right) \tilde{A},
\end{equation*}
\noindent where $\overrightarrow{T}$ is the Dyson time-ordering
operator~\cite{HaeberlenBook}, which places the most recent events to
the right. Then,
\begin{align*}
\frac{ d \hat{\tilde{A}}(t) }{dt} = \hat{\tilde{\mathtensor{L}}}_2
\hat{\tilde{A}}(t).
\end{align*}
The solution is,
\begin{equation*}
\hat{\tilde{A}}(t) = \overrightarrow{T} \exp \left( \int_0^t
 \hat{\tilde{\mathtensor{L}}}_2 (t') dt' \right)
 \hat{\tilde{A}}(0),
\end{equation*}
\noindent where
\begin{equation*}
\hat{\tilde{\mathtensor{L}}}_2 = \overrightarrow{T} \exp \left(- \int_0^t 
\tilde{\mathtensor{L}}_1 (t') dt' \right) \tilde{\mathtensor{L}}_2,
\end{equation*}
\noindent and
\begin{equation*}
\tilde{\mathtensor{L}}_2 = \overrightarrow{T} \exp \left( - \int_0^t d\tau
\mathtensor{L}_0(\tau) \right) \mathtensor{L}_2.
\end{equation*}

\section{Text S3. J Coupling Stereochemical Deviations (DFT) for Amino Acids}

The theoretical investigations into the NMR J couplings in various amino acids were conducted by  DFT using the ORCA {\it ab initio} quantum chemistry software package~\cite{bib:cremer2007}. The amino acids selected for this study are alanine, arginine, aspartic acid, cysteine, glutamic acid, glutamine, glyceraldehyde (non-amino acid), methionine, phenylalanine, serine, threonine, tyrosine and valine. 
Differences in NMR J couplings between ($D, L$) enantiomers were quantified using the J coupling stereochemical deviation, $[J(L)- J(D)]/[J(L) + J(D)]/2$. Nonzero values of this relative difference constitute evidence of chiral selectivity of the scalar coupling.  J couplings between $^1$H and $^{13}$C nuclei were computed. The results are shown in (Supplementary) Figs.~\ref{fig:S2phenyla}-\ref{fig:S13valine} below.  The raw data is provided in the form of tables in Section~``Text S4'' below.

\begin{figure}[h!]
\begin{center}
\includegraphics[width=0.95\textwidth]{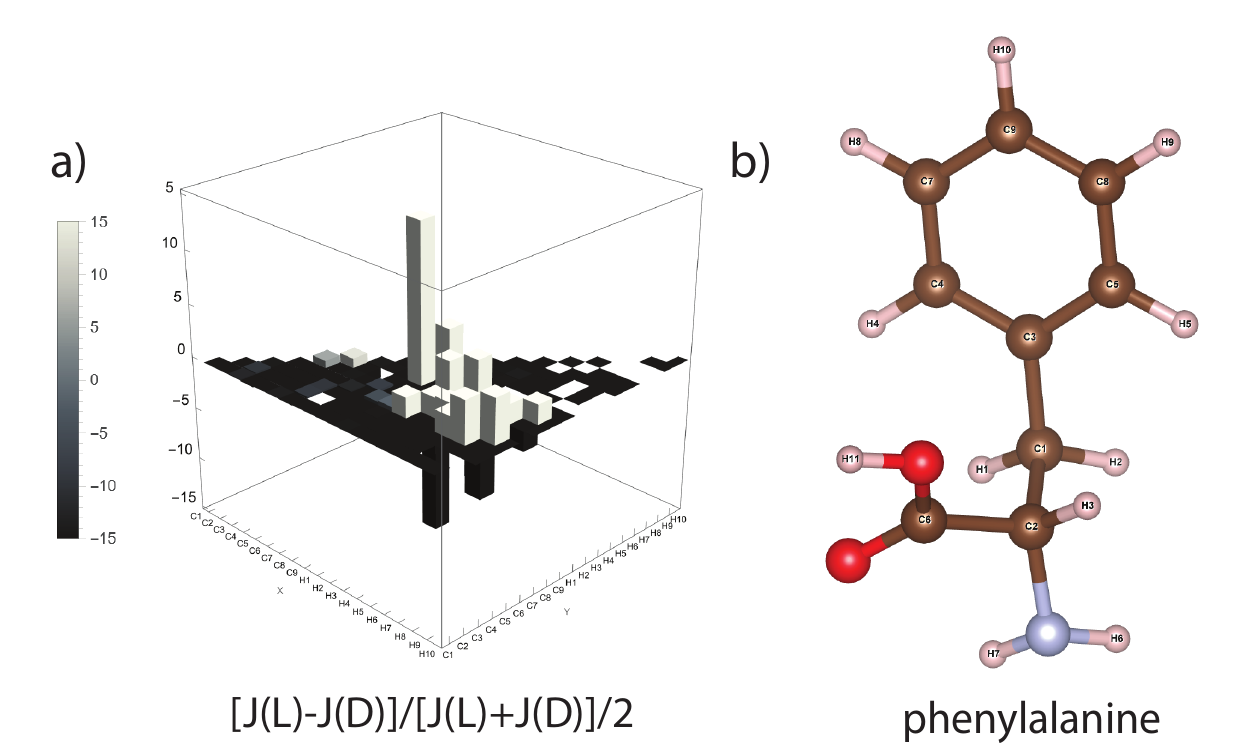}
\end{center}
\caption{J coupling stereochemical deviation, $[J(L)-J(D)]/[J(L) + J(D)]/2$ for phenylalanine. The atom labels are ``C1", ``C2", ``C3", ''C4", ``C5", ``C6",``H1", ``H2", ``H3", ``H4", ``H5", ``H6", ``H7", ``H8", ``H9", ``H10", ``H11".}
\label{fig:S2phenyla}
\end{figure}

\begin{figure}[h!]
\begin{center}
\includegraphics[width=0.95\textwidth]{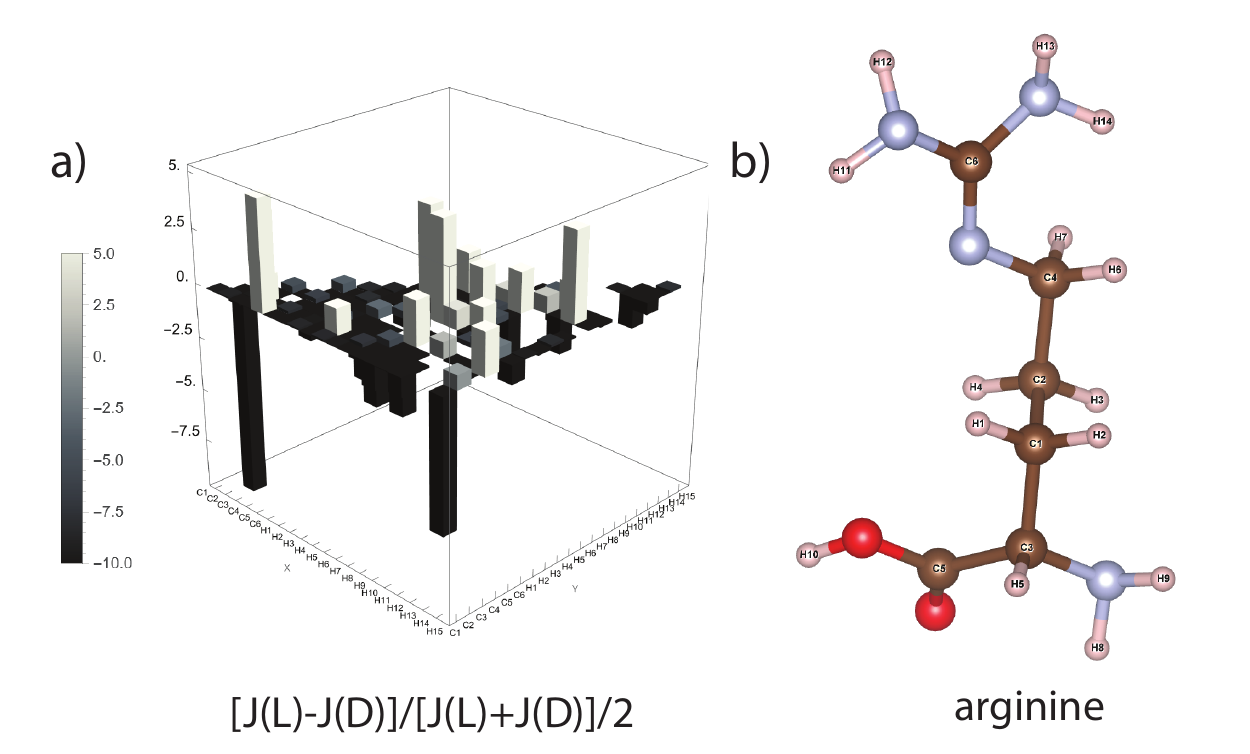}
\end{center}
\caption{J coupling stereochemical deviation, $[J(L)-J(D)]/[J(L)+ J(D)]/2$ for arginine. The atom labels are ``C1", ``C2", ``C3", ``C4", ``C5", ``C6", ``H1", ``H2", ``H3", ``H4", ``H5", \
``H6", ``H7", ``H8", ``H9", ``H10", ``H11", ``H12", ``H13", ``H14".}
\label{fig:S3arginine}
\end{figure}

\begin{figure}[h!]
\begin{center}
\includegraphics[width=0.95\textwidth]{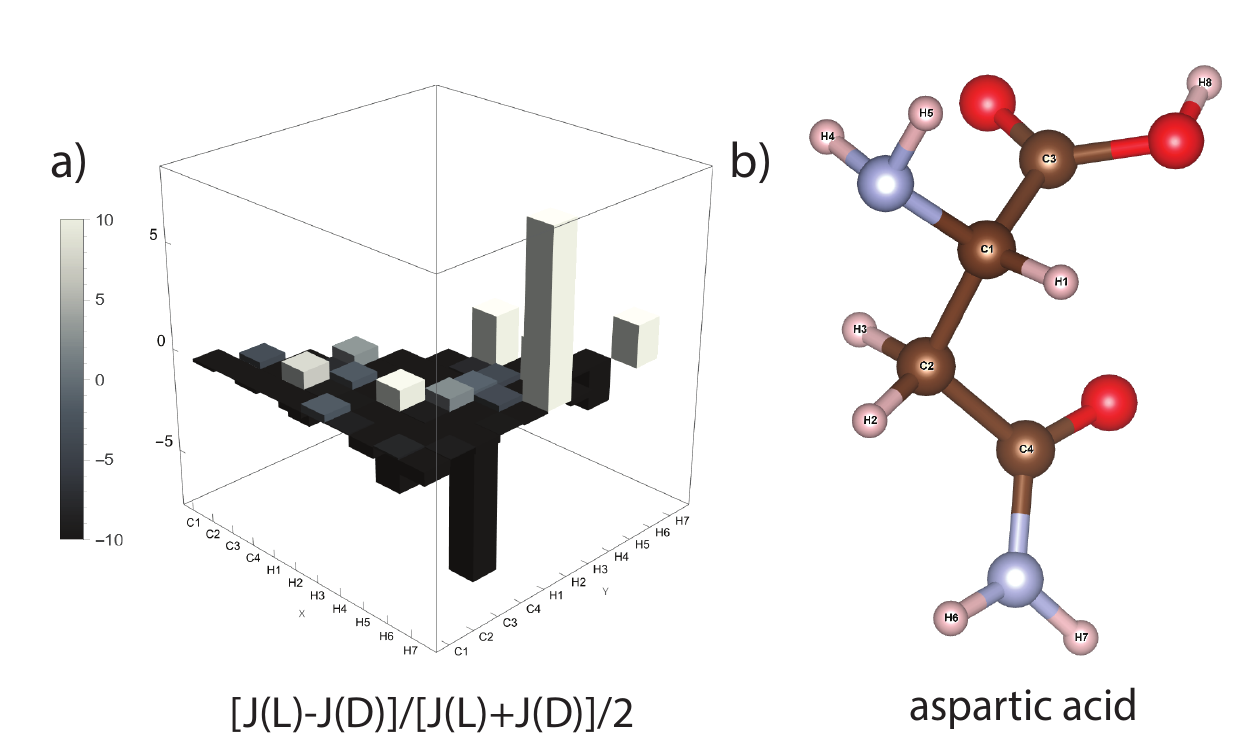}
\end{center}
\caption{J coupling stereochemical deviation, $[J(L)-J(D)]/[J(L)+ J(D)]/2$ for aspartic acid. The atoms labels are ``C1", ``C2", ``C3", ``C4", ``H1", ``H2", ``H3", ``H4", ``H5", ``H6", ``H7".}
\label{fig:S4asparticacid}
\end{figure}

\begin{figure}[h!]
\begin{center}
\includegraphics[width=0.95\textwidth]{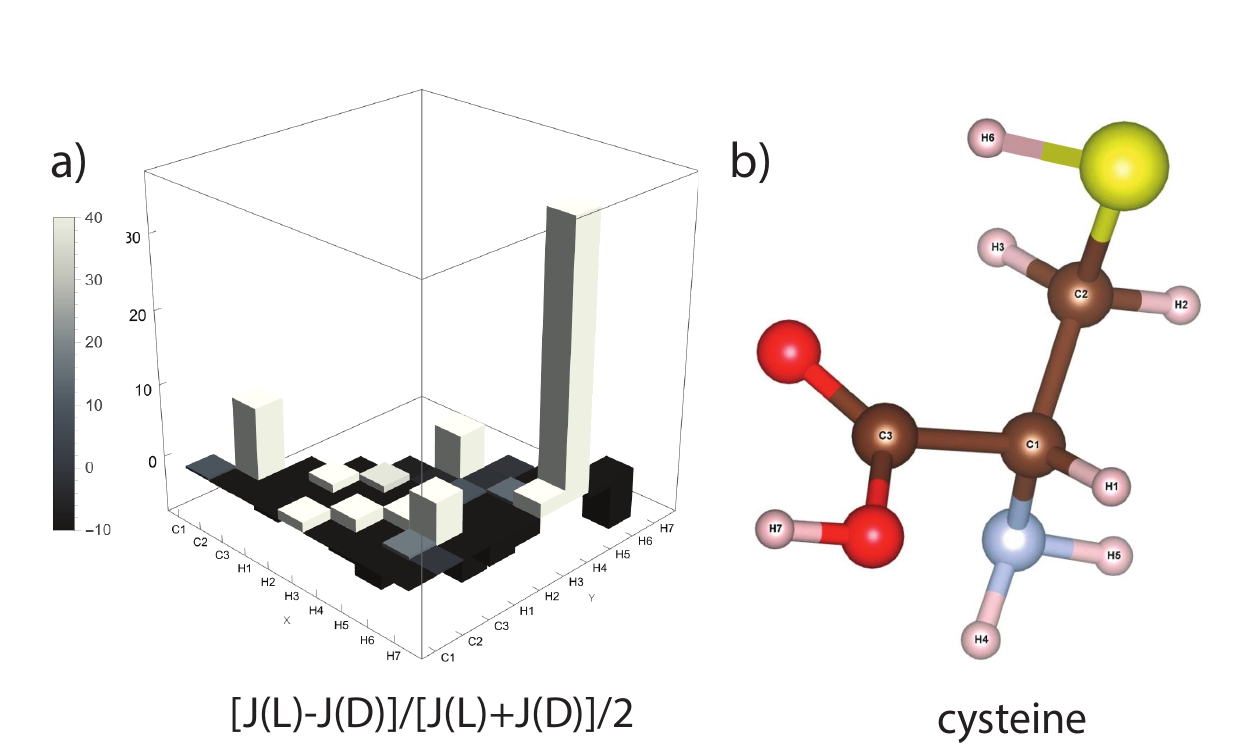}
\end{center}
\caption{J coupling stereochemical deviation, $[J(L)-J(D)]/[J(L)+ J(D)]/2$ for cysteine. The atom labels are ``C1", ``C2", ``C3", ``H1", ``H2", ``H3", ``H4", ``H5", ``H6", ``H7".}
\label{fig:S5cysteine}
\end{figure}

\begin{figure}[h!]
\begin{center}
\includegraphics[width=0.95\textwidth]{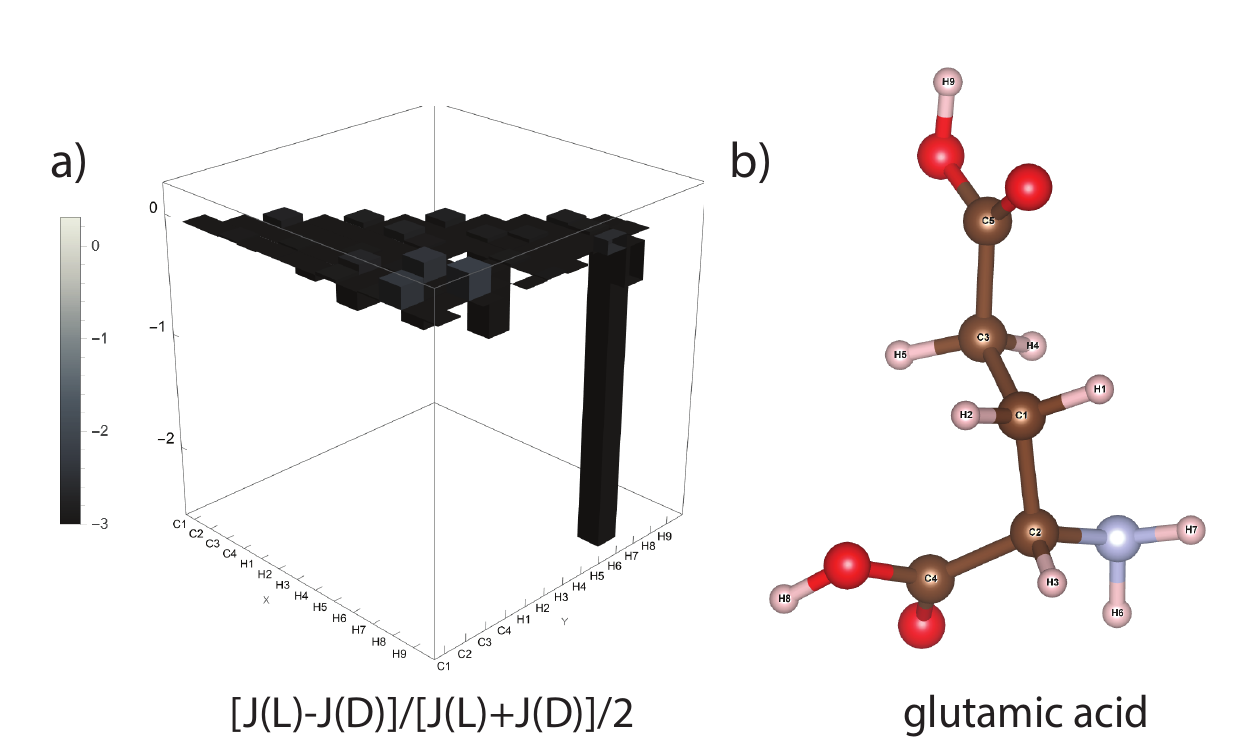}
\end{center}
\caption{J coupling stereochemical deviation, $[J(L)-J(D)]/[J(L)+ J(D)]/2$ for glutamic acid. The atom labels are ``C1", ``C2", ``C3", ``C4", ``C5", ``H1", ``H2", ``H3", ``H4", ``H5", ``H6", ``H7", ``H8", ``H9".}
\label{fig:S6glutamicacid}
\end{figure}

\begin{figure}[h!]
\begin{center}
\includegraphics[width=0.95\textwidth]{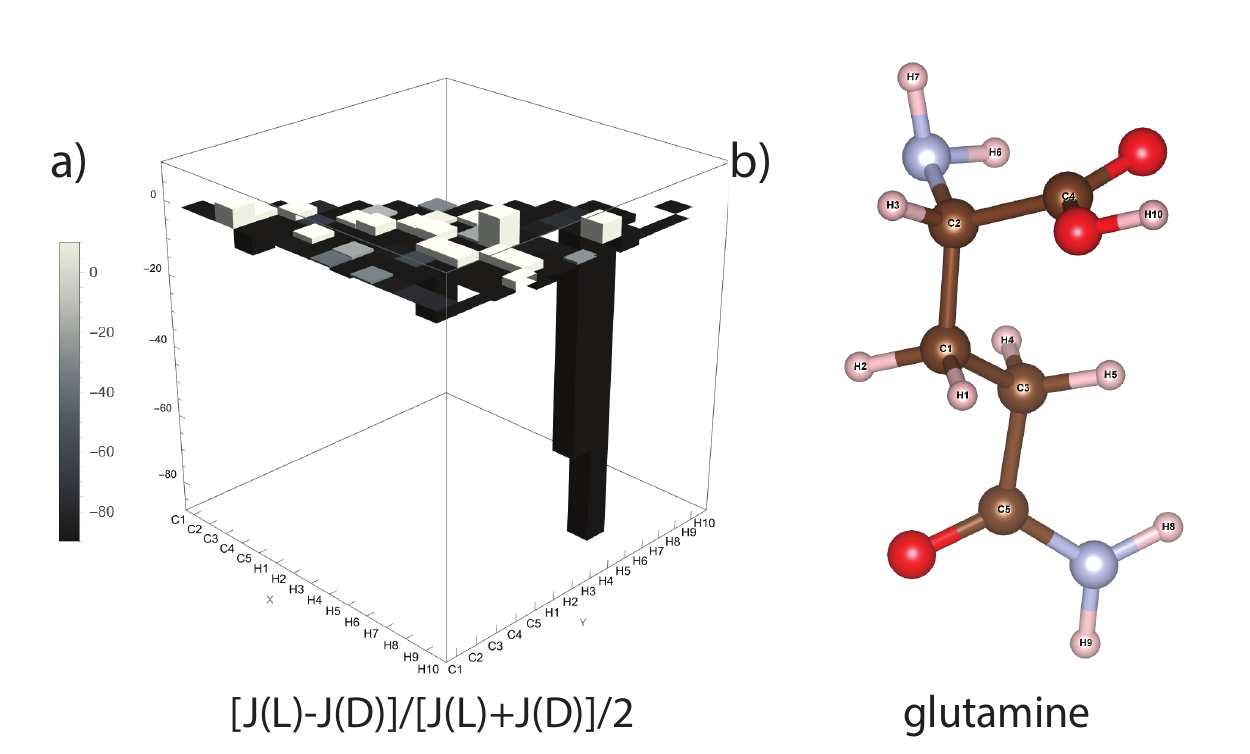}
\end{center}
\caption{J coupling stereochemical deviation, $[J(L)-J(D)]/[J(L)+ J(D)]/2$ for glutamine. The atom labels are ``C1", ``C2", ``C3", ``C4", ``C5", ``H1", ``H2", ``H3", ``H4", ``H5", ``H6", ``H7", ``H8", ``H9", ``H10".}
\label{fig:S7glutamine}
\end{figure}

\begin{figure}[h!]
\begin{center}
\includegraphics[width=0.95\textwidth]{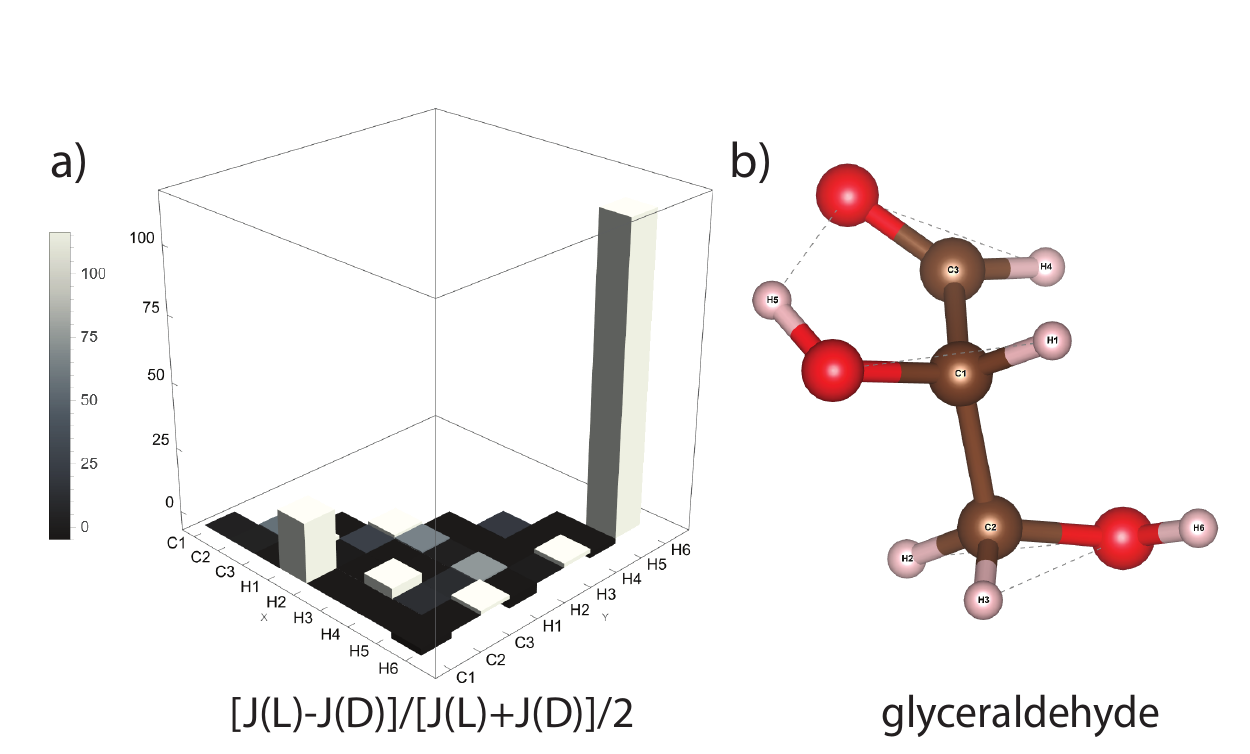}
\end{center}
\caption{J coupling stereochemical deviation, $[J(L)-J(D)]/[J(L)+ J(D)]/2$ for glyceraldehyde. The atom labels are ``C1", ``C2", ``C3", ``H1", ``H2", ``H3", ``H4", ``H5", ``H6".}
\label{fig:S8glyceraldehyde}
\end{figure}

\begin{figure}[h!]
\begin{center}
\includegraphics[width=0.95\textwidth]{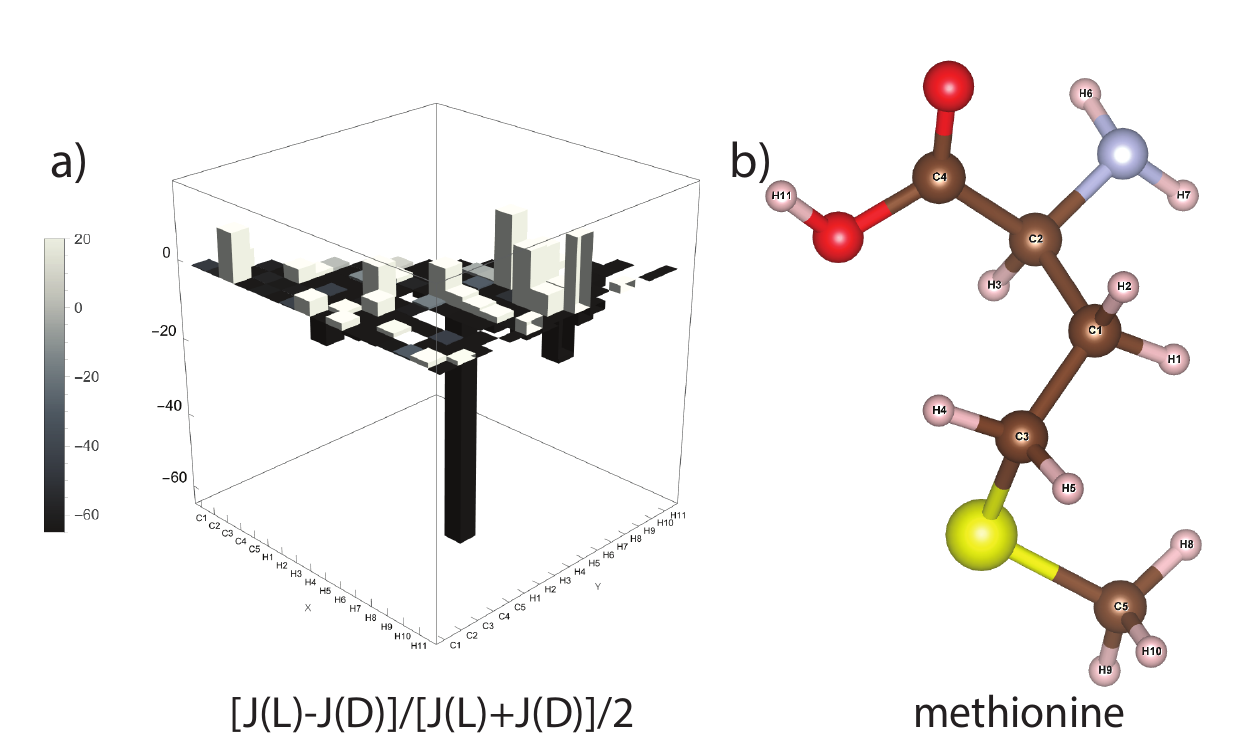}
\end{center}
\caption{J coupling stereochemical deviation, $[J(L)-J(D)]/[J(L)+ J(D)]/2$ for methionine. The atom labels are ``C1", ``C2", ``C3", ``C4", ``C5", ``H1", ``H2", ``H3", ``H4", ``H5", ``H6", ``H7", ``H8", ``H9", ``H10", ``H11".}
\label{fig:S9methionine}
\end{figure}

\begin{figure}[h!]
\begin{center}
\includegraphics[width=0.95\textwidth]{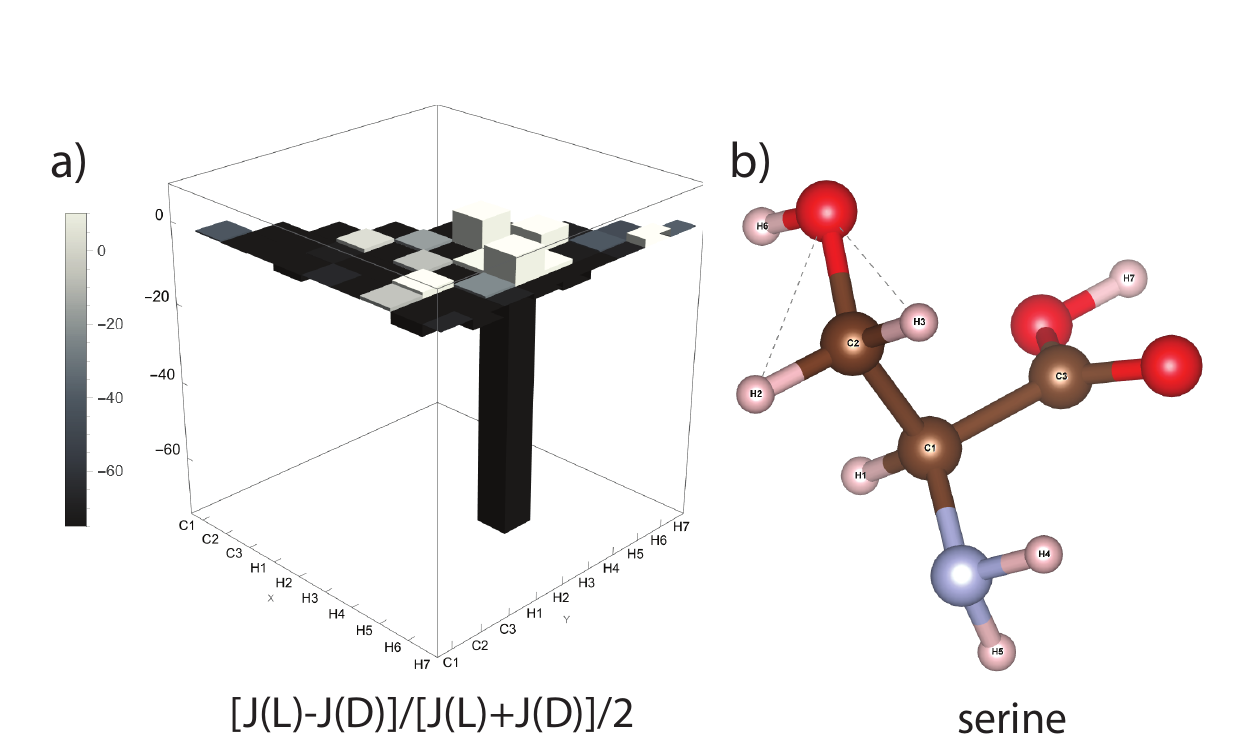}
\end{center}
\caption{J coupling stereochemical deviation, $[J(L)-J(D)]/[J(L)+ J(D)]/2$ for serine. The atom labels are ``C1", ``C2", ``C3", ``H1", ``H2", ``H3", ``H4", ``H5", ``H6", ``H7". }
\label{fig:S10serine}
\end{figure}

\begin{figure}[h!]
\begin{center}
\includegraphics[width=0.95\textwidth]{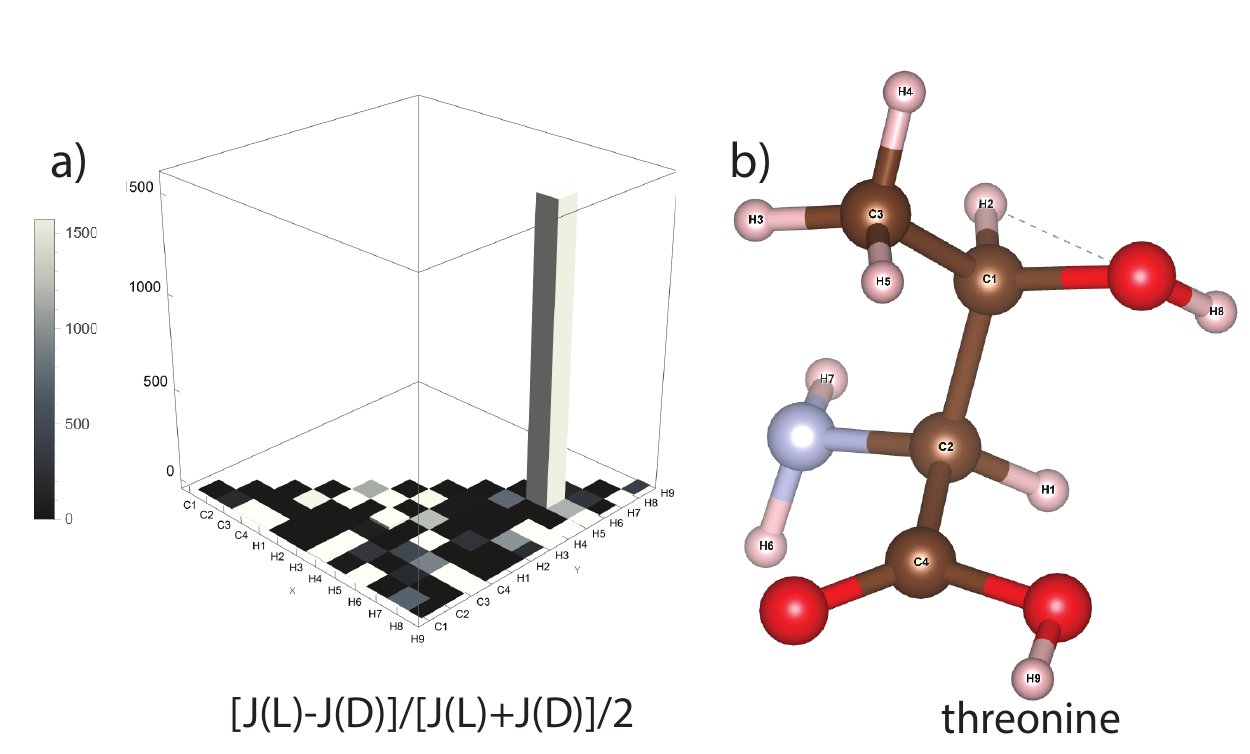}
\end{center}
\caption{J coupling stereochemical deviation, $[J(L)-J(D)]/[J(L)+ J(D)]/2$ for threonine. The atom labels are ``C1", ``C2", ``C3", ``C4", ``H1", ``H2", ``H3", ``H4", ``H5", ``H6", ``H7", ``H8", ``H9". }
\label{fig:S11threonine}
\end{figure}

\begin{figure}[h!]
\begin{center}
\includegraphics[width=0.95\textwidth]{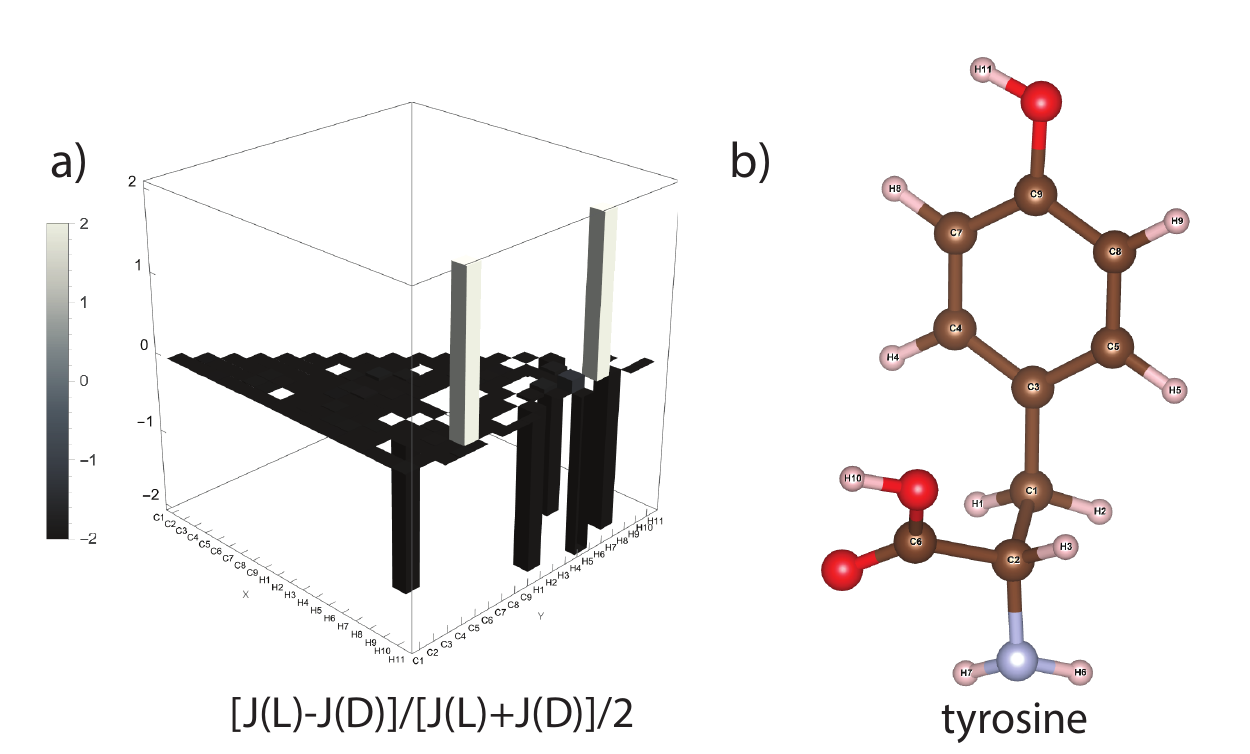}
\end{center}
\caption{J coupling stereochemical deviation, $[J(L)-J(D)]/[J(L)+ J(D)]/2$ for tyrosine. The atom labels are ``C1", ``C2", ``C3", ``C4", ``C5", ``C6", ``C7", ``C8", ``C9", ``H1", ``H2", ``H3", ``H4", ``H5", ``H6", ``H7", ``H8", ``H9", ``H10", ``H11".} 
\label{fig:S12tyrosine}
\end{figure}

\begin{figure}[h!]
\begin{center}
\includegraphics[width=0.95\textwidth]{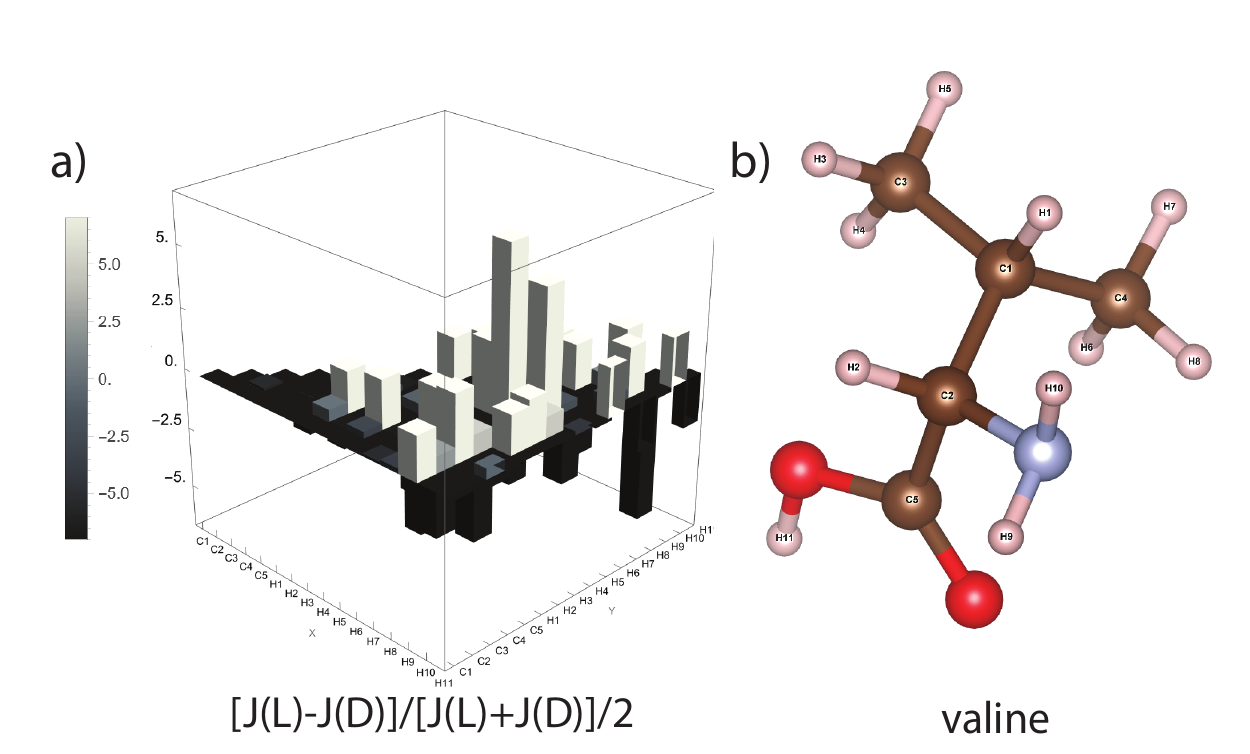}
\end{center}
\caption{J coupling stereochemical deviation, $[J(L)-J(D)]/[J(L)+ J(D)]/2$ for valine. The atom labels are ``C1", ``C2", ``C3", ``C4", ``C5", ``H1", ``H2", ``H3", ``H4", ``H5", ``H6", ``H7", ``H8", ``H9", ``H10", ``H11". }
\label{fig:S13valine}
\end{figure}

\section{Text S4. Raw Data for J Couplings (DFT) in Amino Acids\label{sec:S4}}

This section contains numerical values used to generate plots in the previous section.  Raw J coupling values are also provided. Specifically, we provide numerical values for the 
differences in NMR J couplings between ($D, L$) enantiomers can be quantified using the J coupling stereochemical deviation, $[J(L)-J(D)]/[J(L)+J(D)]/2$. Nonzero values of this relative difference constitute evidence of chiral selectivity of the scalar coupling.  J couplings between $^1$H and $^{13}$C nuclei were computed by DFT using ORCA for the amino acids: alanine, arginine, aspartic acid, cysteine, glutamic acid, glutamine, glyceraldehyde, methionine, phenylalanine, serine, threonine, tyrosine and valine. The results are shown\footnote{In the tables, ``$0.$'' along the diagonal indicates the absence of a J coupling between the same atom (self-coupling).  
Off-diagonal entries marked with $0.$ represent indeterminate J-couplings. These values were flagged by ORCA due to limitations in its ability to reliably calculate certain constants under specific molecular conditions.
``$0.00$'' indicates a number smaller than $0.01$.  Also, ``$-0.01$'' indicates a similarly small number, but negative.} in Supplementary Tables~\ref{tab:S1alanine}-\ref{tab:S13valine}.

\begin{table}
\scriptsize
\centering
\begin{tabular}{ccccccccccc} 
 \text{Alanine} & \text{C1} & \text{C2} & \text{C3} & \text{H1} & \text{H2} & H3 & H4 & H5 & H6 & H7   \\
 \text{C1} & \text{0.} & -0.13 & 0.07 & 0.03 & -0.59  & 0.56 & -0.08 & -0.31 & -0.10 & 0.13 \\
 \text{C2} & -0.13 & \text{0.} & 0.08 & -0.37 & -0.03  & -0.01 & 0.02 & -1.11 & -5.24 & -0.03 \\
 \text{C3} & 0.07 & 0.08 & \text{0.} & 0.24 & -1.67  & 1.77 & -0.02 & 1.64 & 0.86 & -0.01 \\
 \text{H1} & 0.03 & -0.37 & 0.24 & \text{0.} & -0.84  & 0.50 & -0.04 & -1.88 & -0.15 & 7.39 \\
 \text{H2} & -0.59 & -0.03 & -1.67 & -0.84 & \text{0.}  &  0.06 & -0.02 & 2.17 & 0.84 & -0.49 \\
 \text{H3} & 0.56 & -0.01 & 1.77 & 0.50 & 0.06  & \text{0.} & -0.05 & -2.42 & -0.47 & 0.79 \\
 \text{H4} & -0.08 & 0.02 & -0.02 & -0.04 & -0.02 & -0.05 & \text{0.} & 0.5 & 1.17 & -0.96 \\
 \text{H5} & -0.31 & -1.11 & 1.64 & -1.88 & 2.17  & -2.42 & 0.5 & \text{0.} & -0.19 & -2. \\
 \text{H6} & -0.10 & -5.24 & 0.86 & -0.15 & 0.84  & -0.47 & 1.17 & -0.19 & \text{0.} & -0.70 \\
 \text{H7} & 0.13 & -0.03 & -0.01 & 7.39 & -0.49  & 0.79 & -0.96 & -2. & -0.70 & \text{0.} \\
\end{tabular}
\caption{Numerical values of the fractional deviation $[J(L)-J(D)]/[J(L)+J(D)]/2$ for alanine.}
\label{tab:S1alanine}
\end{table}

\begin{table}
\scriptsize
\centering
\begin{tabular}{ccccccccccc}
 \text{D-alanine} & \text{C1} & \text{C2} & \text{C3} & \text{H1} & \text{H2} & \text{H3} & \text{H4} & \text{H5} & \text{H6} & \text{H7} \\
 \text{C1} & 0. & 127.25 & 166.61 & 169.27 & -6.62 & -3.6 & -6.83 & -5.26 & -5.96 & 7.23 \\
 \text{C2} & 127.25 & 0. & -9.22 & -8.51 & 180.78 & 178.40 & 172.88 & 11.30 & 1.39 & -0.45 \\
 \text{C3} & 166.61 & -9.22 & 0. & -6.96 & 9.89 & 0.59 & 2.30 & 1.12 & 1.56 & -7.61 \\
 \text{H1} & 169.27 & -8.51 & -6.96 & 0. & 5.40 & 3.24 & 13.68 & 2.43 & 15.63 & -0.31 \\
 \text{H2} & -6.62 & 180.78 & 9.89 & 5.40 & 0. & -13.56 & -13.69 & 0.02 & -0.27 & -0.15 \\
 \text{H3} & -3.56 & 178.40 & 0.59 & 3.23 & -13.56 & 0. & -13.90 & 3.67 & -0.40 & -0.02 \\
 \text{H4} & -6.83 & 172.88 & 2.30 & 13.68 & -13.69 & -13.90 & 0. & -0.20 & -0.04 & 0.14 \\
 \text{H5} & -5.26 & 11.30 & 1.12 & 2.43 & 0.02 & 3.67 & -0.20 & 0. & -11.71 & -0.08 \\
 \text{H6} & -5.96 & 1.39 & 1.56 & 15.63 & -0.27 & -0.40 & -0.04 & -11.71 & 0. & -0.21 \\
 \text{H7} & 7.23 & -0.45 & -7.61 & -0.31 & -0.15 & -0.03 & 0.14 & -0.08 & -0.21 & 0. \\
\end{tabular}
\caption{Raw values of J couplings in D-alanine.}
\end{table}

\begin{table}
\scriptsize
\centering
\begin{tabular}{ccccccccccc}
 \text{L-alanine} & \text{C1} & \text{C2} & \text{C3} & \text{H1} & \text{H2} & \text{H3} & \text{H4} & \text{H5} & \text{H6} & \text{H7} \\
 \text{C1} & 0. & 111.56 & 179.38 & 175.18 & -3.61 & -6.31 & -6.30 & -3.85 & -5.42 & 8.21 \\
 \text{C2} & 111.56 & 0. & -9.94 & -5.86 & 175.44 & 177.06 & 175.73 & 3.25 & -0.62 & -0.44 \\
 \text{C3} & 179.38 & -9.94 & 0. & -8.84 & 0.90 & 9.85 & 2.26 & 11.27 & 3.90 & -7.56 \\
 \text{H1} & 175.18 & -5.86 & -8.84 & 0. & 2.22 & 5.41 & 13.11 & 0.08 & 13.42 & 0.54 \\
 \text{H2} & -3.61 & 175.44 & 0.90 & 2.22 & 0. & -14.38 & -13.42 & -0.45 & -0.67 & -0.09 \\
 \text{H3} & -6.31 & 177.06 & 9.85 & 5.41 & -14.38 & 0. & -13.23 & -0.35 & -0.25 & -0.06 \\
 \text{H4} & -6.30 & 175.73 & 2.26 & 13.11 & -13.42 & -13.23 & 0. & -0.34 & -0.16 & 0.05 \\
 \text{H5} & -3.85 & 3.25 & 11.27 & 0.08 & -0.45 & -0.35 & -0.34 & 0. & -9.69 & 0. \\
 \text{H6} & -5.42 & -0.62 & 3.90 & 13.42 & -0.67 & -0.25 & -0.16 & -9.69 & 0. & -0.10 \\
 \text{H7} & 8.21 & -0.44 & -7.56 & 0.54 & -0.09 & -0.06 & 0.05 & 0. & -0.10 & 0. \\
\end{tabular}
\caption{Raw values of J couplings in L-alanine.}
\end{table}

\begin{table}
\tiny
\begin{tabular}{ccccccccccc}
 \text{Arginine} & \text{C1} & \text{C2} & \text{C3} & \text{C4} & \text{C5} & \text{C6} & \text{H1} & \text{H2} & \text{H3} & \text{H4}  \\
 \text{C1} & \text{0.} & -0.01 & -0.04 & -0.01 & -0.12 & -0.02 & -0.03 & 0.03 & 0.06 & 0.00  \\
 \text{C2} & -0.01 & \text{0.} & 0.21 & -0.01 & 0.72 & -0.02 & 0.01 & -0.08 & 0.01 & 0.00  \\
 \text{C3} & -0.04 & 0.21 & \text{0.} & -0.27 & -0.09 & \text{0.} & -0.73 & 0.02 & -1.12 & 0.71  \\
 \text{C4} & -0.01 & -0.01 & -0.27 & \text{0.} & -0.45 & -0.03 & 0.85 & -0.61 & 0.04 & -0.01  \\
 \text{C5} & -0.12 & 0.72 & -0.09 & -0.45 & \text{0.} & \text{0.} & -1.11 & -0.11 & 0.06 & -1.61  \\
 \text{C6} & -0.02 & -0.02 & \text{0.} & -0.03 & \text{0.} & \text{0.} & 0.18 & -0.19 & -0.01 & -2. \\
 \text{H1} & -0.03 & 0.01 & -0.73 & 0.85 & -1.11 & 0.18 & \text{0.} & 0.04 & 0.01 & -0.52  \\
\end{tabular}
\caption{Numerical values of the fractional deviation $[J(L)-J(D)]/[J(L)+J(D)]/2$ for arginine.}
\end{table}

\begin{table}
\tiny
\begin{tabular}{ccccccccccc}
 \text{D-arginine} & \text{C1} & \text{C2} & \text{C3} & \text{C4} & \text{C5} & \text{C6} & \text{H1} & \text{H2} & \text{H3} & \text{H4}  \\
 \text{C1} & 0. & 132.23 & 119.47 & -8.47 & -9.08 & -0.157 & 176.79 & 162.55 & -6.05 & -5.64  \\
 \text{C2} & 132.23 & 0. & -5.38 & 143.27 & 0.66 & 8.85 & -6.57 & -6.32 & 160.57 & 167.97  \\
 \text{C3} & 119.47 & -5.38 & 0. & 8.95 & 174.26 & 0. & -7.03 & -6.27 & 2.70 & 1.11  \\
 \text{C4} & -8.47 & 143.27 & 8.95 & 0. & -0.131 & -4.22 & 1.04 & 3.131 & -3.61 & -7.63  \\
 \text{C5} & -9.08 & 0.66 & 174.26 & -0.131 & 0. & 0. & 2.44 & 10.47 & -0.17 & -0.14  \\
 \text{C6} & -0.16 & 8.85 & 0. & -4.22 & 0. & 0. & 0.13 & 0.11 & 1.53 & -0.06 \\
 \text{H1} & 176.79 & -6.57 & -7.03 & 1.04 & 2.44 & 0.13 & 0. & -13.48 & 13.32 & 4.58 \\
\end{tabular}
\caption{Raw values of J couplings in D-arginine.}
\end{table}

\begin{table}
\tiny
\begin{tabular}{ccccccccccc}
 \text{L-arginine} & \text{C1} & \text{C2} & \text{C3} & \text{C4} & \text{C5} & \text{C6} & \text{H1} & \text{H2} & \text{H3} & \text{H4}  \\
 \text{C1} & 0. & 130.34 & 114.85 & -8.36 & -8.05 & -0.15 & 171.75 & 166.93 & -6.43 & -5.66  \\
 \text{C2} & 130.34 & 0. & -6.62 & 141.14 & 1.39 & 8.68 & -6.64 & -5.86 & 162.52 & 168.20  \\
 \text{C3} & 114.85 & -6.62 & 0. & 6.84 & 158.85 & 0. & -3.25 & -6.39 & 0.77 & 2.34  \\
 \text{C4} & -8.36 & 141.14 & 6.84 & 0. & -0.08 & -4.08 & 2.57 & 1.67 & -3.77 & -7.59  \\
 \text{C5} & -8.05 & 1.39 & 158.85 & -0.08 & 0. & 0. & 0.70 & 9.42 & -0.18 & -0.02  \\
 \text{C6} & -0.15 & 8.68 & 0. & -4.08 & 0. & 0. & 0.16 & 0.09 & 1.52 & 0.  \\
 \text{H1} & 171.75 & -6.64 & -3.25 & 2.57 & 0.70 & 0.16 & 0. & -14.04 & 13.47 & 2.70  \\
\end{tabular}
\caption{Raw values of J couplings in L-arginine.}
\end{table}

\begin{table}
\scriptsize
\begin{tabular}{cccccccccccc}
 \text{Aspartic acid} & \text{C1} & \text{C2} & \text{C3} & \text{C4} & \text{H1} & \text{H2} & \text{H3} & \text{H4} & \text{H5} & \text{H6} & \text{H7} \\
 \text{C1} & \text{0.} & -0.07 & 0.03 & -0.22 & -0.01 & 3.67 & 2.14 & -2.24 & -2.23 & 2.08 & 2. \\
 \text{C2} & -0.07 & \text{0.} & -0.01 & -0.01 & 0.25 & 0.08 & -0.04 & 1.79 & -2.16 & -1.30 & 0.01 \\
 \text{C3} & 0.03 & -0.01 & \text{0.} & -1.90 & 0.13 & -0.91 & 0.98 & -1.02 & 1.54 & 0.01 & -0.46 \\
 \text{C4} & -0.22 & -0.01 & -1.90 & \text{0.} & 0.28 & -0.44 & 0.30 & 2.46 & -2.53 & 4.19 & 0.01 \\
 \text{H1} & -0.01 & 0.25 & 0.13 & 0.28 & \text{0.} & 1.18 & -1.15 & -2.31 & 1.40 & 0.16 & -0.12 \\
 \text{H2} & -0.14 & 0.08 & -0.91 & -0.44 & 1.18 & \text{0.} & -0.02 & 1.17 & -3.08 & 2. & 22.24 \\
 \text{H3} & 0.17 & -0.04 & 0.98 & 0.30 & -1.15 & -0.02 & \text{0.} & 3.13 & -0.94 & 0.48 & -17.04 \\
 \text{H4} & -0.05 & 1.79 & -1.02 & 2.46 & -2.31 & 1.17 & 3.13 & \text{0.} & -0.08 & -2.66 & \text{0.} \\
 \text{H5} & -0.09 & -2.16 & 1.54 & -2.53 & 1.40 & -3.08 & -0.94 & -0.08 & \text{0.} & -2. & \text{0.} \\
 \text{H6} & 0.14 & -1.30 & 0.01 & 4.19 & 0.16 & 2. & 0.48 & -2.66 & -2. & \text{0.} & 2. \\
 \text{H7} & -0.20 & 0.01 & -0.46 & 0.01 & -0.12 & 22.24 & -17.05 & \text{0.} & \text{0.} & 2. & \text{0.} \\
\end{tabular}
\caption{Numerical values of the fractional deviation $[J(L)-J(D)]/[J(L)+J(D)]/2$ for aspartic acid.}
\end{table}

\begin{table}
\scriptsize
\begin{tabular}{cccccccccccc}
 \text{D-a. acid} & \text{C1} & \text{C2} & \text{C3} & \text{C4} & \text{H1} & \text{H2} & \text{H3} & \text{H4} & \text{H5} & \text{H6} & \text{H7} \\
 \text{C1} & 0. & 139.48 & 172.68 & -5.65 & 169.25 & 1.75 & 0.25 & 78.22 & 80.38 & -0.14 & 0. \\
 \text{C2} & 139.48 & 0. & -8.33 & 188.81 & -5.38 & 163.72 & 175.59 & 0.53 & 11.10 & 0.97 & 7.48 \\
 \text{C3} & 172.68 & -8.33 & 0. & 0.12 & -9.59 & 10.54 & 3.08 & 11.77 & 1.35 & -7.35 & -0.15 \\
 \text{C4} & -5.65 & 188.81 & 0.12 & 0. & 1.10 & -11.40 & -8.46 & -0.235 & 2.98 & -0.05 & -7.58 \\
 \text{H1} & 169.25 & -5.38 & -9.589 & 1.10 & 0. & 2.86 & 11.08 & 4.77 & 1.44 & -0.66 & -0.11 \\
 \text{H2} & -6.82 & 163.72 & 10.54 & -11.40 & 2.86 & 0. & -17.85 & -0.113 & 0.79 & 0. & -0.84 \\
 \text{H3} & -6.52 & 175.59 & 3.08 & -8.46 & 11.08 & -17.847 & 0. & -0.202 & -0.48 & -0.04 & 0.74 \\
 \text{H4} & -4.63 & 0.53 & 11.77 & -0.24 & 4.77 & -0.113 & -0.20 & 0. & -10.87 & -0.25 & 0. \\
 \text{H5} & -4.71 & 11.10 & 1.35 & 2.98 & 1.44 & 0.79 & -0.48 & -10.87 & 0. & -0.2 & 0. \\
 \text{H6} & 6.43 & 0.97 & -7.35 & -0.05 & -0.66 & 0. & -0.04 & -0.25 & -0.2 & 0. & 0. \\
 \text{H7} & 1.70 & 7.48 & -0.15 & -7.58 & -0.11 & -0.84 & 0.74 & 0. & 0. & 0. & 0. \\
\end{tabular}
\caption{Raw values of J couplings in D-aspartic acid.}
\end{table}

\begin{table}
\scriptsize
\begin{tabular}{cccccccccccc}
 \text{L-a. acid} & \text{C1} & \text{C2} & \text{C3} & \text{C4} & \text{H1} & \text{H2} & \text{H3} & \text{H4} & \text{H5} & \text{H6} & \text{H7} \\
 \text{C1} & 0. & 129.84 & 177.80 & -4.517 & 168.39 & -5.95 & -7.76 & -4.41 & -4.32 & 7.38 & 1.39 \\
 \text{C2} & 129.84 & 0. & -8.26 & 187.36 & -6.90 & 177.36 & 168.44 & 9.42 & -0.44 & 0.20 & 7.52 \\
 \text{C3} & 177.80 & -8.26 & 0. & 0.00 & -10.89 & 3.97 & 8.973 & 3.84 & 10.54 & -7.39 & -0.09 \\
 \text{C4} & -4.52 & 187.36 & 0.003 & 0. & 1.45 & -7.29 & -11.41 & 2.26 & -0.35 & 0.15 & -7.66 \\
 \text{H1} & 168.39 & -6.90 & -10.89 & 1.45 & 0. & 11.144 & 2.99 & -0.35 & 8.14 & -0.77 & -0.09 \\
 \text{H2} & -5.95 & 177.36 & 3.97 & -7.29 & 11.144 & 0. & -17.58 & -0.43 & -0.17 & -0.07 & 1. \\
 \text{H3} & -7.76 & 168.44 & 8.97 & -11.41 & 2.99 & -17.58 & 0. & 0.92 & -0.17 & -0.06 & -0.58 \\
 \text{H4} & -4.41 & 9.42 & 3.84 & 2.26 & -0.35 & -0.43 & 0.92 & 0. & -10.07 & 0.04 & 0. \\
 \text{H5} & -4.32 & -0.44 & 10.54 & -0.35 & 8.14 & -0.17 & -0.17 & -10.07 & 0. & 0. & 0. \\
 \text{H6} & 7.38 & 0.20 & -7.391 & 0.147 & -0.77 & -0.07 & -0.06 & 0.04 & 0. & 0. & -0.04 \\
 \text{H7} & 1.39 & 7.52 & -0.09 & -7.66 & -0.09 & 1. & -0.58 & 0. & 0. & -0.04 & 0. \\
\end{tabular}
\caption{Raw values of J couplings in L-aspartic acid.}
\end{table}

\begin{table}
\scriptsize
\begin{tabular}{ccccccccccc}
 \text{Cysteine} & \text{C1} & \text{C2} & \text{C3} & \text{H1} & \text{H2} & \text{H3} & \text{H4} & \text{H5} & \text{H6} & \text{H7} \\
 \text{C1} & \text{0.} & 0.17 & -0.03 & -0.03 & -0.17 & 0.47 & -0.11 & 0.05 & -2.04 & -0.28 \\
 \text{C2} & 0.17 & \text{0.} & -0.05 & 0.35 & -0.03 & 0.00 & 1.75 & -1.01 & 0.55 & 0.27 \\
 \text{C3} & -0.03 & -0.05 & \text{0.} & 0.12 & 1.33 & -0.89 & -0.22 & -0.02 & 0.30 & -0.08 \\
 \text{H1} & -0.03 & 0.35 & 0.12 & \text{0.} & -1.23 & 0.90 & -1.63 & 0.77 & 6.31 & -4.53 \\
 \text{H2} & -0.17 & -0.03 & 1.33 & -1.23 & \text{0.} & 0.07 & 3.44 & -1.22 & -0.04 & -2. \\
 \text{H3} & 0.47 & 0.00 & -0.89 & 0.90 & 0.07 & \text{0.} & 0.72 & 0.27 & 0.44 & 2. \\
 \text{H4} & -0.11 & 1.75 & -0.22 & -1.63 & 3.44 & 0.72 & \text{0.} & 0.14 & -2. & 0.97 \\
 \text{H5} & 0.05 & -1.01 & -0.02 & 0.77 & -1.22 & 0.27 & 0.14 & \text{0.} & -92.67 & -4.21 \\
 \text{H6} & -2.04 & 0.55 & 0.30 & 6.31 & -0.04 & 0.44 & -2. & -92.67 & \text{0.} & 3.18 \\
 \text{H7} & -0.28 & 0.27 & -0.08 & -4.53 & -2. & 2. & 0.97 & -4.21 & 3.18 & \text{0.} \\
\end{tabular}
\caption{Numerical values of the fractional deviation $[J(L)-J(D)]/[J(L)+J(D)]/2$ for cysteine.}
\end{table}

\begin{table}
\scriptsize
\begin{tabular}{ccccccccccc}
 \text{D-cysteine} &  \text{C1} & \text{C2} & \text{C3} & \text{H1} & \text{H2} & \text{H3} & \text{H4} & \text{H5} & \text{H6} & \text{H7} \\
 \text{C1} & 0. & 124.59 & 182.12 & 176.20 & -6.93 & -3.33 & -5.08 & -4.191 & 1.39 & 8.61 \\
 \text{C2} & 124.59 & 0. & -7.81 & -4.02 & 191.79 & 196.98 & 0.85 & 1.40 & -0.78 & 0.84 \\
 \text{C3} & 182.12 & -7.81 & 0. & -7.46 & 2.35 & 9.22 & 1.56 & 11.69 & -0.141 & -8.00 \\
 \text{H1} & 176.20 & -4.02 & -7.46 & 0. & 11.54 & 3.83 & 15.35 & 2.02 & 0.137 & 1.20 \\
 \text{H2} & -6.93 & 191.79 & 2.35 & 11.54 & 0. & -13.87 & -0.14 & -0.24 & 14.57 & -0.06 \\
 \text{H3} & -3.33 & 196.98 & 9.22 & 3.83 & -13.87 & 0. & -0.19 & -0.10 & 3.72 & 0. \\
 \text{H4} & -5.08 & 0.85 & 1.56 & 15.35 & -0.14 & -0.19 & 0. & -9.07 & -0.05 & -0.05 \\
 \text{H5} & -4.19 & 1.40 & 11.69 & 2.02 & -0.24 & -0.10 & -9.07 & 0. & 0.07 & 0.64 \\
 \text{H6} & 1.39 & -0.78 & -0.14 & 0.14 & 14.57 & 3.72 & -0.05 & 0.07 & 0. & -0.03 \\
 \text{H7} & 8.61 & 0.84 & -8.00 & 1.20 & -0.06 & 0. & -0.05 & 0.64 & -0.03 & 0. \\
\end{tabular}
\caption{Raw values of J couplings in D-cysteine.}
\end{table}

\begin{table}
\scriptsize
\begin{tabular}{ccccccccccc}
 \text{L-cysteine} &  \text{C1} & \text{C2} & \text{C3} & \text{H1} & \text{H2} & \text{H3} & \text{H4} & \text{H5} & \text{H6} & \text{H7} \\
 \text{C1} & 0. & 147.52 & 176.80 & 170.50 & -5.86 & -5.40 & -4.57 & -4.43 & -0.01 & 6.49 \\
 \text{C2} & 147.52 & 0. & -7.44 & -5.73 & 185.64 & 197.27 & 12.98 & 0.46 & -1.37 & 1.10 \\
 \text{C3} & 176.80 & -7.44 & 0. & -8.38 & 11.61 & 3.56 & 1.248 & 11.42 & -0.19 & -7.36 \\
 \text{H1} & 170.50 & -5.73 & -8.38 & 0. & 2.74 & 10.15 & 1.56 & 4.56 & -0.26 & -0.46 \\
 \text{H2} & -5.86 & 185.64 & 11.61 & 2.74 & 0. & -14.93 & 0.51 & -0.06 & 14.04 & 0. \\
 \text{H3} & -5.40 & 197.27 & 3.56 & 10.15 & -14.93 & 0. & -0.41 & -0.13 & 5.84 & -0.07 \\
 \text{H4} & -4.57 & 12.98 & 1.25 & 1.56 & 0.51 & -0.41 & 0. & -10.47 & 0. & -0.15 \\
 \text{H5} & -4.43 & 0.46 & 11.42 & 4.56 & -0.06 & -0.13 & -10.47 & 0. & -0.07 & -0.23 \\
 \text{H6} & -0.01 & -1.37 & -0.19 & -0.26 & 14.04 & 5.84 & 0. & -0.07 & 0. & 0.11 \\
 \text{H7} & 6.49 & 1.10 & -7.36 & -0.46 & 0. & -0.07 & -0.15 & -0.23 & 0.11 & 0. \\
\end{tabular}
\caption{Raw values of J couplings in L-cysteine.}
\end{table}

\begin{table}
\tiny
\begin{tabular}{ccccccccccccccc}
 \text{Glutamic acid} & \text{C1} & \text{C2} & \text{C3} & \text{C4} & \text{C5} & \text{H1} & \text{H2} & \text{H3} & \text{H4} & \text{H5} & \text{H6} & \text{H7} & \text{H8} & \text{H9} \\
 \text{C1} & \text{0.} & -0.01 & 0.01 & 0.03 & -0.03 & 0.00 & 0.00 & 0.02 & 0.27 & -0.28 & 2.15 & -2.14 & -0.99 & 0.02 \\
 \text{C2} & -0.01 & \text{0.} & -0.00 & -0.00 & -0.01 & 0.522823 & -0.55 & 0.01 & 1.03 & -1.06 & 0.10 & -0.13 & -0.00 & \text{0.} \\
 \text{C3} & 0.01 & -0.00 & \text{0.} & 0.07 & 0.00 & 0.06 & -0.05 & 0.01 & 0.01 & -0.01 & 0.25 & -0.32 & -3.33 & -0.01 \\
 \text{C4} & 0.03 & -0.00 & 0.07 & \text{0.} & -0.18 & 1.77 & -1.80 & -0.02 & 1.13 & -0.94 & -0.84 & 0.85 & -0.01 & \text{0.} \\
 \text{C5} & -0.03 & -0.01 & 0.00 & -0.18 & \text{0.} & 1.07 & -1.12 & 0.02 & 0.15 & -0.14 & -2. & 2. & \text{0.} & -0.01 \\
 \text{H1} & 0.00 & 0.52 & 0.06 & 1.77 & 1.07 & \text{0.} & -0.01 & 0.54 & -0.65 & 0.03 & -16.4 & -2.30 & -10.07 & -0.06 \\
 \text{H2} & 0.00 & -0.55 & -0.05 & -1.80 & -1.12 & -0.01 & \text{0.} & -0.53 & -0.04 & 0.62 & 2.31 & 13.79 & 8.73 & 0.17 \\
\end{tabular}
\caption{Numerical values of the fractional deviation $[J(L)-J(D)]/[J(L)+J(D)]/2$ for glutamic acid.}
\end{table}

\begin{table}
\tiny
\colorbox{white}{
\begin{tabular}{ccccccccccccccc}
 \text{D-glutamic acid} & \text{C1} & \text{C2} & \text{C3} & \text{C4} & \text{C5} & \text{H1} & \text{H2} & \text{H3} & \text{H4} & \text{H5} & \text{H6} & \text{H7} & \text{H8} & \text{H9} \\
 \text{C1} & 0. & 122.22 & 141.61 & -8.23 & -5.45 & 171.17 & 171.25 & -6.04 & -5.07 & -6.90 & -0.28 & 7.81 & 0.13 & 1.71 \\
 \text{C2} & 122.23 & 0. & -6.88 & 170.56 & 8.55 & -3.85 & -6.56 & 162.35 & 0.99 & 3.07 & -3.95 & -4.44 & 7.29 & 0. \\
 \text{C3} & 141.61 & -6.88 & 0. & 1.43 & 194.08 & -6.64 & -7.10 & 7.19 & 166.41 & 167.81 & -0.20 & -0.25 & -0.01 & 7.01 \\
 \text{C4} & -8.23 & 170.56 & 1.43 & 0. & -0.15 & 0.68 & 11.13 & -10.79 & -0.04 & -0.13 & 10.34 & 4.07 & -7.59 & 0. \\
 \text{C5} & -5.45 & 8.55 & 194.08 & -0.15 & 0. & 0.74 & 2.37 & 1.10 & -9.70 & -11.46 & -0.02 & 0. & 0. & -7.34 \\
 \text{H1} & 171.17 & -3.85 & -6.64 & 0.68 & 0.74 & 0. & -15.16 & 2.76 & 4.70 & 12.97 & -0.46 & 2.421 & 0.18 & -0.14 \\
 \text{H2} & 171.25 & -6.56 & -7.10 & 11.13 & 2.37 & -15.16 & 0. & 4.82 & 13.47 & 2.52 & -0.171 & 0.34 & -0.11 & -0.13 \\
 \text{H3} & -6.04 & 162.35 & 7.19 & -10.79 & 1.10 & 2.76 & 4.82 & 0. & -0.41 & -0.47 & 7.38 & -0.34 & -0.79 & 0. \\
 \text{H4} & -5.07 & 0.99 & 166.41 & -0.04 & -9.70 & 4.70 & 13.47 & -0.41 & 0. & -19.84 & -0.05 & -0.03 & -0.02 & 0.36 \\
 \text{H5} & -6.90 & 3.07 & 167.81 & -0.13 & -11.46 & 12.97 & 2.52 & -0.47 & -19.84 & 0. & 0.05 & 0.13 & -0.04 & -0.76 \\
 \text{H6} & -0.28 & -3.95 & -0.20 & 10.34 & -0.02 & -0.46 & -0.17 & 7.38 & -0.05 & 0.05 & 0. & -9.27 & 0. & 0. \\
 \text{H7} & 7.81 & -4.44 & -0.25 & 4.07 & 0. & 2.42 & 0.34 & -0.34 & -0.03 & 0.13 & -9.27 & 0. & 0.07 & 0. \\
 \text{H8} & 0.13 & 7.29 & -0.01 & -7.59 & 0. & 0.18 & -0.11 & -0.79 & -0.02 & -0.04 & 0. & 0.069 & 0. & 0. \\
 \text{H9} & 1.71 & 0. & 7.01 & 0. & -7.34 & -0.14 & -0.13 & 0. & 0.36 & -0.76 & 0. & 0. & 0. & 0. \\
\end{tabular}}
\caption{Raw values of J couplings in D-glutamic acid.}
\end{table}

\begin{table}
\tiny
\colorbox{white}{
\begin{tabular}{ccccccccccccccc}
 \text{L-glutamic acid} & \text{C1} & \text{C2} & \text{C3} & \text{C4} & \text{C5} & \text{H1} & \text{H2} & \text{H3} & \text{H4} & \text{H5} & \text{H6} & \text{H7} & \text{H8} & \text{H9} \\
 \text{C1} & 0. & 121.14 & 142.46 & -8.51 & -5.31 & 171.58 & 171.37 & -6.19 & -6.67 & -5.18 & 7.69 & -0.26 & 0.04 & 1.75 \\
 \text{C2} & 121.15 & 0. & -6.87 & 169.81 & 8.48 & -6.58 & -3.75 & 163.40 & 3.08 & 0.95 & -4.37 & -3.90 & 7.28 & 0. \\
 \text{C3} & 142.46 & -6.87 & 0. & 1.53 & 194.15 & -7.05 & -6.76 & 7.26 & 168.25 & 165.49 & -0.26 & -0.18 & 0.00 & 6.96 \\
 \text{C4} & -8.51 & 169.81 & 1.53 & 0. & -0.12 & 11.07 & 0.59 & -10.55 & -0.14 & -0.05 & 4.21 & 10.13 & -7.52 & 0. \\
 \text{C5} & -5.31 & 8.48 & 194.15 & -0.12 & 0. & 2.44 & 0.67 & 1.12 & -11.31 & -9.97 & 0. & -0.02 & 0. & -7.25 \\
 \text{H1} & 171.58 & -6.58 & -7.05 & 11.07 & 2.44 & 0. & -15.06 & 4.81 & 2.39 & 13.42 & 0.36 & -0.17 & -0.12 & -0.13 \\
 \text{H2} & 171.37 & -3.75 & -6.76 & 0.59 & 0.67 & -15.06 & 0. & 2.81 & 12.92 & 4.81 & 2.41 & -0.45 & 0.17 & -0.15 \\
 \text{H3} & -6.19 & 163.40 & 7.26 & -10.55 & 1.12 & 4.81 & 2.81 & 0. & -0.47 & -0.40 & -0.33 & 7.43 & -0.74 & 0. \\
 \text{H4} & -6.67 & 3.08 & 168.25 & -0.14 & -11.31 & 2.39 & 12.92 & -0.47 & 0. & -20.06 & 0.15 & 0.05 & -0.04 & -0.67 \\
 \text{H5} & -5.18 & 0.95 & 165.49 & -0.05 & -9.97 & 13.42 & 4.81 & -0.40 & -20.06 & 0. & -0.03 & -0.06 & -0.01 & 0.16 \\
 \text{H6} & 7.69 & -4.37 & -0.26 & 4.21 & 0. & 0.36 & 2.41 & -0.33 & 0.15 & -0.03 & 0. & -9.31 & 0.07 & 0. \\
 \text{H7} & -0.26 & -3.90 & -0.18 & 10.13 & -0.02 & -0.17 & -0.45 & 7.43 & 0.05 & -0.06 & -9.31 & 0. & 0. & 0. \\
 \text{H8} & 0.04 & 7.28 & 0.00 & -7.52 & 0. & -0.12 & 0.17 & -0.74 & -0.04 & -0.01 & 0.07 & 0. & 0. & 0. \\
 \text{H9} & 1.75 & 0. & 6.96 & 0. & -7.25 & -0.13 & -0.15 & 0. & -0.67 & 0.16 & 0. & 0. & 0. & 0. \\
\end{tabular}}
\caption{Raw values of J couplings in L-glutamic acid.}
\end{table}

\begin{table}
\tiny
\colorbox{white}{
\begin{tabular}{cccccccccccccccc}
 \text{Glutamine} & \text{C1} & \text{C2} & \text{C3} & \text{C4} & \text{C5} & \text{H1} & \text{H2} & \text{H3} & \text{H4} & \text{H5} & \text{H6} & \text{H7} & \text{H8} & \text{H9} & \text{H10} \\
 \text{C1} & \text{0.} & 0.00 & -0.00 & -0.00 & -0.00 & -0.02 & 0.02 & 0.01 & -0.01 & 0.02 & -2.12 & 2.12 & -0.04 & -0.03 & 0.06 \\
 \text{C2} & 0.00 & \text{0.} & 0.00 & 0.00 & 0.01 & -0.85 & 0.86 & -0.00 & -0.82 & 0.83 & 0.15 & -0.11 & \text{0.} & \text{0.} & 0.00 \\
 \text{C3} & -0.00 & 0.00 & \text{0.} & 0.07 & 0.00 & -0.02 & 0.01 & 0.01 & 0.00 & 0.00 & -1.75 & 2.11 & -0.07 & 0.00 & 0.21 \\
 \text{C4} & -0.00 & 0.00 & 0.07 & \text{0.} & 0.05 & -1.76 & 1.77 & 0.02 & 20.86 & -8.88 & 2.49 & -2.53 & \text{0.} & \text{0.} & 0.00 \\
 \text{C5} & -0.00 & 0.01 & 0.00 & 0.05 & \text{0.} & 0.61 & -0.55 & 0.03 & -0.14 & 0.12 & 2. & -2. & -0.02 & 0.01 & \text{0.} \\
 \text{H1} & -0.02 & -0.85 & -0.02 & -1.76 & 0.61 & \text{0.} & -0.00 & -0.72 & -0.02 & -0.43 & 0.09 & 2.55 & -0.13 & -0.10 & -0.14 \\
 \text{H2} & 0.02 & 0.86 & 0.01 & 1.76 & -0.55 & -0.00 & \text{0.} & 0.72 & 0.41 & 0.02 & -2.58 & -0.13 & 0.12 & 0.04 & 0.21 \\
\end{tabular}}
\caption{Numerical values of the fractional deviation $[J(L)-J(D)]/[J(L)+J(D)]/2$ for glutamine.}
\end{table}

\begin{table}
\tiny
\colorbox{white}{
\begin{tabular}{cccccccccccccccc}
 \text{D-glu.} & \text{C1} & \text{C2} & \text{C3} & \text{C4} & \text{C5} & \text{H1} & \text{H2} & \text{H3} & \text{H4} & \text{H5} & \text{H6} & \text{H7} & \text{H8} & \text{H9} & \text{H10} \\
 \text{C1} & 0. & 115.62 & 143.58 & -8.22 & -5.91 & 174.57 & 170.77 & -7.46 & -6.37 & -6.47 & 8.71 & -0.25 & 0.55 & 1.85 & -0.41 \\
 \text{C2} & 115.62 & 0. & -6.95 & 157.71 & 6.75 & -6.62 & -2.61 & 170.04 & 2.07 & 0.91 & -5.19 & -5.89 & 0. & 0. & 7.45 \\
 \text{C3} & 143.58 & -6.95 & 0. & 1.20 & 167.80 & -7.00 & -6.90 & 8.23 & 161.94 & 161.8 & 0.15 & -0.00 & -0.56 & 6.76 & 0.05 \\
 \text{C4} & -8.22 & 157.71 & 1.20 & 0. & -0.17 & 9.85 & 0.61 & -7.80 & 0.13 & -0.17 & -0.44 & 3.92 & 0. & 0. & -7.71 \\
 \text{C5} & -5.91 & 6.75 & 167.80 & -0.17 & 0. & 1.12 & 2.01 & 0.40 & -9.38 & -8.61 & 0. & 0.05 & 3.18 & -2.77 & 0. \\
 \text{H1} & 174.57 & -6.62 & -7.00 & 9.85 & 1.12 & 0. & -15.26 & 5.93 & 13.98 & 4.73 & -0.37 & -0.20 & -0.06 & -0.15 & -0.08 \\
 \text{H2} & 170.77 & -2.61 & -6.90 & 0.61 & 2.01 & -15.26 & 0. & 2.714 & 3.09 & 13.70 & 1.68 & -0.40 & -0.05 & -0.14 & -0.08 \\
\end{tabular}}
\caption{Raw values of J couplings in D-glutamine.}
\end{table}

\begin{table}
\tiny
\colorbox{white}{
\hspace{-0.7in}
\begin{tabular}{cccccccccccccccc}
 \text{L-glu.} & \text{C1} & \text{C2} & \text{C3} & \text{C4} & \text{C5} & \text{H1} & \text{H2} & \text{H3} & \text{H4} & \text{H5} & \text{H6} & \text{H7} & \text{H8} & \text{H9} & \text{H10} \\
 \text{C1} & 0. & 115.91 & 143.14 & -8.20 & -5.89 & 171.51 & 174.04 & -7.51 & -6.30 & -6.57 & -0.26 & 8.68 & 0.53 & 1.80 & -0.43 \\
 \text{C2} & 115.91 & 0. & -6.97 & 157.75 & 6.81 & -2.66 & -6.52 & 169.61 & 0.87 & 2.20 & -6.02 & -5.26 & 0. & 0. & 7.47 \\
 \text{C3} & 143.14 & -6.97 & 0. & 1.28 & 167.90 & -6.88 & -6.95 & 8.30 & 162.22 & 161.92 & 0.01 & 0.11 & -0.52 & 6.77 & 0.06 \\
 \text{C4} & -8.20 & 157.75 & 1.28 & 0. & -0.18 & 0.62 & 9.73 & -7.93 & -0.16 & 0.11 & 4.01 & -0.46 & 0. & 0. & -7.71 \\
 \text{C5} & -5.89 & 6.81 & 167.90 & -0.18 & 0. & 2.1 & 1.14 & 0.41 & -8.194 & -9.69 & 0.05 & 0. & 3.12 & -2.78 & 0. \\
 \text{H1} & 171.51 & -2.66 & -6.88 & 0.62 & 2.1 & 0. & -15.22 & 2.79 & 13.65 & 3.07 & -0.41 & 1.68 & -0.05 & -0.14 & -0.07 \\
 \text{H2} & 174.04 & -6.52 & -6.95 & 9.73 & 1.14 & -15.22 & 0. & 5.79 & 4.71 & 13.99 & -0.21 & -0.36 & -0.06 & -0.14 & -0.09 \\
\end{tabular}}
\caption{Raw values of J couplings in L-glutamine.}
\end{table}

\begin{table}
\scriptsize
\centering
\begin{tabular}{cccccccccc}
 \text{Glyceraldehyde} & \text{C1} & \text{C2} & \text{C3} & \text{H1} & \text{H2} & \text{H3} & \text{H4} & \text{H5} & \text{H6} \\
 \text{C1} & \text{0.} & -0.01 & 0.04 & 0.02 & 1.84 & -0.01 & -0.06 & -0.01 & -4.99 \\
 \text{C2} & -0.01 & \text{0.} & 0.05 & 0.86 & 0.04 & 0.00 & 7.37 & 0.01 & -0.04 \\
 \text{C3} & 0.04 & 0.05 & \text{0.} & -0.10 & 0.30 & -0.33 & 0.03 & 0.14 & 1.13 \\
 \text{H1} & 0.02 & 0.86 & -0.10 & \text{0.} & 1.57 & 0.55 & -0.02 & 0.64 & -1.35 \\
 \text{H2} & 1.84 & 0.04 & 0.30 & 1.57 & \text{0.} & 0.01 & 0.27 & -2.69 & -0.06 \\
 \text{H3} & -0.01 & 0.00 & -0.33 & 0.55 & 0.01 & \text{0.} & 0.21 & 0.03 & 1.52 \\
 \text{H4} & -0.06 & 7.37 & 0.03 & -0.02 & 0.27 & 0.21 & \text{0.} & -0.03 & -6. \\
 \text{H5} & -0.01 & 0.01 & 0.14 & 0.64 & -2.69 & 0.03 & -0.03 & \text{0.} & -4.65 \\
 \text{H6} & -4.99 & -0.04 & 1.13 & -1.35 & -0.06 & 1.52 & -6. & -4.65 & \text{0.} \\
 \end{tabular}
\caption{Numerical values of the fractional deviation $[J(L)-J(D)]/[J(L)+J(D)]/2$ for glyceraldehyde.}
\end{table}

\begin{table}
\scriptsize
\centering
\begin{tabular}{cccccccccc}
 \text{D-glyceraldeyde} & \text{C1} & \text{C2} & \text{C3} & \text{H1} & \text{H2} & \text{H3} & \text{H4} & \text{H5} & \text{H6} \\
 \text{C1} & 0. & 133.69 & 152.46 & 170.72 & -0.19 & -5.11 & 25.77 & -3.87 & 1.54 \\
 \text{C2} & 133.69 & 0. & -8.32 & -3.82 & 177.45 & 187.05 & 0.16 & 3.54 & -4.82 \\
 \text{C3} & 152.46 & -8.32 & 0. & -6.39 & 2.57 & 10.64 & 223.53 & 3.88 & -0.07 \\
 \text{H1} & 170.72 & -3.82 & -6.40 & 0. & 1.37 & 3.84 & -0.66 & 1.57 & -0.70 \\
 \text{H2} & -0.19 & 177.45 & 2.57 & 1.37 & 0. & -13.24 & -0.04 & 1.69 & 15.05 \\
 \text{H3} & -5.11 & 187.05 & 10.64 & 3.84 & -13.24 & 0. & -0.35 & -0.20 & 0.71 \\
 \text{H4} & 25.77 & 0.16 & 223.53 & -0.66 & -0.04 & -0.35 & 0. & 1.51 & -0.07 \\
 \text{H5} & -3.87 & 3.54 & 3.88 & 1.57 & 1.69 & -0.20 & 1.511 & 0. & 0.23 \\
 \text{H6} & 1.54 & -4.82 & -0.07 & -0.70 & 15.05 & 0.71 & -0.072 & 0.23 & 0. \\
\end{tabular}
\caption{Raw values of J couplings in D-glyceraldehyde.}
\end{table}

\begin{table}
\scriptsize
\centering
\begin{tabular}{cccccccccc}
 \text{L-glyceraldehyde} & \text{C1} & \text{C2} & \text{C3} & \text{H1} & \text{H2} & \text{H3} & \text{H4} & \text{H5} & \text{H6} \\
 \text{C1} & 0. & 132.82 & 158.41 & 174.11 & -4.69 & -5.08 & 24.36 & -3.84 & -0.66 \\
 \text{C2} & 132.82 & 0. & -8.78 & -9.53 & 184.95 & 187.88 & -0.28 & 3.59 & -4.65 \\
 \text{C3} & 158.41 & -8.78 & 0. & -5.81 & 3.48 & 7.66 & 231.02 & 4.46 & -0.25 \\
 \text{H1} & 174.11 & -9.53 & -5.81 & 0. & 11.40 & 6.73 & -0.65 & 3.05 & -0.14 \\
 \text{H2} & -4.69 & 184.95 & 3.48 & 11.40 & 0. & -13.35 & -0.05 & -0.25 & 14.18 \\
 \text{H3} & -5.08 & 187.88 & 7.66 & 6.73 & -13.35 & 0. & -0.43 & -0.21 & 5.20 \\
 \text{H4} & 24.36 & -0.28 & 231.02 & -0.65 & -0.05 & -0.43 & 0. & 1.47 & 0.04 \\
 \text{H5} & -3.84 & 3.59 & 4.46 & 3.05 & -0.25 & -0.21 & 1.47 & 0. & -0.09 \\
 \text{H6} & -0.66 & -4.65 & -0.25 & -0.14 & 14.18 & 5.20 & 0.04 & -0.09 & 0. \\
\end{tabular}
\caption{Raw values of J couplings in L-glyceraldehyde.}
\end{table}

\begin{table}
\tiny
\centering
\begin{tabular}{ccccccccc}
 \text{Methionine} & \text{C1} & \text{C2} & \text{C3} & \text{C4} & \text{C5} & \text{H1} & \text{H2} & \text{H3}  \\
 \text{C1} & \text{0.} & 0.21 & 0.00 & -0.33 & 0.60 & 0.07 & -0.06 & 0.56  \\
 \text{C2} & 0.21 & \text{0.} & -0.36 & -0.02 & 0.25 & 0.03 & 0.01 & -0.07  \\
 \text{C3} & 0.00 & -0.36 & \text{0.} & 1.79 & -0.15 & 0.68 & -0.63 & 1.11  \\
 \text{C4} & -0.33 & -0.02 & 1.79 & \text{0.} & \text{0.} & 0.76 & -0.69 & 0.23  \\
 \text{C5} & 0.60 & 0.25 & -0.15 & \text{0.} & \text{0.} & 0.57 & 0.29 & -1.06  \\
 \text{H1} & 0.07 & 0.03 & 0.68 & 0.76 & 0.57 & \text{0.} & 0.10 & -1.70  \\
 \text{H2} & -0.06 & 0.01 & -0.63 & -0.69 & 0.29 & 0.10 & \text{0.} & 0.90  \\
\end{tabular}
\caption{Numerical values of the fractional deviation $[J(L)-J(D)]/[J(L)+J(D)]/2$ for methionine.}
\end{table}

\begin{table}
\tiny
\centering
\begin{tabular}{ccccccccc}
 \text{D-methionine} & \text{C1} & \text{C2} & \text{C3} & \text{C4} & \text{C5} & \text{H1} & \text{H2} & \text{H3}  \\
 \text{C1} & 0. & 104.42 & 138.28 & -8.18 & 0.84 & 167.30 & 172.23 & -4.79  \\
 \text{C2} & 104.42 & 0. & -7.22 & 174.89 & -0.12 & -6.04 & -6.06 & 178.22  \\
 \text{C3} & 138.28 & -7.22 & 0. & 0.08 & -1.07 & -3.99 & -8.52 & 1.48  \\
 \text{C4} & -8.18 & 174.89 & 0.08 & 0. & 0. & 2.57 & 8.86 & -8.26  \\
 \text{C5} & 0.84 & -0.12 & -1.07 & 0. & 0. & 0.13 & 0.22 & 0.03  \\
 \text{H1} & 167.30 & -6.04 & -3.99 & 2.57 & 0.13 & 0. & -14.71 & 11.12  \\
 \text{H2} & 172.23 & -6.06 & -8.52 & 8.86 & 0.22 & -14.71 & 0. & 3.78  \\
\end{tabular}
\caption{Raw values of J couplings in D-methionine.}
\end{table}

\begin{table}
\tiny
\centering
\begin{tabular}{ccccccccc}
 \text{L-methionine} & \text{C1} & \text{C2} & \text{C3} & \text{C4} & \text{C5} & \text{H1} & \text{H2} & \text{H3}  \\
 \text{C1} & 0. & 128.33 & 138.70 & -5.89 & 1.56 & 179.24 & 162.75 & -8.54   \\
 \text{C2} & 128.33 & 0. & -5.00 & 171.07 & -0.16 & -6.21 & -6.11 & 165.85   \\
 \text{C3} & 138.70 & -5.00 & 0. & 1.35 & -0.91 & -8.12 & -4.42 & 5.16   \\
 \text{C4} & -5.89 & 171.07 & 1.35 & 0. & 0. & 5.72 & 4.30 & -10.46   \\
 \text{C5} & 1.56 & -0.16 & -0.91 & 0. & 0. & 0.24 & 0.30 & 0.01   \\
 \text{H1} & 179.24 & -6.21 & -8.12 & 5.72 & 0.24 & 0. & -16.249 & 0.9   \\
 \text{H2} & 162.75 & -6.11 & -4.42 & 4.30 & 0.30 & -16.25 & 0. & 10.03   \\
\end{tabular}
\caption{Raw values of J couplings in L-methionine.}
\end{table}

\begin{table}
\tiny
\centering
\begin{tabular}{ccccccccccc}
 \text{Phenylalanine} & \text{C1} & \text{C2} & \text{C3} & \text{C4} & \text{C5} & \text{C6} & \text{C7} & \text{C8} & \text{C9} & \text{H1}  \\
 \text{C1} & \text{0.} & -0.01 & -0.00 & -0.24 & 0.24 & 0.00 & 0.02 & -0.02 & 0. & 0.01  \\
 \text{C2} & -0.01 & \text{0.} & 0.00 & 0.15 & -0.15 & 0.00 & -0.06 & 0.06 & \text{0.} & -0.04  \\
 \text{C3} & -0.00 & 0.00 & \text{0.} & -0.00 & 0.00 & -0.02 & -0.04 & 0.04 & 0.00 & -0.18  \\
 \text{C4} & -0.24 & 0.15 & -0.00 & \text{0.} & -0.01 & 0.53 & -0.01 & -0.03 & 0.04 & 0.09  \\
 \text{C5} & 0.24 & -0.15 & 0.00 & -0.01 & \text{0.} & -0.51 & 0.03 & 0.01 & -0.04 & 0.16  \\
 \text{C6} & 0.00 & 0.00 & -0.02 & 0.53 & -0.51 & \text{0.} & -0.11 & 0.14 & \text{0.} & -1.26  \\
 \text{C7} & 0.02 & -0.06 & -0.04 & -0.01 & 0.03 & -0.11 & \text{0.} & -0.00 & 0.00 & -1.45  \\
\end{tabular}
\caption{Numerical values of the fractional deviation $[J(L)-J(D)]/[J(L)+J(D)]/2$ for phenylalanine.}
\end{table}

\begin{table}
\tiny
\centering
\begin{tabular}{ccccccccccc}
 \text{D-ph.} & \text{C1} & \text{C2} & \text{C3} & \text{C4} & \text{C5} & \text{C6} & \text{C7} & \text{C8} & \text{C9} & \text{H1}  \\
 \text{C1} & 0. & 94.48 & 153.16 & -4.33 & -3.40 & -7.30 & 3.51 & 3.58 & -1.17 & 170.41   \\
 \text{C2} & 94.48 & 0. & -7.72 & 2.10 & 2.44 & 177.46 & -1.40 & -1.32 & 0. & -7.13   \\
 \text{C3} & 153.16 & -7.72 & 0. & 249.36 & 248.63 & 0.54 & -3.35 & -3.22 & 8.65 & -7.99   \\
 \text{C4} & -4.33 & 2.10 & 249.36 & 0. & -3.88 & -0.05 & 275.77 & 8.16 & -3.16 & 4.07   \\
 \text{C5} & -3.40 & 2.44 & 248.63 & -3.88 & 0. & -0.09 & 7.96 & 273.85 & -3.28 & 6.31   \\
 \text{C6} & -7.30 & 177.46 & 0.54 & -0.05 & -0.09 & 0. & 0.04 & 0.03 & 0. & 9.04   \\
 \text{C7} & 3.51 & -1.40 & -3.35 & 275.77 & 7.96 & 0.04 & 0. & -2.20 & 265.94 & -0.63   \\
\end{tabular}
\caption{Raw values of J couplings in D-phenylalanine.}
\end{table}

\begin{table}
\tiny
\begin{tabular}{ccccccccccc}
 \text{L-ph.} &  \text{C1} & \text{C2} & \text{C3} & \text{C4} & \text{C5} & \text{C6} & \text{C7} & \text{C8} & \text{C9} & \text{H1}  \\
 \text{C1} & 0.00 & 94.00 & 153.04 & -3.41 & -4.33 & -7.33 & 3.58 & 3.51 & -1.17 & 172.23   \\
 \text{C2} & 93.99 & 0.00 & -7.75 & 2.43 & 2.10 & 177.57 & -1.32 & -1.40 & 0.00 & -6.86   \\
 \text{C3} & 153.04 & -7.75 & 0.00 & 248.85 & 249.40 & 0.53 & -3.23 & -3.35 & 8.65 & -6.65   \\
 \text{C4} & -3.41 & 2.43 & 248.85 & 0.00 & -3.85 & -0.09 & 273.88 & 7.95 & -3.28 & 4.46   \\
 \text{C5} & -4.33 & 2.10 & 249.40 & -3.85 & 0.00 & -0.06 & 8.16 & 275.73 & -3.16 & 7.44   \\
 \text{C6} & -7.33 & 177.57 & 0.53 & -0.09 & -0.06 & 0.00 & 0.03 & 0.04 & 0.00 & 2.06   \\
 \text{C7} & 3.57 & -1.32 & -3.23 & 273.88 & 8.16 & 0.03 & 0.00 & -2.20 & 267.17 & -0.10   \\
\end{tabular}
\caption{Raw values of J couplings in L-phenylalanine.}
\end{table}

\begin{table}
\centering
\begin{tabular}{ccccccccccc}
 \text{Serine} & \text{C1} & \text{C2} & \text{C3} & \text{H1} & \text{H2} & \text{H3} & \text{H4} & \text{H5} & \text{H6} & \text{H7} \\
 \text{C1} & 0.00 & 0.17 & -0.06 & -0.04 & 0.54 & 0.06 & 0.09 & 0.51 & -1.65 & 0.14 \\
 \text{C2} & 0.17 & 0.00 & -0.18 & -0.00 & 0.04 & -0.01 & -2.24 & 1.14 & -0.01 & -2.91 \\
 \text{C3} & -0.06 & -0.18 & 0.00 & -0.57 & 0.88 & -1.26 & -0.38 & -2.04 & 0.27 & 0.08 \\
 \text{H1} & -0.04 & -0.00 & -0.57 & 0.00 & -0.48 & 0.78 & 0.11 & 2.14 & -1.45 & 2.13 \\
 \text{H2} & 0.54 & 0.04 & 0.88 & -0.48 & 0.00 & 0.03 & 0.23 & 0.74 & 1.61 & 1.83 \\
 \text{H3} & 0.06 & -0.01 & -1.26 & 0.78 & 0.03 & 0.00 & -0.44 & -0.51 & 0.07 & -128.00 \\
 \text{H4} & 0.09 & -2.24 & -0.38 & 0.11 & 0.23 & -0.44 & 0.00 & 0.25 & -0.22 & 0.56 \\
 \text{H5} & 0.51 & 1.14 & -2.04 & 2.14 & 0.74 & -0.51 & 0.25 & 0.00 & 1.34 & 2.00 \\
 \text{H6} & -1.65 & -0.01 & 0.27 & -1.45 & 1.61 & 0.07 & -0.22 & 1.34 & 0.00 & -1.21 \\
 \text{H7} & 0.14 & -2.91 & 0.08 & 2.13 & 1.83 & -128.00 & 0.56 & 2.00 & -1.21 & 0.00 \\
 \end{tabular}
\caption{Numerical values of the fractional deviation $[J(L)-J(D)]/[J(L)+J(D)]/2$ for serine.}
\end{table}

\begin{table}
\scriptsize
\centering
\begin{tabular}{ccccccccccc}
 \text{D-serine} & \text{C1} & \text{C2} & \text{C3} & \text{H1} & \text{H2} & \text{H3} & \text{H4} & \text{H5} & \text{H6} & \text{H7} \\
 \text{C1} & 0.00 & 104.06 & 180.41 & 176.84 & -1.52 & -5.54 & -4.93 & -2.84 & 1.24 & 6.61 \\
 \text{C2} & 104.06 & 0.00 & -9.32 & -8.32 & 178.77 & 184.34 & -1.25 & 3.72 & -4.49 & -0.44 \\
 \text{C3} & 180.41 & -9.32 & 0.00 & -8.88 & 3.18 & 9.28 & 4.48 & 10.84 & -0.07 & -7.22 \\
 \text{H1} & 176.84 & -8.32 & -8.88 & 0.00 & 10.24 & 4.99 & 12.81 & -0.16 & -0.27 & -0.02 \\
 \text{H2} & -1.52 & 178.77 & 3.18 & 10.24 & 0.00 & -12.13 & -0.15 & -0.23 & 0.19 & 0.01 \\
 \text{H3} & -5.54 & 184.34 & 9.28 & 4.99 & -12.13 & 0.00 & -0.20 & -0.36 & 14.93 & 0.06 \\
 \text{H4} & -4.93 & -1.25 & 4.48 & 12.81 & -0.15 & -0.20 & 0.00 & -8.57 & -0.03 & -0.09 \\
 \text{H5} & -2.84 & 3.72 & 10.84 & -0.16 & -0.23 & -0.36 & -8.57 & 0.00 & -0.04 & 0.00 \\
 \text{H6} & 1.24 & -4.49 & -0.07 & -0.27 & 0.19 & 14.93 & -0.03 & -0.04 & 0.00 & -0.16 \\
 \text{H7} & 6.61 & -0.44 & -7.22 & -0.02 & 0.01 & 0.06 & -0.09 & 0.00 & -0.16 & 0.00 \\
\end{tabular}
\caption{Raw values of J couplings in D-serine.}
\end{table}

\begin{table}
\scriptsize
\centering
\begin{tabular}{ccccccccccc}
 \text{L-serine} & \text{C1} & \text{C2} & \text{C3} & \text{H1} & \text{H2} & \text{H3} & \text{H4} & \text{H5} & \text{H6} & \text{H7} \\
 \text{C1} & 0.00 & 123.13 & 170.37 & 170.10 & -2.65 & -5.86 & -5.42 & -4.77 & 0.12 & 7.64 \\
 \text{C2} & 123.13 & 0.00 & -7.76 & -8.31 & 186.71 & 182.41 & 0.07 & 13.50 & -4.46 & 0.08 \\
 \text{C3} & 170.37 & -7.76 & 0.00 & -4.93 & 8.17 & 2.11 & 3.06 & -0.11 & -0.09 & -7.81 \\
 \text{H1} & 170.10 & -8.31 & -4.93 & 0.00 & 6.25 & 11.44 & 14.34 & 4.69 & -0.04 & 0.60 \\
 \text{H2} & -2.65 & 186.71 & 8.17 & 6.25 & 0.00 & -12.44 & -0.19 & -0.51 & 1.80 & 0.32 \\
 \text{H3} & -5.86 & 182.41 & 2.11 & 11.44 & -12.44 & 0.00 & -0.13 & -0.22 & 15.99 & -0.06 \\
 \text{H4} & -5.42 & 0.07 & 3.06 & 14.34 & -0.19 & -0.13 & 0.00 & -11.04 & -0.02 & -0.16 \\
 \text{H5} & -4.77 & 13.50 & -0.11 & 4.69 & -0.51 & -0.22 & -11.04 & 0.00 & -0.21 & -0.04 \\
 \text{H6} & 0.12 & -4.46 & -0.09 & -0.04 & 1.80 & 15.99 & -0.02 & -0.21 & 0.00 & -0.04 \\
 \text{H7} & 7.64 & 0.08 & -7.81 & 0.60 & 0.32 & -0.06 & -0.16 & -0.04 & -0.04 & 0.00 \\
\end{tabular}
\caption{Raw values of J couplings in L-serine.}
\end{table}

\begin{table}
\scriptsize
\centering
\begin{tabular}{cccccccccccccc}
 \text{Threonine} & \text{C1} & \text{C2} & \text{C3} & \text{C4} & \text{H1} & \text{H2} & \text{H3} & \text{H4} & \text{H5} & \text{H6} & \text{H7} & \text{H8} & \text{H9} \\
 \text{C1} & 0.00 & 0.11 & 0.05 & -2.42 & -5.61 & -0.02 & -0.76 & 1.62 & -0.18 & 3.75 & 1.64 & -0.91 & -1.55 \\
 \text{C2} & 0.11 & 0.00 & 0.62 & 0.29 & 5.01 & 0.74 & -1.69 & 1.98 & 0.85 & -0.07 & 1.72 & -1.60 & 0.20 \\
 \text{C3} & 0.05 & 0.62 & 0.00 & 3.13 & -0.58 & 0.22 & 0.03 & 0.01 & -0.01 & -0.36 & -1.80 & 1.81 & 2.00 \\
 \text{C4} & -2.42 & 0.29 & 3.13 & 0.00 & -8.38 & 3.58 & -2.30 & -0.95 & -1.90 & 0.64 & -1.53 & -2.05 & 0.68 \\
 \text{H1} & -5.61 & 5.01 & -0.58 & -8.38 & 0.00 & 2.24 & -2.15 & -1.17 & 3.56 & -0.53 & 0.93 & -1.17 & 0.72 \\
 \text{H2} & -0.02 & 0.74 & 0.22 & 3.58 & 2.24 & 0.00 & -1.08 & 1.31 & 0.06 & -1.51 & 0.05 & -0.87 & 1.43 \\
 \text{H3} & -0.76 & -1.69 & 0.03 & -2.30 & -2.15 & -1.08 & 0.00 & -0.04 & -0.16 & 3.49 & 3.12 & 2.63 & 2.00 \\
 \text{H4} & 1.62 & 1.98 & 0.01 & -0.95 & -1.17 & 1.31 & -0.04 & 0.00 & -0.01 & -1.09 & 3.24 & -1.81 & 0.00 \\
 \text{H5} & -0.18 & 0.85 & -0.01 & -1.90 & 3.56 & 0.06 & -0.16 & -0.01 & 0.00 & -0.95 & -25.36 & -1.14 & 2.00 \\
 \text{H6} & 3.75 & -0.07 & -0.36 & 0.64 & -0.53 & -1.51 & 3.49 & -1.09 & -0.95 & 0.00 & -0.09 & -1.38 & -1.97 \\
 \text{H7} & 1.64 & 1.72 & -1.80 & -1.53 & 0.93 & 0.05 & 3.12 & 3.24 & -25.36 & -0.09 & 0.00 & 8.91 & 2.00 \\
 \text{H8} & -0.91 & -1.60 & 1.81 & -2.05 & -1.17 & -0.87 & 2.63 & -1.81 & -1.14 & -1.38 & 8.91 & 0.00 & -0.79 \\
 \text{H9} & -1.55 & 0.20 & 2.00 & 0.68 & 0.72 & 1.43 & 2.00 & 0.00 & 2.00 & -1.97 & 2.00 & -0.79 & 0.00 \\
 \end{tabular}
\caption{Numerical values of the fractional deviation $[J(L)-J(D)]/[J(L)+J(D)]/2$ for threonine.}
\end{table}

\begin{table}
\scriptsize
\colorbox{white}{
\hspace{-0.7in}
\begin{tabular}{cccccccccccccc}
 \text{D-threonine} &  \text{C1} & \text{C2} & \text{C3} & \text{C4} & \text{H1} & \text{H2} & \text{H3} & \text{H4} & \text{H5} & \text{H6} & \text{H7} & \text{H8} & \text{H9} \\
 \text{C1} & 0.00 & 97.40 & 141.88 & 79.49 & 17.04 & 186.12 & -6.07 & -0.76 & -9.84 & -0.47 & 0.42 & -11.15 & -2.34 \\
 \text{C2} & 97.40 & 0.00 & -3.99 & 129.45 & -71.73 & -1.28 & 6.63 & 0.05 & 1.24 & -4.85 & -0.32 & 6.55 & 5.69 \\
 \text{C3} & 141.88 & -3.99 & 0.00 & -0.45 & 11.35 & -4.19 & 170.84 & 172.58 & 176.55 & -0.14 & -0.41 & 0.48 & 0.00 \\
 \text{C4} & 79.49 & 129.45 & -0.45 & 0.00 & 13.77 & -2.45 & 1.54 & -0.53 & 2.34 & 3.31 & 10.45 & 10.95 & -3.89 \\
 \text{H1} & 17.04 & -71.73 & 11.35 & 13.77 & 0.00 & -0.32 & 2.39 & 4.59 & 0.11 & 10.56 & 3.54 & -0.35 & 0.13 \\
 \text{H2} & 186.12 & -1.28 & -4.19 & -2.45 & -0.32 & 0.00 & 5.68 & 1.01 & 11.81 & -0.18 & -0.34 & 1.92 & 0.03 \\
 \text{H3} & -6.07 & 6.63 & 170.84 & 1.54 & 2.39 & 5.68 & 0.00 & -14.95 & -15.74 & -0.02 & -0.05 & -0.54 & 0.00 \\
 \text{H4} & -0.76 & 0.05 & 172.58 & -0.53 & 4.59 & 1.01 & -14.95 & 0.00 & -13.96 & 0.08 & 0.03 & -0.90 & 0.00 \\
 \text{H5} & -9.84 & 1.24 & 176.55 & 2.34 & 0.11 & 11.81 & -15.74 & -13.96 & 0.00 & -0.14 & -0.17 & -0.50 & 0.00 \\
 \text{H6} & -0.47 & -4.85 & -0.14 & 3.31 & 10.56 & -0.18 & -0.02 & 0.08 & -0.14 & 0.00 & -10.99 & -0.38 & -0.14 \\
 \text{H7} & 0.42 & -0.32 & -0.41 & 10.45 & 3.54 & -0.34 & -0.05 & 0.03 & -0.17 & -10.99 & 0.00 & 0.04 & 0.00 \\
 \text{H8} & -11.15 & 6.55 & 0.48 & 10.95 & -0.35 & 1.92 & -0.54 & -0.90 & -0.50 & -0.38 & 0.04 & 0.00 & -0.21 \\
 \text{H9} & -2.34 & 5.69 & 0.00 & -3.89 & 0.13 & 0.03 & 0.00 & 0.00 & 0.00 & -0.14 & 0.00 & -0.21 & 0.00 \\
\end{tabular}}
\caption{Raw values of J couplings in D-threonine.}
\end{table}

\begin{table}
\scriptsize
\colorbox{white}{
\hspace{-0.7in}
\begin{tabular}{cccccccccccccc}
 \text{L-threonine} & \text{C1} & \text{C2} & \text{C3} & \text{C4} & \text{H1} & \text{H2} & \text{H3} & \text{H4} & \text{H5} & \text{H6} & \text{H7} & \text{H8} & \text{H9} \\
 \text{C1} & 0.00 & 108.73 & 149.19 & -7.48 & -8.08 & 183.24 & -2.73 & -7.20 & -8.21 & 1.54 & 4.27 & -4.15 & -0.30 \\
 \text{C2} & 108.73 & 0.00 & -7.60 & 172.85 & 166.93 & -2.76 & 0.55 & 9.38 & 3.06 & -4.54 & -4.21 & 0.72 & 6.97 \\
 \text{C3} & 149.19 & -7.60 & 0.00 & 2.05 & 6.21 & -5.23 & 175.27 & 174.38 & 174.24 & -0.09 & -0.02 & 9.74 & 0.06 \\
 \text{C4} & -7.48 & 172.85 & 2.05 & 0.00 & -8.46 & 8.64 & -0.11 & -0.19 & 0.06 & 6.46 & 1.38 & -0.14 & -7.87 \\
 \text{H1} & -8.08 & 166.93 & 6.21 & -8.46 & 0.00 & 5.52 & -0.08 & 1.21 & -0.39 & 6.13 & 9.69 & -0.09 & 0.27 \\
 \text{H2} & 183.24 & -2.76 & -5.23 & 8.64 & 5.52 & 0.00 & 1.69 & 4.83 & 12.52 & -0.03 & -0.35 & 0.76 & 0.17 \\
 \text{H3} & -2.73 & 0.55 & 175.27 & -0.11 & -0.08 & 1.69 & 0.00 & -14.32 & -13.45 & 0.08 & 0.24 & 3.96 & -0.05 \\
 \text{H4} & -7.20 & 9.38 & 174.38 & -0.19 & 1.21 & 4.83 & -14.32 & 0.00 & -13.79 & 0.02 & -0.14 & -0.05 & 0.00 \\
 \text{H5} & -8.21 & 3.06 & 174.24 & 0.06 & -0.39 & 12.52 & -13.45 & -13.79 & 0.00 & -0.05 & 0.15 & -0.14 & -0.02 \\
 \text{H6} & 1.54 & -4.54 & -0.09 & 6.46 & 6.13 & -0.03 & 0.08 & 0.02 & -0.05 & 0.00 & -10.00 & -0.07 & -0.00 \\
 \text{H7} & 4.27 & -4.21 & -0.02 & 1.38 & 9.69 & -0.35 & 0.24 & -0.14 & 0.15 & -10.00 & 0.00 & -0.06 & -0.14 \\
 \text{H8} & -4.15 & 0.72 & 9.74 & -0.14 & -0.09 & 0.76 & 3.96 & -0.05 & -0.14 & -0.07 & -0.06 & 0.00 & -0.09 \\
 \text{H9} & -0.30 & 6.97 & 0.06 & -7.87 & 0.27 & 0.17 & -0.05 & 0.00 & -0.02 & -0.00 & -0.14 & -0.09 & 0.00 \\
\end{tabular}}
\caption{Raw values of J couplings in L-threonine.}
\end{table}

\begin{table}
\tiny
\begin{tabular}
{ccccccccccc}
 \text{Tyrosine} & \text{C1} & \text{C2} & \text{C3} & \text{C4} & \text{C5} & \text{C6} & \text{C7} & \text{C8} & \text{C9} & \text{H1}   \\
 \text{C1} & \text{0.00} & -0.00 & -0.00 & -0.23 & 0.23 & 0.01 & 0.02 & -0.02 & 0.00 & 0.01   \\
 \text{C2} & -0.00 & \text{0.00} & -0.00 & 0.20 & -0.20 & -0.00 & -0.10 & 0.09 & \text{0.00} & -0.01   \\
 \text{C3} & -0.00 & -0.00 & \text{0.00} & 0.01 & -0.00 & -0.03 & -0.03 & 0.03 & 0.00 & -0.17   \\
 \text{C4} & -0.23 & 0.20 & 0.01 & \text{0.00} & -0.01 & 0.67 & 0.01 & 0.02 & 5.13 & 0.10   \\
 \text{C5} & 0.23 & -0.20 & -0.00 & -0.01 & \text{0.00} & -0.70 & -0.02 & -0.01 & -5.12 & 0.16   \\
 \text{C6} & 0.01 & -0.00 & -0.03 & 0.67 & -0.70 & \text{0.00} & 0.06 & -0.05 & \text{0.00} & -1.27   \\
 \text{C7} & 0.02 & -0.10 & -0.03 & 0.01 & -0.02 & 0.06 & \text{0.00} & -0.00 & -0.09 & -1.22   \\
\end{tabular}
\caption{Numerical values of the fractional deviation $[J(L)-J(D)]/[J(L)+J(D)]/2$ for tyrosine.}
\end{table}

\begin{table}
\tiny
\centering
\begin{tabular}{ccccccccccc}
 \text{D-tyrosine} & \text{C1} & \text{C2} & \text{C3} & \text{C4} & \text{C5} & \text{C6} & \text{C7} & \text{C8} & \text{C9} & \text{H1}   \\
 \text{C1} & 0. & 95.80 & 155.50 & -3.92 & -3.13 & -7.19 & 3.78 & 3.85 & -1.20 & 169.93 \\
 \text{C2} & 95.80 & 0. & -7.67 & 2.04 & 2.48 & 177.62 & -1.39 & -1.26 & 0. & -6.94   \\
 \text{C3} & 155.50 & -7.67 & 0. & 250.04 & 251.19 & 0.58 & -3.69 & -3.59 & 8.58 & -8.04   \\
 \text{C4} & -3.92 & 2.04 & 250.04 & 0. & -3.26 & -0.06 & 289.06 & 5.78 & 0.79 & 4.01   \\
 \text{C5} & -3.13 & 2.48 & 251.19 & -3.26 & 0. & -0.12 & 5.90 & 291.00 & -1.82 & 6.38   \\
 \text{C6} & -7.19 & 177.62 & 0.58 & -0.06 & -0.12 & 0. & 0.04 & 0.04 & 0. & 9.14   \\
 \text{C7} & 3.78 & -1.39 & -3.69 & 289.06 & 5.90 & 0.04 & 0. & 3.27 & 286.75 & -0.66   \\
\end{tabular}
\caption{Raw values of J couplings in D-tyrosine.}
\end{table}

\begin{table}
\tiny
\begin{tabular}
{ccccccccccc}
 \text{L-tyrosine} & \text{C1} & \text{C2} & \text{C3} & \text{C4} & \text{C5} & \text{C6} & \text{C7} & \text{C8} & \text{C9} & \text{H1}   \\
 \text{C1} & 0. & 95.34 & 155.42 & -3.11 & -3.93 & -7.23 & 3.85 & 3.79 & -1.20 & 171.64  \\
 \text{C2} & 95.34 & 0. & -7.64 & 2.49 & 2.03 & 177.59 & -1.26 & -1.38 & 0. & -6.88  \\
 \text{C3} & 155.42 & -7.64 & 0. & 251.47 & 250.07 & 0.57 & -3.59 & -3.69 & 8.59 & -6.80 \\
 \text{C4} & -3.11 & 2.49 & 251.47 & 0. & -3.24 & -0.12 & 290.91 & 5.90 & -1.81 & 4.44   \\
 \text{C5} & -3.93 & 2.03 & 250.07 & -3.24 & 0. & -0.06 & 5.79 & 289.08 & 0.80 & 7.50   \\
 \text{C6} & -7.23 & 177.59 & 0.57 & -0.12 & -0.06 & 0. & 0.04 & 0.04 & 0. & 2.05   \\
 \text{C7} & 3.85 & -1.26 & -3.59 & 290.91 & 5.79 & 0.04 & 0. & 3.27 & 262.31 & -0.16   \\
\end{tabular}
\caption{Raw values of J couplings in L-tyrosine.}
\end{table}

\begin{table}
\tiny
\begin{tabular}{ccccccccccccccccc}
 \text{Valine} & \text{C1} & \text{C2} & \text{C3} & \text{C4} & \text{C5} & \text{H1} & \text{H2} & \text{H3} & \text{H4} & \text{H5} & \text{H6} & \text{H7} & \text{H8} & \text{H9} & \text{H10} & \text{H11} \\
 \text{C1} & \text{0.} & -0.00 & -0.03 & 0.03 & -0.00 & 0.00 & 0.00 & -0.12 & 0.09 & -0.05 & -0.10 & 0.05 & 0.12 & 2.17 & -2.11 & 0.03 \\
 \text{C2} & -0.00 & \text{0.} & -0.29 & 0.30 & -0.00 & 0.00 & 0.00 & 0.35 & -0.67 & 0.14 & 0.66 & -0.14 & -0.34 & 0.16 & -0.16 & 0.00 \\
 \text{C3} & -0.03 & -0.29 & \text{0.} & -0.00 & -2.22 & -0.05 & -1.05 & -0.05 & 0.05 & 0.00 & -0.03 & -0.01 & 0.05 & 2.33 & -0.20 & 1.62 \\
 \text{C4} & 0.03 & 0.30 & -0.00 & \text{0.} & 2.21 & 0.04 & 1.05 & -0.05 & 0.04 & 0.01 & -0.05 & -0.00 & 0.05 & 0.19 & -2.29 & -1.59 \\
 \text{C5} & -0.00 & -0.00 & -2.22 & 2.21 & \text{0.} & -0.00 & -0.00 & 0.72 & 1.49 & 0.61 & -1.61 & -0.57 & -0.69 & -0.48 & 0.47 & 0.00 \\
 \text{H1} & 0.00 & 0.00 & -0.05 & 0.04 & -0.00 & \text{0.} & 0.01 & 0.02 & 0.07 & 0.10 & -0.07 & -0.10 & -0.02 & 4.32 & -4.22 & -0.04 \\
 \text{H2} & 0.00 & 0.00 & -1.05 & 1.05 & -0.00 & 0.01 & \text{0.} & -0.31 & -0.61 & -1.53 & 0.65 & 1.50 & 0.37 & -1.88 & 1.96 & -0.09 \\
\end{tabular}
\caption{Numerical values of the fractional deviation $[J(L)-J(D)]/[J(L)+J(D)]/2$ for valine.}
\end{table}

\begin{table}
\tiny
\colorbox{white}{
\hspace{-0.7in}\begin{tabular}{ccccccccccccccccc}
 \text{D-valine} & \text{C1} & \text{C2} & \text{C3} & \text{C4} & \text{C5} & \text{H1} & \text{H2} & \text{H3} & \text{H4} & \text{H5} & \text{H6} & \text{H7} & \text{H8} & \text{H9} & \text{H10} & \text{H11} \\
 \text{C1} & 0. & 108.01 & 127.74 & 124.28 & -7.34 & 165.19 & -6.18 & -5.72 & -6.30 & -5.07 & -6.92 & -4.82 & -5.08 & -0.26 & 6.78 & -0.41 \\
 \text{C2} & 108.01 & 0. & -6.81 & -5.04 & 169.87 & -7.25 & 164.59 & 2.01 & 2.00 & 10.36 & 1.00 & 11.88 & 2.86 & -3.58 & -4.21 & 7.03 \\
 \text{C3} & 127.74 & -6.81 & 0. & -7.76 & 1.81 & -7.22 & 6.10 & 177.56 & 169.34 & 170.24 & 2.69 & 1.22 & 10.72 & -0.09 & -0.29 & 0.03 \\
 \text{C4} & 124.28 & -5.04 & -7.76 & 0. & -0.09 & -6.91 & 1.90 & 11.25 & 2.59 & 1.21 & 177.83 & 170.78 & 169.63 & -0.24 & 1.26 & 0.36 \\
 \text{C5} & -7.34 & 169.87 & 1.81 & -0.09 & 0. & 9.52 & -9.76 & -0.09 & -0.03 & -0.11 & -0.19 & -0.21 & -0.18 & 8.34 & 4.92 & -7.62 \\
 \text{H1} & 165.19 & -7.25 & -7.22 & -6.91 & 9.52 & 0. & 3.23 & 2.65 & 13.23 & 4.31 & 14.17 & 4.79 & 2.70 & -0.25 & 0.72 & -0.10 \\
 \text{H2} & -6.18 & 164.59 & 6.10 & 1.90 & -9.76 & 3.23 & 0. & -0.24 & -0.31 & 1.45 & -0.16 & 0.20 & -0.16 & 8.32 & 0.08 & -0.45 \\
\end{tabular}}
\caption{Raw values of J couplings in D-valine.}
\end{table}

\begin{table}
\tiny
\colorbox{white}{
\hspace{-0.7in}
\begin{tabular}{ccccccccccccccccc}
 \text{L-val.} & \text{C1} & \text{C2} & \text{C3} & \text{C4} & \text{C5} & \text{H1} & \text{H2} & \text{H3} & \text{H4} & \text{H5} & \text{H6} & \text{H7} & \text{H8} & \text{H9} & \text{H10} & \text{H11} \\
 \text{C1} & 0. & 107.75 & 124.48 & 127.99 & -7.32 & 165.43 & -6.20 & -5.10 & -6.92 & -4.82 & -6.29 & -5.07 & -5.74 & 6.48 & -0.18 & -0.42 \\
 \text{C2} & 107.75 & 0. & -5.11 & -6.84 & 169.87 & -7.26 & 164.81 & 2.86 & 1.00 & 11.91 & 2.00 & 10.37 & 2.02 & -4.19 & -3.58 & 7.03 \\
 \text{C3} & 124.48 & -5.11 & 0. & -7.75 & -0.10 & -6.89 & 1.90 & 169.67 & 177.71 & 170.81 & 2.60 & 1.22 & 11.26 & 1.23 & -0.24 & 0.35 \\
 \text{C4} & 127.99 & -6.84 & -7.75 & 0. & 1.87 & -7.22 & 6.15 & 10.73 & 2.70 & 1.23 & 169.34 & 170.34 & 177.63 & -0.29 & -0.09 & 0.04 \\
 \text{C5} & -7.32 & 169.87 & -0.10 & 1.87 & 0. & 9.47 & -9.75 & -0.19 & -0.19 & -0.21 & -0.02 & -0.12 & -0.09 & 5.10 & 7.98 & -7.62 \\
 \text{H1} & 165.43 & -7.26 & -6.89 & -7.22 & 9.47 & 0. & 3.26 & 2.70 & 14.19 & 4.78 & 13.26 & 4.32 & 2.66 & 0.69 & -0.26 & -0.09 \\
 \text{H2} & -6.20 & 164.81 & 1.90 & 6.15 & -9.75 & 3.26 & 0. & -0.18 & -0.16 & 0.19 & -0.32 & 1.45 & -0.24 & 0.25 & 8.52 & -0.41 \\
\end{tabular}}
\caption{Raw values of J couplings in L-valine.}
\label{tab:S13valine}
\end{table}

{\bf References}\par
\bibliographystyle{naturemag}
\bibliography{refs.bib}

\end{document}